\documentclass[12pt]{article}
\pdfoutput=1
\usepackage{cite}

\usepackage[height=8.85in,width=6.45in]{geometry}
\usepackage{booktabs}
\usepackage{epsfig}
\usepackage{amsfonts}
\usepackage{amscd}
\usepackage{latexsym}
\usepackage{amsmath,amssymb}
\usepackage{verbatim}
\usepackage{setspace}
\usepackage[dvipsnames]{xcolor}
\usepackage{fancyhdr}
\usepackage{cite}
\usepackage[svgnames]{xcolor}
\usepackage[colorlinks,citecolor=DarkGreen,linkcolor=FireBrick,linktocpage]{hyperref}
\usepackage{hyperref}
\usepackage{slashed}
\usepackage{multirow}

\usepackage{preamble}
\usepackage{hyperref}
\usepackage{graphicx,color,subcaption}
\usepackage{cite}
\usepackage{mciteplus}
\usepackage{bbm}
\usepackage[english]{babel}
\usepackage{amsthm}
\usepackage{soul}
\usepackage{dsfont}
\usepackage{tcolorbox}

\def\ie{\begin{equation}\begin{aligned}}
\def\fe{\end{aligned}\end{equation}}
\def\NCB{\mathrm{NCB}}
\def\CB{\mathrm{CB}}

\usepackage{mathtools}

\numberwithin{equation}{section}
\hypersetup{pdfborder=0 0 0}
\tcbuselibrary{theorems, skins, breakable}
\newtcbtheorem[number within=section]{claim}{Claim}{
  enhanced,
  colback=violet!5!white,
  colframe=violet!50!black,
  fonttitle=\bfseries,
  attach boxed title to top left={yshift=-2mm, xshift=4mm},
  boxed title style={
    colback=violet!50!black,
    colframe=violet!50!black,
    rounded corners
  },
  rounded corners,
}{cl}

\usepackage{tikz}
\usetikzlibrary{arrows.meta,calc,decorations.markings}
\usetikzlibrary{patterns.meta}
\tikzset{
  silh/.style={black!70, line width=0.6pt},   
  endcap/.style={tline, pattern={Lines[angle=45,distance=2pt,line width=0.3pt]},
                 pattern color=black!45},     
}

\usepackage{enumitem}

\definecolor{ifacecol}{HTML}{D55E00}      
\definecolor{ncbfill}{RGB}{222,233,247}   
\definecolor{sufill}{RGB}{226,240,222}    
\tikzset{
  ncb/.style={fill=ncbfill},
  su/.style={fill=sufill},
  frame/.style={black!55, dashed, line width=0.5pt},
  iface/.style={ifacecol, line width=1.4pt},   
  tline/.style={line width=1.4pt},             
}

\tikzset{
  bstrip/.style={pattern={Lines[angle=45,distance=2pt,line width=0.3pt]},
                 pattern color=black!65},   
}
\tikzset{
  midarrow/.style={postaction={decorate}, decoration={markings,
      mark=at position 0.5 with {\arrow{Stealth[length=6pt,width=5pt]}}}},
}

\tikzset{
arcarrows/.style={arcarrowsat={0.25}{0.75}},   
  arcarrowsat/.style 2 args={postaction={decorate}, decoration={markings,
      mark=at position #1 with {\arrow{Stealth[length=6pt,width=5pt]}},
      mark=at position #2 with {\arrow{Stealth[length=6pt,width=5pt]}}}},
}

\makeatletter
\DeclareFontFamily{OMX}{MnSymbolE}{}
\DeclareSymbolFont{MnLargeSymbols}{OMX}{MnSymbolE}{m}{n}
\SetSymbolFont{MnLargeSymbols}{bold}{OMX}{MnSymbolE}{b}{n}
\DeclareFontShape{OMX}{MnSymbolE}{m}{n}{
    <-6>  MnSymbolE5
   <6-7>  MnSymbolE6
   <7-8>  MnSymbolE7
   <8-9>  MnSymbolE8
   <9-10> MnSymbolE9
  <10-12> MnSymbolE10
  <12->   MnSymbolE12
}{}
\DeclareFontShape{OMX}{MnSymbolE}{b}{n}{
    <-6>  MnSymbolE-Bold5
   <6-7>  MnSymbolE-Bold6
   <7-8>  MnSymbolE-Bold7
   <8-9>  MnSymbolE-Bold8
   <9-10> MnSymbolE-Bold9
  <10-12> MnSymbolE-Bold10
  <12->   MnSymbolE-Bold12
}{}

\let\llangle\@undefined
\let\rrangle\@undefined
\DeclareMathDelimiter{\llangle}{\mathopen}%
                     {MnLargeSymbols}{'164}{MnLargeSymbols}{'164}
\DeclareMathDelimiter{\rrangle}{\mathclose}%
                     {MnLargeSymbols}{'171}{MnLargeSymbols}{'171}
\makeatother
\begin{document}
 	
\begin{titlepage}

\title{All Symmetries and Boundaries of the Two-Dimensional Free Scalar from Continuous Flat-Gauging}

\author{Yichul Choi$^1$, Ho Tat Lam$^{2,3}$, Brandon C. Rayhaun$^1$}

\address{${}^1${\small School of Natural Sciences, Institute for Advanced Study, Princeton, NJ, USA}}
\address{${}^2${\small Leinweber Institute for Theoretical Physics, Stanford University, Stanford, CA, USA}}
\address{${}^3${\small Department of Physics and Astronomy, University of Southern California, Los Angeles, CA, USA}}

\abstract{

We propose a remote detectability criterion for two-dimensional conformal field theories, which asserts that any local operator outside of the product of the left- and right-moving Virasoro algebras can be detected by some topological line defect. We discuss its relationship to existing results in the conformal net literature. We then use this criterion to argue that the known $SU(2)$ moduli space of unitary conformal boundary conditions and the $SO(4)$ moduli space of unitary topological line defects of the $SU(2)_1$ Wess-Zumino-Witten model completely capture all unitary conformal boundaries and unitary topological lines of the theory. Invoking the fact that the non-compact boson conformal field theory can be obtained from $SU(2)_1$ by ``flat-gauging'' a $U(1)$ symmetry, we are also able to completely classify the unitary conformal boundaries and unitary topological line defects of the two-dimensional free scalar field. In particular, we find a 4-dimensional moduli space of topological line defects, most of which were not known previously.

We support our proposed classifications by writing down concrete Lagrangian descriptions of all of the boundaries and lines of the non-compact boson. For every boundary condition, we calculate the corresponding boundary state, and for each pair of boundary conditions, we calculate the corresponding interval Hilbert space. We verify that the annulus partition functions of the theory respect open/closed duality. We also calculate the fusion rules of the topological line defects, their action on local operators, their action on the boundary conditions by parallel fusion, and their corresponding twisted-sector Hilbert spaces. 
}

\end{titlepage}

\eject

\setcounter{tocdepth}{3}

\tableofcontents

\section{Introduction}

The modern viewpoint on symmetries in quantum field theory is that they are implemented by topological defects \cite{Gaiotto:2014kfa}. In a theory with two spacetime dimensions, one is led to study the algebraic properties of its topological \emph{line} defects, which assemble into a mathematical structure known as a tensor category (see e.g.\ \cite{Bhardwaj:2017xup,Chang:2018iay} for pedagogical reviews for physicists).

It is natural to attempt to elucidate the full symmetry structure of a quantum field theory. That is, given some model, one can ask: what is its full category of topological line defects? 

For the $c<1$ unitary minimal models, the answer to this question is understood, at least in principle. An important simplifying feature of this setting is that the minimal models are all rational when organized with respect to their left- and right-moving Virasoro algebras. In particular, this implies that each possesses only finitely many simple topological line defects. Mathematically, their topological lines are described by a structure known as a \emph{fusion category}. For example, the topological line defects of the diagonal minimal models are all recovered as Verlinde lines \cite{Verlinde:1988sn}. The topological line defects of the remaining minimal models can be calculated by applying the ideas of \cite{Fuchs:2002cm}, using the fact that the non-diagonal minimal models can be obtained as orbifolds of the diagonal ones by  invertible or non-invertible symmetries \cite{Kirillov:2001ti,Ostrik:2001xnt,Nakayama:2024msv,Tanaka:2025qou}.

By contrast, to the best of the authors' knowledge, there is not a single compact, unitary 2D conformal field theory with central charge $c>1$ for which the full category of topological line defects is known. Part of the difficulty is that any conformal field theory with $c\geq 1$ is irrational with respect to its left- and right-moving Virasoro algebras. In turn, one expects every such theory to possess infinitely many topological line defects. The precise mathematical formulation of infinite symmetries is still being debated, see e.g.\ \cite{Freed:2009qp,Marin-Salvador:2025stc,Stockall:2025ngz,Jia:2025vrj,Jia:2026vcr} for prior work in this direction. However, most of the results of this paper are intended to be agnostic about which axiomatization is ultimately equilibrated on. Having said this, we occasionally take inspiration from the conformal net literature, which suggests bicommutant categories \cite{Henriques:2015xxa,HenriquesPenneys:2017bicommutant,Henriques:2017conformalnets,HenriquesPenneys:2023commutants,HenriquesNiveditaPenneys:2024complete,Nivedita:2026wws} as a mathematical arena in which our calculations might eventually be made rigorous.

The claim that $c\geq 1$ conformal field theories have infinitely many topological line defects appears to be at odds with the oft-repeated mantra that a generic interacting conformal field theory has no symmetries beyond conformal invariance. One possible reconciliation of these two belief systems is to observe that the topological line defects of a theory with $c\geq 1$ may have infinite quantum dimension, a phenomenon one does not encounter when working only with finite or invertible symmetries. Thus, one may have a revised hope: in a generic interacting theory, most of the infinitely many topological line defects are expected to have infinite quantum dimension.

The $c=1$ conformal moduli space \cite{Ginsparg:1987eb} is an edge case.\footnote{See \cite{Gannon:2026ihc,Carpi:2026itu,Eberhardt:2026qqm} for recent work on the classification of {$c=1$} unitary 2D CFTs.} The theories which live here exhibit much of the richness of $c>1$ conformal field theory (infinitely many topological line defects, many of which have infinite quantum dimension) but, as we will see, certain miracles make calculations far more tractable. In addition to serving as a laboratory for testing general ideas, the $c=1$ moduli space harbors some of the simplest quantum field theories known to humankind---namely, compact and non-compact free scalar fields---which, despite their simplicity, have not yet had their basic observables (like topological line defects and conformal boundaries) fully elucidated. One of the central goals of this paper is to introduce tools for closing this gap.

How does one know when one has found all of the topological line defects or conformal boundary conditions of a theory? To help address this question, we propose a remote detectability criterion, which is partially inspired by rigorous theorems in the conformal net literature \cite{Bischoff:2022fxf}.\footnote{These theorems are about unital, completely positive maps of conformal nets which, at least in the case of finite symmetries, furnish a hypergroup \cite{Bischoff:2016jmy}. Our formulation of remote detectability relies on a recent physical interpretation of such maps \cite{Gannon:2026ttf}, which we sketch in Section \ref{sec:SU(2)1}.} Remote detectability \cite{Lan:2018vjb,Lan:2018bui} is a physical hypothesis, articulated primarily for gapped theories, which asserts that every topological excitation must be detectable remotely by some other topological excitation. The unitarity of the modular S-matrix, i.e.\ the braiding non-degeneracy of anyons in a 3D topological quantum field theory (TQFT), is an example (see e.g.\ Appendix E of \cite{Kitaev:2005hzj}). We advocate that a variation of remote detectability holds in 2D conformal field theories as well. \\

\noindent \textbf{Remote detectability I.} \emph{Every local operator of a unitary 2D conformal field theory  outside of the product of the left- and right-moving Virasoro subalgebras is detected by some topological line defect.}\\

Here, we say that a topological line defect ``detects'' a local operator if the local operator is not transparent to the defect. A local operator is transparent to a topological line defect if the line can be dragged through the operator without changing the value of any correlation function.

\begin{center}
\tikzset{every picture/.style={line width=0.75pt}}   
\begin{tikzpicture}[x=0.75pt,y=0.75pt,yscale=-1,xscale=1]
\draw  [draw opacity=0][fill={rgb, 255:red, 0; green, 0; blue, 0 }  ,fill opacity=1 ] (38.01,70.46) .. controls (38.01,68.81) and (39.35,67.47) .. (41,67.47) .. controls (42.65,67.47) and (43.98,68.81) .. (43.98,70.46) .. controls (43.98,72.11) and (42.65,73.45) .. (41,73.45) .. controls (39.35,73.45) and (38.01,72.11) .. (38.01,70.46) -- cycle ;
\draw [line width=1.5]    (21.33,10.33) -- (21.33,109.97) ;
\draw [line width=1.5]    (131,10.64) .. controls (132.33,41.97) and (154.33,39.3) .. (164.33,45.3) .. controls (174.33,51.3) and (177,69.3) .. (165.67,77.97) .. controls (154.33,86.64) and (131.67,80.64) .. (131.67,109.97) ;
\draw  [draw opacity=0][fill={rgb, 255:red, 0; green, 0; blue, 0 }  ,fill opacity=1 ] (148.01,69.79) .. controls (148.01,68.14) and (149.35,66.81) .. (151,66.81) .. controls (152.65,66.81) and (153.98,68.14) .. (153.98,69.79) .. controls (153.98,71.44) and (152.65,72.78) .. (151,72.78) .. controls (149.35,72.78) and (148.01,71.44) .. (148.01,69.79) -- cycle ;
\draw (37.33,47.4) node [anchor=north west][inner sep=0.75pt]    {$\mathcal{O}( z,\overline{z})$};
\draw (89.33,60.07) node [anchor=north west][inner sep=0.75pt]    {$=$};
\draw (112,47.4) node [anchor=north west][inner sep=0.75pt]    {$\mathcal{O}( z,\overline{z})$};
\end{tikzpicture}
\end{center}

Remote detectability provides a diagnostic for checking whether one has found all the topological line defects of a theory. If the unitary topological line defects one has found are not able to detect some local operator outside of the Virasoro subalgebras, then there are more unitary lines to be found. On the other hand, if they are able to detect all of the local operators outside of the Virasoro subalgebras, remote detectability does not, on its own, allow one to conclude that there are not additional lines. To establish completeness, one requires a kind of converse to remote detectability, which informally asserts that anything smaller than the \emph{full} category of unitary topological line defects will be unable to detect some local operator. There is an analog of this statement for three-dimensional topological field theories: if $\mathcal{C}$ is a proper fusion subcategory of the full category of anyons, then there exists at least one non-trivial anyon of the theory which (double) braids trivially with every anyon in $\mathcal{C}$.\footnote{Mathematically, this follows from the fact that, if $\mathcal{D}$ is a modular fusion category and $\mathcal{C}$ is a fusion subcategory, then $\dim(\mathcal{C})\dim(\mathcal{C}')=\dim(\mathcal{D})$, where $\mathcal{C}'$ is the centralizer of $\mathcal{C}$ in $\mathcal{D}$ \cite{DGNO2010}.} Our claim is that a version of this is true for 2D conformal field theories as well.\\

\noindent\textbf{Remote detectability II.} \emph{If $\mathcal{C}$ is a topologically closed, proper tensor subcategory of the full category of unitary topological line defects, then there is some local operator outside of the Virasoro subalgebras which is transparent to every topological line defect in $\mathcal{C}$.} \\

In short, if the lines one has found are unitary, closed under fusion, topologically closed, and detect every local operator outside of the Virasoro subalgebras then, according to remote detectability, those lines are exhaustive.

We pause to emphasize that, although we strongly believe that some version of remote detectability holds---for example, we discuss its relationship to rigorous theorems \cite{Bischoff:2022fxf} in the conformal net literature in Section \ref{sec:SU(2)1}---our formulation of this physical principle should be understood as a tentative proposal whose details should be sharpened and shaped as our understanding of topological line defects improves. One subtlety is that, as we alluded to earlier, the categorical formalism for infinite symmetry in quantum field theory is still being developed, and hence remote detectability is not yet mathematically crisp enough to be stated as a conjecture. Relatedly, in our proposal, we are assuming that the space of unitary lines admits some kind of topology: since we will mainly invoke remote detectability on Lie group symmetries below, this is not an issue for the applications we explore in this paper, but this point deserves better justification for general 2D conformal field theories.

As a simple example, consider the 2D Ising CFT. This theory has an obvious $\mathbb{Z}_2$ symmetry, but  the energy operator $\varepsilon_{\frac12,\frac12}$ is neutral under it. Remote detectability I therefore tells us that there must be additional unitary topological line defects in the theory. It is a celebrated fact that the Ising CFT does have an additional topological line: namely, a non-invertible duality defect \cite{Frohlich:2004ef,Frohlich:2006ch} which flips the sign of $\varepsilon_{\frac12,\frac12}$ when it is dragged through it. Thus, the $\mathbb{Z}_2$ symmetry together with the duality line are able to detect all primary operators of the Ising CFT. Remote detectability II then tells us that these are \emph{all} the unitary topological line defects of the theory. 

As another example, one can apply this criterion to the $SU(2)_1$ Wess-Zumino-Witten (WZW) model. One straightforwardly checks (see \eqref{eqn:SU(2)1HSdecomp}) that the only operators which transform trivially under the $SO(4)\cong (SU(2)_L\times SU(2)_R)/\mathbb{Z}_2$ unitary symmetry group of the theory are the operators in the Virasoro subalgebras. Thus, remote detectability gives us the following.\footnote{This example reveals the importance of incorporating unitarity in the remote detectability criterion. The full symmetry group of $SU(2)_1$ is $\mathrm{Aut}(SU(2)_1)=(SL(2,\mathbb{C})\times SL(2,\mathbb{C}))/\mathbb{Z}_2$, but the subgroup which preserves the unitary structure of $SU(2)_1$ is $\mathrm{Aut}^\dagger(SU(2)_1) =(SU(2)_L\times SU(2)_R)/\mathbb{Z}_2$. So $\mathrm{Aut}^\dagger(SU(2)_1)$ generates a proper tensor subcategory of the full category of (not necessarily unitary) topological line defects of $SU(2)_1$, and yet there is no local operator outside of the Virasoro algebra which is uncharged under $\mathrm{Aut}^\dagger(SU(2)_1).$ }\textsuperscript{\textcolor{red!70!black}{,}}\footnote{The reason the qualifier ``topologically closed'' appears in the remote detectability criterion is that a dense, proper subgroup of the symmetry group $(SU(2)_L\times SU(2)_R)/\mathbb{Z}_2$ would otherwise be a counterexample: it is not the full symmetry category, and yet it detects every local operator outside of the Virasoro algebras.}\\

\noindent\textbf{Symmetries of $SU(2)_1$.} \emph{Every simple, unitary topological line defect of the $SU(2)_1$ Wess-Zumino-Witten model corresponds to an element of its $SO(4)\cong (SU(2)_L\times SU(2)_R)/\mathbb{Z}_2$ symmetry group. In particular, the $SU(2)_1$ conformal field theory has no simple non-invertible unitary topological line defects. } \\

\noindent See e.g.\ \cite{Fuchs:2007tx,Bischoff:2022fxf,Moller:2024xtt,Gannon:2026ttf} for previous results along these lines. 

The fact that the unitary symmetries of $SU(2)_1$ are all invertible is a remarkable property which, to the authors' knowledge, is not shared by any other theory. For example, if one applies the remote detectability criterion to any other WZW model, one can easily convince oneself that the obvious symmetries (combinations of invertible symmetries and Verlinde lines) are never complete. It is natural to speculate that the $SU(2)_1$ WZW model is the \emph{only} theory which has no non-invertible unitary symmetries. Much of the work we do in this paper and in our follow-ups proceeds by exploiting this ``groupiness'' of $SU(2)_1$.

Before moving on, we note that a variation of remote detectability also applies to conformal boundary conditions in many cases. For example, in a diagonal rational conformal field theory, one can use the fact that conformal boundary conditions are in one-to-one correspondence with topological line defects supported on the maximal chiral algebra \cite{Kapustin:2010if,Gannon:2026ttf}. Topological line defects of a chiral algebra are also controlled by (a version of) remote detectability, as we describe in Section \ref{sec:SU(2)1}, and hence so are conformal boundary conditions of diagonal rational conformal field theories. Applying this to $SU(2)_1$, one reproduces the known conformal moduli space of boundary conditions \cite{Recknagel:1998ih,Gaberdiel:2001zq,Gaberdiel:2001xm}, albeit from a very different conceptual starting point. \\

\noindent{\textbf{Boundaries of $SU(2)_1$.}} \emph{The moduli space of simple, unitary conformal boundary conditions of the $SU(2)_1$ Wess-Zumino-Witten model is given by $SU(2)$. Each point corresponds to a Dirichlet boundary condition in the description of the theory as a sigma model into the group manifold of $SU(2)$.}\\

So much for the symmetries and boundaries of the $SU(2)_1$ WZW model.  What can we say about the topological line defects and conformal boundaries of the remaining $c=1$ unitary conformal field theories? We will focus in what follows on the non-compact boson conformal field theory as a simple example on which to illustrate our methods, though everything we say can be broadly generalized, in particular to any unitary $c=1$ conformal field theory.  

The key fact we will make use of is that the non-compact boson can be obtained by flat gauging the $U(1)$ winding symmetry of the $SU(2)_1$ WZW model \cite{Argurio:2024ewp,Argurio:2026txf,Jia:2026tfh}.\footnote{As is well-known, the $SU(2)_1$ WZW model is equivalent to the free compact boson conformal field theory at the self-dual radius. By the $U(1)$ ``winding'' symmetry, we mean the winding symmetry of this compact boson. By T-duality, one can equivalently consider flat gauging the $U(1)$ momentum symmetry of the compact boson. See Section \ref{subsec:Lagflatgauging}.} The term ``flat gauging'' (a.k.a.\ continuous orbifolding) \cite{Gaberdiel:2011aa,Fredenhagen:2012bw,Gaberdiel:2014vca,Brennan:2024fgj,Antinucci:2024zjp,Argurio:2024ewp,Bonetti:2024cjk,Stockall:2025ngz,Oguz:2025ftx,Paznokas:2025auw,Apruzzi:2025hvs,Argurio:2026txf,Jia:2026tsl,Jia:2026tfh} deserves further explanation. When one gauges the $U(1)$ symmetry of a quantum field theory $\mathcal{Q}$ in the usual sense, one thinks of $\mathcal{Q}$ as a matter sector and introduces a dynamical  $U(1)$ gauge field which is coupled to $\mathcal{Q}$ through its Noether current and is summed over in the path integral. For example, if one gauges the $U(1)$ winding symmetry of $SU(2)_1$ in the usual sense, the gauge coupling flows and the theory is gapped out at long distances \cite{Gawedzki:1988hq,Gawedzki:1988nj}. By contrast, in flat gauging, one carries out a similar procedure but sums only over flat (i.e., zero curvature) gauge fields in the path integral. We review the Lagrangian description of this procedure via BF theory in more detail in Section \ref{sec:NCBLag}. 

Flat gauging a continuous symmetry behaves in many ways like orbifolding \cite{Dixon:1985jw,Dixon:1986jc} a finite symmetry. For example, when applied to a conformal field theory, the central charge is preserved (unlike standard gauging, where it generally decreases). Another way in which the two are similar is that a flat gauged theory inherits a dual/quantum symmetry from the ungauged theory which acts on the twisted sectors, in the same sense as \cite{Vafa:1989ih}. Furthermore, this quantum symmetry can be orbifolded to reproduce the original theory. In a word, flat gauging, like finite orbifolding, is a lossless operation, and so we expect that any observables of a flat gauged theory can be constructed in terms of observables of the original theory, and vice versa.

\begin{figure}[t]
    \begin{center}
\tikzset{every picture/.style={line width=0.75pt}} 

\begin{tikzpicture}[x=0.75pt,y=0.75pt,yscale=-1,xscale=1]

\draw   (184.37,13.33) .. controls (201.95,21.42) and (213.8,32.89) .. (213.8,48.35) .. controls (213.8,108.23) and (36.05,108.23) .. (36.05,83.57) .. controls (36.05,58.91) and (213.8,58.91) .. (213.8,118.8) .. controls (213.8,178.68) and (36.05,178.68) .. (36.05,154.02) .. controls (36.05,129.36) and (213.8,129.36) .. (213.8,189.25) .. controls (213.8,249.13) and (36.05,249.13) .. (36.05,224.47) .. controls (36.05,200.34) and (206.3,199.83) .. (213.56,255.95) ;
\draw  [draw opacity=0][fill={rgb, 255:red, 208; green, 2; blue, 27 }  ,fill opacity=0.7 ] (35.42,153.79) .. controls (35.42,134.34) and (74.93,118.57) .. (123.66,118.57) .. controls (172.4,118.57) and (211.91,134.34) .. (211.91,153.79) .. controls (211.91,173.25) and (172.4,189.02) .. (123.66,189.02) .. controls (74.93,189.02) and (35.42,173.25) .. (35.42,153.79) -- cycle ;
\draw   (184.37,13.33) .. controls (201.95,21.42) and (213.8,32.89) .. (213.8,48.35) .. controls (213.8,108.23) and (36.05,108.23) .. (36.05,83.57) .. controls (36.05,58.91) and (213.8,58.91) .. (213.8,118.8) .. controls (213.8,152.11) and (158.79,166.89) .. (109.97,169.2) ;
\draw    (35.42,153.79) .. controls (38.57,164.8) and (66.3,170.97) .. (110.43,169.2) ;
\draw  [draw opacity=0][fill={rgb, 255:red, 0; green, 0; blue, 0 }  ,fill opacity=1 ] (120.68,153.79) .. controls (120.68,152.14) and (122.01,150.81) .. (123.66,150.81) .. controls (125.31,150.81) and (126.65,152.14) .. (126.65,153.79) .. controls (126.65,155.44) and (125.31,156.78) .. (123.66,156.78) .. controls (122.01,156.78) and (120.68,155.44) .. (120.68,153.79) -- cycle ;
\draw [color={rgb, 255:red, 155; green, 155; blue, 155 }  ,draw opacity=1 ]   (62.8,119.2) .. controls (65.07,118.58) and (66.52,119.41) .. (67.14,121.68) .. controls (67.77,123.95) and (69.22,124.78) .. (71.49,124.15) .. controls (73.76,123.53) and (75.21,124.36) .. (75.83,126.63) .. controls (76.46,128.9) and (77.91,129.73) .. (80.18,129.1) .. controls (82.45,128.48) and (83.9,129.31) .. (84.52,131.58) .. controls (85.14,133.85) and (86.59,134.68) .. (88.86,134.05) .. controls (91.13,133.43) and (92.58,134.26) .. (93.21,136.53) .. controls (93.83,138.8) and (95.28,139.63) .. (97.55,139.01) .. controls (99.82,138.38) and (101.27,139.21) .. (101.9,141.48) .. controls (102.52,143.75) and (103.97,144.58) .. (106.24,143.96) -- (109.91,146.05) -- (116.86,150.01) ;
\draw [shift={(118.6,151)}, rotate = 209.68] [color={rgb, 255:red, 155; green, 155; blue, 155 }  ,draw opacity=1 ][line width=0.75]    (10.93,-3.29) .. controls (6.95,-1.4) and (3.31,-0.3) .. (0,0) .. controls (3.31,0.3) and (6.95,1.4) .. (10.93,3.29)   ;
\draw [color={rgb, 255:red, 155; green, 155; blue, 155 }  ,draw opacity=1 ]   (140.4,34.4) .. controls (142.07,32.73) and (143.73,32.73) .. (145.4,34.4) .. controls (147.07,36.07) and (148.73,36.07) .. (150.4,34.4) .. controls (152.07,32.73) and (153.73,32.73) .. (155.4,34.4) .. controls (157.07,36.07) and (158.73,36.07) .. (160.4,34.4) .. controls (162.07,32.73) and (163.73,32.73) .. (165.4,34.4) .. controls (167.07,36.07) and (168.73,36.07) .. (170.4,34.4) .. controls (172.07,32.73) and (173.73,32.73) .. (175.4,34.4) .. controls (177.07,36.07) and (178.73,36.07) .. (180.4,34.4) .. controls (182.07,32.73) and (183.73,32.73) .. (185.4,34.4) .. controls (187.07,36.07) and (188.73,36.07) .. (190.4,34.4) -- (192.6,34.4) -- (200.6,34.4) ;
\draw [shift={(202.6,34.4)}, rotate = 180] [color={rgb, 255:red, 155; green, 155; blue, 155 }  ,draw opacity=1 ][line width=0.75]    (10.93,-3.29) .. controls (6.95,-1.4) and (3.31,-0.3) .. (0,0) .. controls (3.31,0.3) and (6.95,1.4) .. (10.93,3.29)   ;
\draw [color={rgb, 255:red, 155; green, 155; blue, 155 }  ,draw opacity=1 ]   (139.8,42.8) .. controls (142.09,42.27) and (143.51,43.15) .. (144.04,45.44) .. controls (144.57,47.74) and (145.99,48.62) .. (148.29,48.09) .. controls (150.58,47.56) and (152,48.44) .. (152.53,50.73) .. controls (153.06,53.02) and (154.48,53.91) .. (156.77,53.38) .. controls (159.06,52.85) and (160.48,53.73) .. (161.02,56.02) .. controls (161.55,58.31) and (162.97,59.19) .. (165.26,58.66) .. controls (167.56,58.13) and (168.98,59.01) .. (169.51,61.31) .. controls (170.04,63.6) and (171.46,64.48) .. (173.75,63.95) .. controls (176.04,63.42) and (177.46,64.31) .. (177.99,66.6) .. controls (178.53,68.89) and (179.95,69.77) .. (182.24,69.24) -- (183.31,69.91) -- (190.1,74.14) ;
\draw [shift={(191.8,75.2)}, rotate = 211.93] [color={rgb, 255:red, 155; green, 155; blue, 155 }  ,draw opacity=1 ][line width=0.75]    (10.93,-3.29) .. controls (6.95,-1.4) and (3.31,-0.3) .. (0,0) .. controls (3.31,0.3) and (6.95,1.4) .. (10.93,3.29)   ;
\draw [color={rgb, 255:red, 155; green, 155; blue, 155 }  ,draw opacity=1 ]   (215,109.6) .. controls (214.15,111.8) and (212.63,112.48) .. (210.43,111.63) .. controls (208.23,110.78) and (206.71,111.46) .. (205.86,113.66) .. controls (205.01,115.86) and (203.49,116.54) .. (201.29,115.69) .. controls (199.09,114.84) and (197.57,115.52) .. (196.72,117.72) .. controls (195.87,119.92) and (194.35,120.6) .. (192.15,119.75) .. controls (189.95,118.9) and (188.43,119.58) .. (187.59,121.78) .. controls (186.74,123.98) and (185.22,124.66) .. (183.02,123.81) .. controls (180.82,122.97) and (179.3,123.65) .. (178.45,125.85) .. controls (177.6,128.05) and (176.08,128.73) .. (173.88,127.88) .. controls (171.68,127.03) and (170.16,127.71) .. (169.31,129.91) -- (166.54,131.14) -- (159.23,134.39) ;
\draw [shift={(157.4,135.2)}, rotate = 336.04] [color={rgb, 255:red, 155; green, 155; blue, 155 }  ,draw opacity=1 ][line width=0.75]    (10.93,-3.29) .. controls (6.95,-1.4) and (3.31,-0.3) .. (0,0) .. controls (3.31,0.3) and (6.95,1.4) .. (10.93,3.29)   ;

\draw (11.2,103.2) node [anchor=north west][inner sep=0.75pt]   [align=left] {Neumann};
\draw (216,99.2) node [anchor=north west][inner sep=0.75pt]   [align=left] {Generalized Neumann};
\draw (78,26.6) node [anchor=north west][inner sep=0.75pt]   [align=left] {Dirichlet};

\end{tikzpicture}
\caption{The space of unitary conformal boundary conditions of the non-compact boson CFT forms a disk. At the origin of the disk sits the Neumann boundary condition, whereas a point on the boundary of the disk corresponds to an infinite superposition of an equidistant lattice of Dirichlet boundary conditions. A generic point in the interior of the disk corresponds to a generalized Neumann boundary condition.}\label{fig:ncb_bdy}
\end{center}
\end{figure}
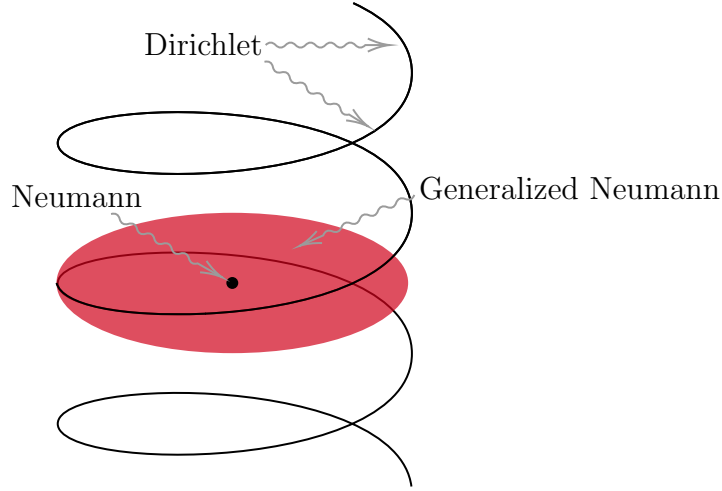

We describe in detail in Section \ref{sec:NCBAbs} how to construct the boundary conditions and topological line defects of the non-compact boson conformal field theory from those of $SU(2)_1$ using this flat gauging perspective. In the process, we obtain physical interpretations of certain mathematical formulae of Ostrik \cite{Ostrik:2002ohv}.\footnote{For our more mathematically inclined readers, we are referring to Ostrik's characterization of the $A$-$B$-bimodules in the fusion category $\mathrm{Vec}_G^\omega$, where $A$ and $B$ are two algebra objects. Strictly speaking, Ostrik's results apply to finite groups $G$, and we focus on the case that $G=U(1)$ in this paper, but the same idea we invoke applies to general $G$. This will be spelled out more explicitly elsewhere.} Our constructions make heavy use of the existence of topological interfaces, known as gauging interfaces \cite{Fuchs:2002cm,Bhardwaj:2017xup,Choi:2021kmx}, between a theory and its orbifolds. Flat gauging of non-abelian Lie group symmetries will be further developed in \cite{WIP}.

For example, every boundary condition of a flat gauged theory is obtained by fusing the corresponding gauging interface onto some boundary condition of the original theory, as in \eqref{eqn:GaugingIntFuseBdy}. More precisely, fusing a gauging interface onto a boundary generally produces a \emph{non-simple} boundary of the flat gauged theory, and one must extract the simple summands (sometimes called fractional branes in the string theory literature). We describe how to carry this out in the case of flat $U(1)$ gauging in Section \ref{subsec:NCBconformalbdy} (the general case will be treated elsewhere). This results in the following classification of unitary conformal boundary conditions of the non-compact boson (see Section \ref{subsec:NCBconformalbdy} for details). \\

\noindent\textbf{Boundaries of the non-compact boson.}\emph{ Every unitary, simple conformal boundary condition of the non-compact boson is either Dirichlet $\mathbf{D}_{\Phi_0}$, labeled by the location of the brane $\Phi_0\in\mathbb{R}$ in the target space, or generalized Neumann $\mathbf{N}_a$, labeled by a parameter $a\in\mathbb{C}$ satisfying $|a|<1$. The standard Neumann boundary condition corresponds to $a=0$, and the generalized Neumann boundaries dissolve into an infinite direct sum of Dirichlet boundary conditions in the limit as $|a|\to 1$, 
\begin{align}
    \mathbf{N}_a\xrightarrow{a\to e^{i\alpha}}\bigoplus_{Q\in\mathbb{Z}}\mathbf{D}_{\alpha-2\pi Q}.
\end{align}
The Dirichlet boundary conditions have finite $g$-function, while the generalized Neumann boundary conditions all have infinite $g$-function. The moduli space formed by these boundary conditions is visualized in Figure \ref{fig:ncb_bdy}.} \\

The generalized Neumann boundary condition $\mathbf{N}_a$ is, by our definition, the one obtained by fusing the $U(1)$ flat gauging interface between $SU(2)_1$ and the non-compact boson onto the boundary condition of $SU(2)_1$ corresponding to the group element $g=\left(\begin{smallmatrix} a & r(a) \\ -r(a) & \bar a \end{smallmatrix}\right)$, where $r(a)=\sqrt{1-|a|^2}$.
These were found previously \cite{Callan:1993mw,Callan:1994ub,Polchinski:1994my} as deformations of the Neumann boundary condition $\mathbf{N}_0$ using exactly marginal boundary vertex operators.\footnote{Boundary conditions of the non-compact boson were also explored in \cite{Recknagel:1998ih}, though it appears there was an error in extracting the boundary moduli space from their expressions for the boundary states.} The moduli space of Figure \ref{fig:ncb_bdy} was also argued to be the complete set of unitary conformal boundary conditions of the non-compact boson more recently in \cite{Gannon:2026ttf} using very different methods. 

In addition to confirming the proposed classification of  \cite{Gannon:2026ttf}, we subject these boundaries to a number of detailed consistency checks in Section \ref{sec:bdyabst}. For example, for each boundary $\mathbf{D}_{\Phi_0}$ and $\mathbf{N}_a$, we calculate the corresponding boundary states $|\mathbf{D}_{\Phi_0}\rangle$ and $|\mathbf{N}_a\rangle$ in the $S^1$ Hilbert space which represent the boundaries in the closed string channel \cite{Cardy:1989ir,Cardy:2004hm}. We also calculate the interval Hilbert spaces between every pair of boundary conditions. We then confirm that the annulus partition functions between every pair of boundaries respect open/closed duality: that is, every annulus partition function can be represented equivalently as a trace over an interval Hilbert space, or as an overlap between two boundary states. We also give a Lagrangian description of these boundaries in Section \ref{subsec:NCBbdyLag}.

\begin{figure}
    \begin{center}
        \input{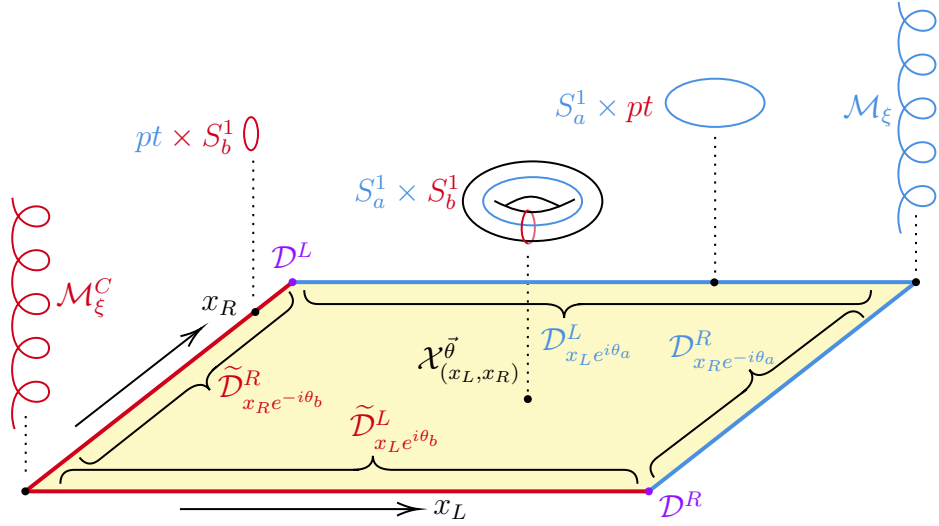}
        \caption{The moduli space of simple topological line defects of the non-compact boson, visualized as a singular $T^2$ fibration over a square. At the boundary of the square, the $T^2$ fiber degenerates in ways that we depict in the figure. The entire moduli space is smooth, except at the bottom left and top right corners of the square, where the fiber ``unfurls'' to a real line. The notations which underlie this figure are explained in Section \ref{subsubsec:modulispacelines}.}\label{fig:stackyS4}
    \end{center}
\end{figure}

Finally, we perform an analogous calculation of the full moduli space of unitary topological line defects of the non-compact boson conformal field theory. To the best of our knowledge, most of the symmetries we uncover were not known previously in the literature (though see below for the relationship to known boundaries of two non-compact bosons by folding). The resulting moduli space can be described as follows. \\

\noindent\textbf{Symmetries of the non-compact boson.} \emph{Consider a simple, unitary topological line defect $\mathcal{X}$ of the non-compact boson conformal field theory, and suppose that $\mathcal{X}$ does not implement one of the invertible symmetries in $\mathbb{R}\rtimes \mathbb{Z}_2^{\mathrm{C}}$.\footnote{The $\mathbb{R}$ corresponds to the shift symmetry $\Phi\to \Phi+\xi$ of the non-compact boson $\Phi$, and $\mathbb{Z}_2^{\mathrm{C}}$ is the charge conjugation symmetry which acts as $\Phi\to -\Phi$.} The moduli space of unitary topological line defects which are continuously connected to $\mathcal{X}$ is  
\begin{align}\label{eqn:S4modulispace}
    S^4\cong U(1)\backslash SO(4)/U(1).
\end{align} 
Every simple, unitary, topological line defect of the non-compact boson appears as a summand of a line in \eqref{eqn:S4modulispace}. At a generic point on this moduli space, the corresponding topological line defect is simple and has infinite quantum dimension. At two particular $S^1$ submanifolds of $S^4$, the topological line defects dissolve into an infinite direct sum of invertible lines, each with quantum dimension equal to $1$. This is visualized in Figure \ref{fig:stackyS4}.}  \\

A particular one-dimensional slice of this $S^4$ moduli space of topological lines appears in \cite{Callan:1995em} as a family of boundary conditions of the $c=2$ theory of two non-compact bosons. More specifically, this one-parameter family of lines, after folding and then performing a change of variables, corresponds to the boundary conditions appearing in Equation (2.1) of \cite{Callan:1995em} with  particular ``magic'' values of the couplings given by $(\alpha,\beta)=(\frac12,\frac12)$.\footnote{The couplings $(\alpha,\beta)=(\frac12,-\frac12)$ are related to $(\alpha,\beta)=(\frac12,\frac12)$ by a charge conjugation symmetry.} These boundaries have a beautiful string theoretic interpretation as describing a D-brane in spacetime with a magnetic field activated for the $U(1)$ gauge field on its world-volume.

Again, we carry out a number of detailed calculations related to these lines. For example, we compute their fusion rules, as well as how they act on boundary conditions by parallel fusion in Section \ref{subsec:abstract_top_lines}. We also compute their action on local operators, and their corresponding twisted-sector Hilbert spaces in Section \ref{subsec:explicit_con_lines}. We give completely Lagrangian descriptions of these topological line defects in Section \ref{subsec:Lag_lines}. Along the way, we also give a Lagrangian description of the $SO(4)$ symmetry lines of the $SU(2)_1$ WZW model in its formulation as a compact boson theory (as opposed to a sigma model into the group manifold of $SU(2)$).

The remainder of this paper is organized as follows. In Section \ref{sec:SU(2)1}, we discuss the remote detectability criterion for 2D conformal field theories and use it to classify topological lines and conformal boundaries of $SU(2)_1$. In Secion \ref{sec:NCBAbs}, we use gauging interfaces to translate the classification of lines and boundaries in $SU(2)_1$ to a classification of lines and boundaries in the non-compact boson conformal field theory. In Section \ref{sec:NCBLag}, we give various Lagrangian descriptions of this procedure. In Section \ref{sec:explicitNCB}, we calculate the boundary states, the action of topological lines on local operators, and the spectra of boundary-condition-changing operators as well as defect operators attached to topological lines.

\section{Moduli Spaces of Boundaries and Lines of the \texorpdfstring{$SU(2)_1$}{SU(2)1} WZW model}\label{sec:SU(2)1}

As explained in the introduction, our strategy is to convert our control over the $SU(2)_1$ WZW model into control over the non-compact boson theory, using the fact that the latter can be obtained from the former via flat gauging. As such, we spend some time reviewing the conformal boundary conditions and topological line defects of the $SU(2)_1$ CFT.

Although much of what we say is known in the literature, some of our arguments for the known facts are new. For example, we give a new argument that the $SU(2)$ worth of unitary conformal boundary conditions constructed in the literature exhausts \emph{all} unitary conformal boundary conditions of the $SU(2)_1$ WZW model. Similarly, we argue that all simple unitary topological line defects of the $SU(2)_1$ WZW model are invertible, and hence are captured by the group $SO(4)$.

More details on the boundaries and lines of $SU(2)_1$ are reviewed in Appendix \ref{app:su2}.

\subsection{Classification of Conformal Boundaries}\label{subsec:CBbdy}

Let us begin by studying the unitary conformal boundary conditions of the $SU(2)_1$ WZW model. 
We will give an argument that the simple unitary conformal boundary conditions of the theory are completely classified by elements of $SU(2)$. 
We make crucial use of the relationship between boundary conditions of a CFT and topological line defects supported on its chiral algebra \cite{Kapustin:2010if,Gannon:2026ttf}. The boundary conformal manifold of $SU(2)_1$ has appeared in many earlier papers (see e.g.\ \cite{Recknagel:1998ih,Gaberdiel:2001zq,Gaberdiel:2001xm}), and this section can be seen as providing a complementary treatment. 

\begin{figure}
\begin{center}
    \tikzset{every picture/.style={line width=0.75pt}} 

\begin{tikzpicture}[x=0.75pt,y=0.75pt,yscale=-1,xscale=1,scale=.8]

\draw  [draw opacity=0][fill={rgb, 255:red, 184; green, 233; blue, 134 }  ,fill opacity=.6 ][line width=1.5]  (224,120.9) -- (224,11) -- (320,58.1) -- (320,168) -- cycle ;
\draw [line width=1.5]  [dash pattern={on 1.69pt off 2.76pt}]  (17.87,120.9) -- (224,120.9) ;
\draw [line width=1.5]  [dash pattern={on 1.69pt off 2.76pt}]  (17.87,11) -- (224,11) ;
\draw [line width=1.5]  [dash pattern={on 1.69pt off 2.76pt}]  (111.55,168) -- (317.67,168) ;
\draw [line width=1.5]  [dash pattern={on 1.69pt off 2.76pt}]  (111.55,58.1) -- (317.67,58.1) ;
\draw  [draw opacity=0][fill={rgb, 255:red, 184; green, 233; blue, 134 }  ,fill opacity=.6 ][line width=1.5]  (17.87,120.9) -- (17.87,11) -- (113.87,58.1) -- (113.87,168) -- cycle ;
\draw    (332,103) -- (379.5,103) ;
\draw [shift={(381.5,103)}, rotate = 180] [color={rgb, 255:red, 0; green, 0; blue, 0 }  ][line width=0.75]    (10.93,-3.29) .. controls (6.95,-1.4) and (3.31,-0.3) .. (0,0) .. controls (3.31,0.3) and (6.95,1.4) .. (10.93,3.29)   ;
\draw  [draw opacity=0][fill={rgb, 255:red, 184; green, 233; blue, 134 }  ,fill opacity=.6 ][line width=1.5]  (404,122.9) -- (404,13) -- (500,60.1) -- (500,170) -- cycle ;

\draw (250.87,81.4) node [anchor=north west][inner sep=0.75pt]    {$\widehat{\mathfrak{su}}( 2)_{1}$};
\draw (120.67,72.4) node [anchor=north west][inner sep=0.75pt]    {$SU( 2)_{1} \ \text{CS}$};
\draw (41.87,81.4) node [anchor=north west][inner sep=0.75pt]    {$\overline{\widehat{\mathfrak{su}}( 2)}_{1}$};
\draw (392,87.4) node [anchor=north west][inner sep=0.75pt]    {$SU( 2)_{1} \ \text{WZW}$};

\end{tikzpicture}
    \caption{The 3D TQFT presentation of the $SU(2)_1$ WZW model.}\label{fig:su(2)1sandwich}
\end{center}
\end{figure}
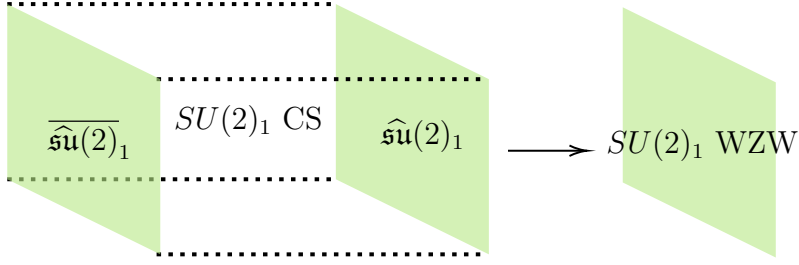

We start by recalling certain features of the 3D TQFT approach to 2D rational CFTs \cite{Witten:1988hf,Elitzur:1989nr,Moore:1988qv,Fuchs:2002cm,Kapustin:2010if}. The starting point is the observation that the $SU(2)_1$ WZW model can be obtained by compactifying the $SU(2)_1$ Chern-Simons theory on an interval, with the $\widehat{\mathfrak{su}}(2)_1$ chiral algebra boundary condition \cite{Witten:1988hf,Elitzur:1989nr} imposed on one end of the interval, and its complex conjugate imposed on the other. This is depicted schematically in Figure \ref{fig:su(2)1sandwich}. 

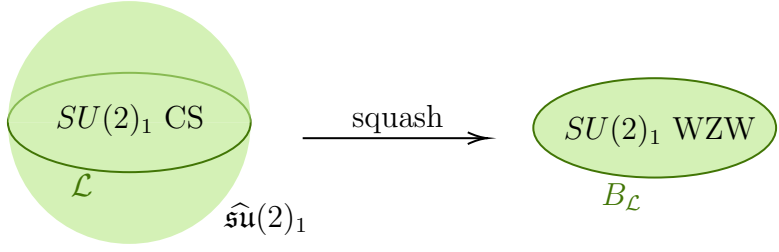
\begin{figure}
    \begin{center}
        \tikzset{every picture/.style={line width=0.75pt}} 

\begin{tikzpicture}[x=0.75pt,y=0.75pt,yscale=-1,xscale=1,scale=.9]

\draw    (176,82) -- (277,82) ;
\draw [shift={(279,82)}, rotate = 180] [color={rgb, 255:red, 0; green, 0; blue, 0 }  ][line width=0.75]    (10.93,-3.29) .. controls (6.95,-1.4) and (3.31,-0.3) .. (0,0) .. controls (3.31,0.3) and (6.95,1.4) .. (10.93,3.29)   ;
\draw  [color={rgb, 255:red, 65; green, 117; blue, 5 }  ,draw opacity=1 ][fill={rgb, 255:red, 184; green, 233; blue, 134 }  ,fill opacity=.6 ] (304.62,76.25) .. controls (304.62,60.95) and (334.92,48.55) .. (372.31,48.55) .. controls (409.69,48.55) and (440,60.95) .. (440,76.25) .. controls (440,91.55) and (409.69,103.95) .. (372.31,103.95) .. controls (334.92,103.95) and (304.62,91.55) .. (304.62,76.25) -- cycle ;
\draw  [draw opacity=0][fill={rgb, 255:red, 184; green, 233; blue, 134 }  ,fill opacity=.6 ] (79.32,140.95) .. controls (41.93,140.95) and (11.62,110.64) .. (11.62,73.25) -- (147.02,73.25) .. controls (147.02,110.64) and (116.71,140.95) .. (79.32,140.95) -- cycle ;
\draw  [color={rgb, 255:red, 65; green, 117; blue, 5 }  ,draw opacity=1 ] (11.62,73.25) .. controls (11.62,57.95) and (41.92,45.55) .. (79.31,45.55) .. controls (116.69,45.55) and (147,57.95) .. (147,73.25) .. controls (147,88.55) and (116.69,100.95) .. (79.31,100.95) .. controls (41.92,100.95) and (11.62,88.55) .. (11.62,73.25) -- cycle ;
\draw  [draw opacity=0][fill={rgb, 255:red, 184; green, 233; blue, 134 }  ,fill opacity=.6 ] (79.32,5.55) .. controls (116.71,5.55) and (147.02,35.86) .. (147.02,73.25) -- (11.62,73.25) .. controls (11.62,35.86) and (41.93,5.55) .. (79.32,5.55) -- cycle ;

\draw (45,100.4) node [anchor=north west][inner sep=0.75pt]  [color={rgb, 255:red, 65; green, 117; blue, 5 }  ,opacity=1 ]  {$\mathcal{L}$};
\draw (131,116.4) node [anchor=north west][inner sep=0.75pt]    {$\widehat{\mathfrak{su}}( 2)_{1}$};
\draw (202,62) node [anchor=north west][inner sep=0.75pt]   [align=left] {squash};
\draw (341,105.4) node [anchor=north west][inner sep=0.75pt]  [color={rgb, 255:red, 65; green, 117; blue, 5 }  ,opacity=1 ]  {$B_{\mathcal{L}}$};
\draw (320,67.4) node [anchor=north west][inner sep=0.75pt]    {$SU( 2)_{1} \ \text{WZW}$};
\draw (36.67,61.4) node [anchor=north west][inner sep=0.75pt]    {$SU( 2)_{1} \ \text{CS}$};

\end{tikzpicture}
    \end{center}
    \caption{Boundary conditions of the $SU(2)_1$ WZW model correspond to topological line defects supported on the $\widehat{\mathfrak{su}}(2)_1$ chiral algebra boundary condition of $SU(2)_1$ Chern-Simons theory.}\label{fig:su(2)1bdysquash}
\end{figure}

There is a generalization of Figure \ref{fig:su(2)1sandwich} which will help us understand boundary conditions \cite{Kapustin:2010if,Gannon:2026ttf}. Suppose one studies $SU(2)_1$ Chern-Simons theory on a solid ball $B^{(3)}$ with the $\widehat{\mathfrak{su}}(2)_1$ chiral algebra boundary condition imposed on the boundary $\partial B^{(3)}=S^2$, and with a unitary topological line defect $\mathcal{L}$ wrapping the equator of this boundary $S^2$, as in Figure \ref{fig:su(2)1bdysquash}. If one squashes the ball $B^{(3)}$ in the vertical direction, one goes over to a disk geometry $D^{(2)}$. Away from the boundary of the disk, the model is described by the dimensional reduction of $SU(2)_1$ Chern-Simons theory on an interval which, by the comments of the previous paragraph, is simply the $SU(2)_1$ WZW model. At the boundary of the disk, we engineer some boundary condition which is determined by the topological line defect $\mathcal{L}$, and which we accordingly call $\mathbf{B}_{\mathcal{L}}$. The results of \cite{Kapustin:2010if,Gannon:2026ttf} imply that the map
\begin{align}
    \mathcal{L}\mapsto \mathbf{B}_\mathcal{L}
\end{align}
affords a one-to-one correspondence between:
\begin{enumerate}[label=\arabic*)]
    \item unitary topological line operators supported on the $\widehat{\mathfrak{su}}(2)_1$ chiral algebra boundary of the $SU(2)_1$ Chern-Simons TQFT, and,
    \item unitary boundary conditions of the full $SU(2)_1$ WZW 2D CFT.
\end{enumerate}  
 A simple way to understand this correspondence is to note that conformal boundary conditions in a CFT obey $T(z)=\bar T(\bar z)$ on the boundary. Upon unfolding, a boundary condition becomes a line defect, and the condition on the energy-momentum tensor implies that it is continuous across the line defect, i.e. the line is topological.

Thus, classifying unitary conformal boundaries of the $SU(2)_1$ WZW model is equivalent to classifying topological line operators supported on the $\widehat{\mathfrak{su}}(2)_1$ chiral algebra, thought of as a boundary condition of $SU(2)_1$ Chern-Simons TQFT. We take up this latter problem. 

Now, by integrating exponentials of conserved spin-1 Noether currents, we obtain an $SU(2)$ worth of topological line operators supported on the $\widehat{\mathfrak{su}}(2)_1$ chiral algebra. We label these as $\mathcal{L}_g$, with $g\in SU(2)$. We can ask whether $\widehat{\mathfrak{su}}(2)_1$ has any remaining elementary unitary topological lines which are not among the $\mathcal{L}_g$. 

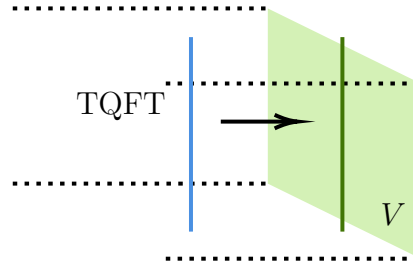
\begin{figure}
\begin{center}
    \tikzset{every picture/.style={line width=0.75pt}} 

\begin{tikzpicture}[x=0.75pt,y=0.75pt,yscale=-1,xscale=1,scale=.8]

\draw  [draw opacity=0][fill={rgb, 255:red, 184; green, 233; blue, 134 }  ,fill opacity=.6 ][line width=1.5]  (168.2,119.9) -- (168.2,10) -- (264.2,57.1) -- (264.2,167) -- cycle ;
\draw [line width=1.5]  [dash pattern={on 1.69pt off 2.76pt}]  (8.2,10) -- (168.2,10) ;
\draw [line width=1.5]  [dash pattern={on 1.69pt off 2.76pt}]  (104.2,167) -- (264.2,167) ;
\draw [line width=1.5]  [dash pattern={on 1.69pt off 2.76pt}]  (8.2,119.9) -- (168.2,119.9) ;
\draw [line width=1.5]  [dash pattern={on 1.69pt off 2.76pt}]  (104.2,57.1) -- (264.2,57.1) ;
\draw [line width=1.5]    (138.7,81) -- (183.7,81) ;
\draw [shift={(186.7,81)}, rotate = 180] [color={rgb, 255:red, 0; green, 0; blue, 0 }  ][line width=1.5]    (14.21,-4.28) .. controls (9.04,-1.82) and (4.3,-0.39) .. (0,0) .. controls (4.3,0.39) and (9.04,1.82) .. (14.21,4.28)   ;
\draw [color={rgb, 255:red, 65; green, 117; blue, 5 }  ,draw opacity=1 ][line width=1.5]    (215,28) -- (215,150) ;
\draw [color={rgb, 255:red, 74; green, 144; blue, 226 }  ,draw opacity=1 ][line width=1.5]    (120,28) -- (120,150) ;

\draw (237.2,130.4) node [anchor=north west][inner sep=0.75pt]    {$V$};
\draw (46,61) node [anchor=north west][inner sep=0.75pt]   [align=left] {TQFT};

\end{tikzpicture}
    \caption{We impose that the category $\mathcal{C}$ contains every topological line operator obtained by pushing a bulk anyon to the boundary.}\label{fig:bulktoboundary}
\end{center}
\end{figure}

This can be answered by invoking the following remote detectability criterion, which is adapted from the conformal net literature. Specifically, we reinterpret Theorem 4.15 and Section 7 of \cite{Bischoff:2022fxf} using the physical description of quantum operations (called hypergroup symmetries when they are sufficiently nice, e.g.\ finite) given in \cite{Gannon:2026ttf}. 

Suppose $V$ is a chiral algebra boundary condition of a 3D TQFT with central charge $c$, and denote its Virasoro subalgebra by $\mathrm{Vir}_c\subset V$. Let us say that a tensor category $\mathcal{C}$ of unitary topological line defects supported on $V$ is \emph{good} if: 
\begin{enumerate}[label=\arabic*)]
    \item $\mathcal{C}$ contains all the lines obtained by pushing anyons in the bulk onto the boundary, as in Figure \ref{fig:bulktoboundary},
    \item $\mathcal{C}$ is topologically closed, 
    \item $\mathcal{C}$ is closed under fusion, and closed under taking orientation reversals of lines.
\end{enumerate}

\noindent \textbf{Proposal.}  A good tensor category $\mathcal{C}$ of unitary topological line defects of $V$ is the \emph{full} category of unitary lines of $V$ if and only if a local operator $\mathcal{O}(z)$ being transparent to every line in $\mathcal{C}$ implies that it resides in the Virasoro subalgebra, $\mathcal{O}(z)\in \mathrm{Vir}_c\subset V$.

\begin{figure}
    \begin{center}
        \tikzset{every picture/.style={line width=0.75pt}} 

\begin{tikzpicture}[x=0.75pt,y=0.75pt,yscale=-1,xscale=1,scale=.8]

\draw  [draw opacity=0][fill={rgb, 255:red, 184; green, 233; blue, 134 }  ,fill opacity=.6 ][line width=1.5]  (82,117.9) -- (82,8) -- (178,55.1) -- (178,165) -- cycle ;
\draw [line width=1.5]  [dash pattern={on 1.69pt off 2.76pt}]  (7.5,8) -- (82,8) ;
\draw [line width=1.5]  [dash pattern={on 1.69pt off 2.76pt}]  (92.5,165) -- (178,165) ;
\draw [line width=1.5]  [dash pattern={on 1.69pt off 2.76pt}]  (7.5,117.9) -- (82,117.9) ;
\draw [line width=1.5]  [dash pattern={on 1.69pt off 2.76pt}]  (92.5,55.1) -- (178,55.1) ;
\draw [color={rgb, 255:red, 65; green, 117; blue, 5 }  ,draw opacity=1 ][line width=1.5]    (112.8,17) -- (112.8,139) ;
\draw  [draw opacity=0][fill={rgb, 255:red, 184; green, 233; blue, 134 }  ,fill opacity=.6 ][line width=1.5]  (325.2,117.9) -- (325.2,8) -- (421.2,55.1) -- (421.2,165) -- cycle ;
\draw [line width=1.5]  [dash pattern={on 1.69pt off 2.76pt}]  (250.7,8) -- (325.2,8) ;
\draw [line width=1.5]  [dash pattern={on 1.69pt off 2.76pt}]  (335.7,165) -- (421.2,165) ;
\draw [line width=1.5]  [dash pattern={on 1.69pt off 2.76pt}]  (250.7,117.9) -- (325.2,117.9) ;
\draw  [fill={rgb, 255:red, 0; green, 0; blue, 0 }  ,fill opacity=1 ] (131.2,95.6) .. controls (131.2,94.38) and (132.18,93.4) .. (133.4,93.4) .. controls (134.62,93.4) and (135.6,94.38) .. (135.6,95.6) .. controls (135.6,96.82) and (134.62,97.8) .. (133.4,97.8) .. controls (132.18,97.8) and (131.2,96.82) .. (131.2,95.6) -- cycle ;
\draw  [fill={rgb, 255:red, 0; green, 0; blue, 0 }  ,fill opacity=1 ] (371.8,97.2) .. controls (371.8,95.98) and (372.78,95) .. (374,95) .. controls (375.22,95) and (376.2,95.98) .. (376.2,97.2) .. controls (376.2,98.42) and (375.22,99.4) .. (374,99.4) .. controls (372.78,99.4) and (371.8,98.42) .. (371.8,97.2) -- cycle ;
\draw [color={rgb, 255:red, 65; green, 117; blue, 5 }  ,draw opacity=1 ][line width=1.5]    (358.8,17) .. controls (364.8,56.6) and (402.4,62.2) .. (401.6,89) .. controls (400.8,115.8) and (364,105.4) .. (352.4,140.2) ;
\draw [line width=1.5]  [dash pattern={on 1.69pt off 2.76pt}]  (335.7,55.1) -- (421.2,55.1) ;

\draw (151,128.4) node [anchor=north west][inner sep=0.75pt]    {$V$};
\draw (21.8,57) node [anchor=north west][inner sep=0.75pt]   [align=left] {TQFT};
\draw (394.2,128.4) node [anchor=north west][inner sep=0.75pt]    {$V$};
\draw (265,57) node [anchor=north west][inner sep=0.75pt]   [align=left] {TQFT};
\draw (196,76.4) node [anchor=north west][inner sep=0.75pt]    {$=$};
\draw (120,71.4) node [anchor=north west][inner sep=0.75pt]    {$\mathcal{O}( z)$};
\draw (355.6,73) node [anchor=north west][inner sep=0.75pt]    {$\mathcal{O}( z)$};

\end{tikzpicture}
        \caption{The chiral algebra $V^{\mathcal{C}}$ consists of the local operators $\mathcal{O}(z)\in V$ which are transparent to every boundary topological line in $\mathcal{C}$.}\label{fig:VC}
    \end{center}
\end{figure}
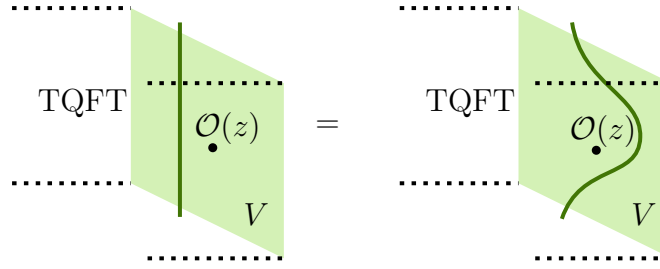

Equivalently a good tensor category $\mathcal{C}$ of unitary topological line defects of $V$ is the full category of unitary lines of $V$ if and only if
\begin{align}\label{eqn:VCvir}
    V^{\mathcal{C}}=\mathrm{Vir}_c,
\end{align}
where $V^{\mathcal{C}}$ is the subalgebra of local operators in $V$ which are transparent to all the lines in $\mathcal{C}$, as in Figure \ref{fig:VC}.

This proposal can be understood heuristically as a kind of remote detectability criterion for chiral algebras. Indeed, it asserts in particular that any local operator which could conceivably be charged with respect to a topological line defect---that is, any local operator outside of the Virasoro algebra---is in fact so. We emphasize that this proposal should be understood at a physics level of a rigor. It will inevitably be refined as our mathematical understanding of infinite symmetries sharpens.

\begin{figure}
    \begin{center}
        \tikzset{every picture/.style={line width=0.75pt}} 

\begin{tikzpicture}[x=0.75pt,y=0.75pt,yscale=-1,xscale=1,scale=.85]

\draw  [draw opacity=0][fill={rgb, 255:red, 184; green, 233; blue, 134 }  ,fill opacity=.6 ][line width=1.5]  (76.2,118.9) -- (76.2,9) -- (172.2,56.1) -- (172.2,166) -- cycle ;
\draw [line width=1.5]  [dash pattern={on 1.69pt off 2.76pt}]  (5.5,9) -- (76.2,9) ;
\draw [line width=1.5]  [dash pattern={on 1.69pt off 2.76pt}]  (100.5,166) -- (172.2,166) ;
\draw [line width=1.5]  [dash pattern={on 1.69pt off 2.76pt}]  (5.5,118.9) -- (76.2,118.9) ;
\draw [line width=1.5]  [dash pattern={on 1.69pt off 2.76pt}]  (100.5,56.1) -- (172.2,56.1) ;
\draw  [draw opacity=0][fill={rgb, 255:red, 80; green, 227; blue, 194 }  ,fill opacity=.6 ] (3.98,72.1) -- (73.81,72) -- (173.71,114) -- (103.88,114.1) -- cycle ;
\draw [color={rgb, 255:red, 208; green, 2; blue, 27 }  ,draw opacity=1 ][line width=1.5]    (76,72) -- (172.1,114) ;
\draw  [draw opacity=0][fill={rgb, 255:red, 184; green, 233; blue, 134 }  ,fill opacity=.6 ][line width=1.5]  (306.2,120.9) -- (306.2,11) -- (402.2,58.1) -- (402.2,168) -- cycle ;
\draw [line width=1.5]  [dash pattern={on 1.69pt off 2.76pt}]  (235.5,11) -- (306.2,11) ;
\draw [line width=1.5]  [dash pattern={on 1.69pt off 2.76pt}]  (330.5,168) -- (402.2,168) ;
\draw [line width=1.5]  [dash pattern={on 1.69pt off 2.76pt}]  (235.5,120.9) -- (306.2,120.9) ;
\draw [line width=1.5]  [dash pattern={on 1.69pt off 2.76pt}]  (330.5,58.1) -- (402.2,58.1) ;
\draw  [draw opacity=0][fill={rgb, 255:red, 80; green, 227; blue, 194 }  ,fill opacity=.6 ] (233.98,74.1) -- (303.81,74) -- (403.71,116) -- (333.88,116.1) -- cycle ;
\draw [color={rgb, 255:red, 208; green, 2; blue, 27 }  ,draw opacity=1 ][line width=1.5]    (306,74) -- (402.1,116) ;
\draw [color={rgb, 255:red, 74; green, 144; blue, 226 }  ,draw opacity=1 ][line width=1.5]    (233.98,74.1) -- (333.88,116.1) ;
\draw    (269,84) -- (313,84) ;
\draw [shift={(315,84)}, rotate = 180] [color={rgb, 255:red, 0; green, 0; blue, 0 }  ][line width=0.75]    (10.93,-3.29) .. controls (6.95,-1.4) and (3.31,-0.3) .. (0,0) .. controls (3.31,0.3) and (6.95,1.4) .. (10.93,3.29)   ;
\draw  [draw opacity=0][fill={rgb, 255:red, 184; green, 233; blue, 134 }  ,fill opacity=.6 ][line width=1.5]  (498.2,120.9) -- (498.2,11) -- (594.2,58.1) -- (594.2,168) -- cycle ;
\draw [line width=1.5]  [dash pattern={on 1.69pt off 2.76pt}]  (427.5,11) -- (498.2,11) ;
\draw [line width=1.5]  [dash pattern={on 1.69pt off 2.76pt}]  (522.5,168) -- (594.2,168) ;
\draw [line width=1.5]  [dash pattern={on 1.69pt off 2.76pt}]  (427.5,120.9) -- (498.2,120.9) ;
\draw [line width=1.5]  [dash pattern={on 1.69pt off 2.76pt}]  (522.5,58.1) -- (594.2,58.1) ;
\draw [color={rgb, 255:red, 144; green, 19; blue, 254 }  ,draw opacity=1 ][line width=1.5]    (498,74) -- (594.1,116) ;

\draw (146.2,126.4) node [anchor=north west][inner sep=0.75pt]    {$V$};
\draw (115.2,71.5) node [anchor=north west][inner sep=0.75pt]  [color={rgb, 255:red, 208; green, 2; blue, 27 }  ,opacity=1 ]  {$X_{k}$};
\draw (15,27.4) node [anchor=north west][inner sep=0.75pt]    {$\mathrm{TQFT}$};
\draw (80,84.4) node [anchor=north west][inner sep=0.75pt]    {$\mathcal{S}_{k}$};
\draw (376.2,128.4) node [anchor=north west][inner sep=0.75pt]    {$V$};
\draw (344.2,73.5) node [anchor=north west][inner sep=0.75pt]  [color={rgb, 255:red, 208; green, 2; blue, 27 }  ,opacity=1 ]  {$X_{k}$};
\draw (245,29.4) node [anchor=north west][inner sep=0.75pt]    {$\mathrm{TQFT}$};
\draw (311,87.4) node [anchor=north west][inner sep=0.75pt]    {$\mathcal{S}_{k}$};
\draw (250,88.4) node [anchor=north west][inner sep=0.75pt]  [color={rgb, 255:red, 74; green, 144; blue, 226 }  ,opacity=1 ]  {$\underline{L}$};
\draw (568.2,128.4) node [anchor=north west][inner sep=0.75pt]    {$V$};
\draw (536.2,73.5) node [anchor=north west][inner sep=0.75pt]  [color={rgb, 255:red, 144; green, 19; blue, 254 }  ,opacity=1 ]  {$L$};
\draw (437,29.4) node [anchor=north west][inner sep=0.75pt]    {$\mathrm{TQFT}$};
\draw (405,88.4) node [anchor=north west][inner sep=0.75pt]    {$\leftrightharpoons $};

\end{tikzpicture}
        \caption{A quantum operation is interpreted as a bulk topological surface defect $\mathcal{S}_k$ terminating on a boundary topological line junction $X_k$, as on the left. A given quantum operation gives rise to several different boundary topological line operators $L$, which are in correspondence with different choices of boundaries $\underline{L}$ for the surface $\mathcal{S}_k$, as on the right. }\label{fig:quantumoperations}
    \end{center}
\end{figure}
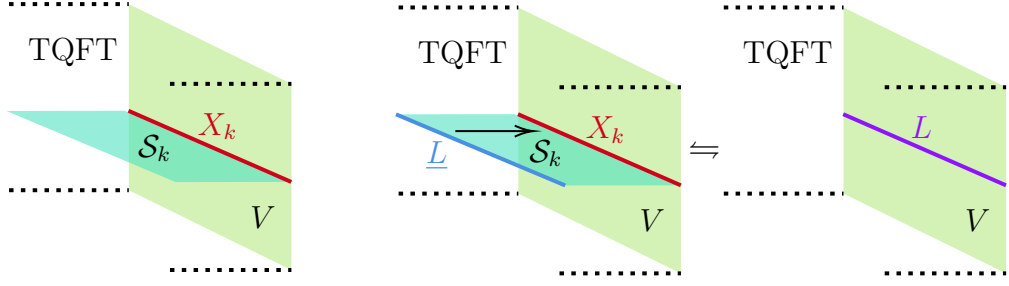

Let us briefly sketch where this proposal comes from. Theorem 4.15 of \cite{Bischoff:2022fxf} is phrased in terms of the notion of quantum operations on a conformal net, which were given a physical interpretation in \cite{Gannon:2026ttf}. A quantum operation can be thought of as a kind of non-invertible generalization of the notion of an automorphism of a chiral algebra. In terms of topological defects, a quantum operation $k$ is implemented by a bulk topological surface operator $\mathcal{S}_k$ terminating on a boundary topological line junction $X_k$, as in the left of Figure \ref{fig:quantumoperations}.\footnote{There is a mild ambiguity in the choice of junction $X_k$, but it will not affect any of the discussion which follows.} Theorem 4.15 of \cite{Bischoff:2022fxf} can then be interpreted more physically as asserting that every boundary local operator in $V$ outside of the Virasoro algebra is charged under some quantum operation.

There is a close relationship between quantum operations and boundary topological line defects. More specifically, every quantum operation $k$ defines a category $\mathcal{B}_k$ of boundary topological line operators. One way to think about $\mathcal{B}_k$ is that it is the collection of all boundary lines which act on genuine boundary local operators in the same way that $k$ does (up to an overall multiplicative factor).\footnote{Though, the different lines in $\mathcal{B}_k$ will act differently on non-genuine boundary local operators, e.g.\ boundary local operators attached to an anyon protruding into the bulk.} Its objects can be described as follows: for every topological line operator $\underline{L}$ which can bound the surface $\mathcal{S}_k$, one obtains a topological line $L$ supported on $V$ by fusing $\underline{L}$ onto $X_k$, as in the right-hand side of Figure \ref{fig:quantumoperations}. Conversely, every $L$ can be inflated into a triple $(\underline{L},\mathcal{S}_k,X_k)$, with $\mathcal{S}_k,X_k$ defining a quantum operation. 

Thus, if one has discovered all quantum operations $\mathrm{QuOp}(V)$, then one can reconstruct the full tensor category of unitary topological line operators as 
\begin{align}
    \mathcal{C}=\bigoplus_{k\in \mathrm{QuOp}(V)}\mathcal{B}_k.
\end{align}
In particular, Theorem 4.15 of \cite{Bischoff:2022fxf} then says that, if $\mathcal{C}$ is the full tensor category of unitary topological line operators, then $V^{\mathcal{C}}=\mathrm{Vir}_c$. This gives the forward direction of our proposal. The Galois theory of Section 7 of op.\ cit., after some translation, says that any other good tensor category of boundary topological lines will preserve a \emph{larger} chiral algebra, i.e.\ will leave some local operators undetected. This gives the reverse direction.

Let us apply this proposal to the $\widehat{\mathfrak{su}}(2)_1$ chiral algebra, letting $\mathcal{C}$ be the category whose elementary lines are the $\mathcal{L}_g$ with $g\in SU(2)$. We note that the bulk fundamental Wilson line of $SU(2)_1$ Chern-Simons theory, which is the only non-trivial bulk anyon, maps to the $\mathbb{Z}_2$ central element inside of $SU(2)$ when it is pushed onto the $\widehat{\mathfrak{su}}(2)_1$ boundary, so $\mathcal{C}$ satisfies the hypotheses of our proposal.

Next, note that $\widehat{\mathfrak{su}}(2)_1$ decomposes into representations of $SU(2)$ as (see also \eqref{eq:su2_to_vir})
\begin{align}
    \widehat{\mathfrak{su}}(2)_1 \cong \bigoplus_{J\in\mathbb{Z}_{\geq 0}}R_J\otimes \mathcal{V}_{J},
\end{align}
where $\mathcal{V}_J$ is the $c=1$ Virasoro representation of conformal weight $h=J^2$, and $R_J$ is the spin-$J$ representation of $SU(2)$ with dimension $\dim(R_J)=2J+1$. In particular, the only states transforming in the trivial representation $R_0$ of $SU(2)$ reside in $\mathcal{V}_0\cong \mathrm{Vir}_{c=1}$. That is, 
\begin{align}
    \widehat{\mathfrak{su}}(2)_1^{SU(2)}\cong \mathrm{Vir}_{c=1},
\end{align}
and we conclude from \eqref{eqn:VCvir} that there are no further topological line operators of $\widehat{\mathfrak{su}}(2)_1$ outside of $SU(2)$. 

By Figure \ref{fig:su(2)1bdysquash}, this means that every elementary unitary boundary condition of the full $SU(2)_1$ WZW model is of the form $\mathbf{B}_g\equiv \mathbf{B}_{\mathcal{L}_g}$, for some $g\in SU(2)$. 
These can be thought of as Dirichlet boundary conditions in the description of $SU(2)_1$ as a non-linear sigma model into the group manifold of $SU(2)$ \cite{Novikov:1982hamiltonian,Polyakov:1983nonabelian,Witten:1984nonabelian}. We refer to Appendix \ref{app:boundaries} for further details behind these boundary conditions.

Before moving on, we pause to appreciate the importance of unitarity in this proposal. Indeed, $\widehat{\mathfrak{su}}(2)_1$ certainly admits further topological line operators that are not unitary. For example, by exponentiating Noether currents with arbitrary complex coefficients, we obtain an $SL(2,\mathbb{C})$ worth of topological line operators, extending the $SU(2)$ identified above. In fact, the $SU(2)_1$ WZW model also admits an $SL(2,\mathbb{C})$ worth of (not necessarily unitary) boundary conditions \cite{Gaberdiel:2001xm}. It is an interesting question to determine whether $SL(2,\mathbb{C})$ covers all topological lines of the $\widehat{\mathfrak{su}}(2)_1$ chiral algebra (equivalently, whether it covers all boundary conditions of the $SU(2)_1$ WZW CFT). We leave this interesting question to the future.

It is also important that we impose that $\mathcal{C}$ is topologically closed in our notion of a ``good'' tensor category of unitary topological line defects. Indeed, otherwise, we could take $\mathcal{C}$ to be generated by the topological line defects corresponding to an arbitrary dense proper subgroup $G\subset SU(2)$. Such a category would satisfy \eqref{eqn:VCvir}, and yet not be the full category of unitary topological line defects of $\widehat{\mathfrak{su}}(2)_1$.  

\subsection{Classification of Topological Lines}\label{subsec:SU(2)1lines}

Let us next move on to describe the unitary topological line defects of the $SU(2)_1$ WZW model.

Just as for the $\widehat{\mathfrak{su}}(2)_1$ chiral algebra, the $SU(2)_1$ CFT has both left- and right-moving conserved Noether currents, which when exponentiated give rise to an internal $SO(4)\cong (SU(2)_L\times SU(2)_R)/\mathbb{Z}_2$ symmetry group (see e.g.\ \cite{Fuchs:2007tx}). We argue that this generates the full tensor category of unitary topological line operators in two ways (see also \cite{Moller:2024xtt} for an earlier attempt).

The first approach is to simply invoke the obvious analog of the proposal which we formulated in Section \ref{subsec:CBbdy}, generalized so that it applies to full 2D CFTs with both left- and right-movers.\footnote{In this version of the proposal, we assume that the 2D CFT is defined relative to an \emph{invertible} 3D bulk TQFT.} It reads: 

\noindent \textbf{Proposal.} A good tensor category $\mathcal{C}$ of unitary topological line defects of a 2D CFT $\mathcal{Q}$ is the \emph{full} category of such lines if and only if $\mathcal{Q}^{\mathcal{C}}=\mathrm{Vir}_c\otimes \overline{\mathrm{Vir}}_{\bar c}$, i.e.\ if and only if every local operator outside of the Virasoro subalgebras is charged under some topological line defect.  

Here, $c$ and $\bar{c}$ are the left and right central charges of $\mathcal{Q}$, respectively, and $\mathcal{Q}^\mathcal{C}$ denotes the set of local operators of $\mathcal{Q}$ which are invariant under all the topological lines in $\mathcal{C}$.
This proposal has the same remote detectability interpretation we provided in the previous subsection in the chiral case. We also note in passing that it has a SymTFT interpretation. Indeed, for a chosen symmetry category $\mathcal{C}$, the subalgebra $\mathcal{Q}^{\mathcal{C}}$ of operators which are invariant under $\mathcal{C}$ can be thought of as defining the ``physical boundary condition'' of the SymTFT. What this proposal says then is that if one chooses the \emph{maximal} unitary symmetry category of a 2D CFT, then the physical boundary condition is the product of the left- and right-moving Virasoro algebras, and the SymTFT is roughly the (double of the) Virasoro TQFT \cite{Collier:2023fwi}. 

It is straightforward to apply this proposal to $\mathcal{Q}$ the $SU(2)_1$ WZW CFT and $\mathcal{C}$ the category of lines generating the $SO(4)$ symmetry.
Indeed, by inspecting the Hilbert space of the theory (see Appendix \ref{sec:su2_spec}), 
\begin{equation}\label{eqn:SU(2)1HSdecomp}
    \mathcal{H}_{SU(2)_1} \cong \bigoplus_{\substack{J,\bar{J}\in \frac{1}{2}\mathbb{Z}_{\geq 0}\\J-\bar{J}\in\mathbb{Z}}} R_J \otimes R_{\bar{J}} \otimes \mathcal{V}_{J}\otimes \overline{\mathcal{V}}_{\bar{J}} \,,
\end{equation}
we see that the only operators transforming in the trivial representation of $SO(4)$ live in $\mathcal{V}_0\otimes \overline{\mathcal{V}}_0 = \mathrm{Vir}_{c=1}\otimes \overline{\mathrm{Vir}}_{\bar c=1}$. Thus, the proposal above tells us that $SO(4)$ captures the full tensor category of unitary topological line defects of the theory. 

Remarkably, a corollary of this is that the doubled $c=1$ Virasoro TQFT corresponds to the SymTFT of an (anomalous) $(SU(2)_L\times SU(2)_R)/\mathbb{Z}_2$ symmetry, which can in turn be described by a level $(k_L,k_R)=(1,-1)$ three-dimensional $(SU(2)_L\times SU(2)_R)/\mathbb{Z}_2$ Chern-Simons BF theory (cf.\ \cite{Apruzzi:2025hvs}). This observation will be treated in more detail in another paper.

Another approach is to directly calculate the topological lines of $SU(2)_1$ by gluing together the symmetries of its left- and right-moving chiral algebras. Indeed, a special case of the results of \cite{Gannon:2026ttf} is that, if $\mathcal{Q}$ is a diagonal rational CFT such that the left-moving chiral algebra hosts a tensor category $\mathcal{C}_L$ of unitary topological line operators, and the right-moving chiral algebra hosts a tensor category $\mathcal{C}_R$ of unitary topological line operators, then the category of lines in the full theory is given by the \emph{relative Deligne product over $\mathcal{B}$}, 
\begin{align}
    \mathbf{Sym}^\dagger(\mathcal{Q})=\mathcal{C}_L\boxtimes_{\mathcal{B}}\mathcal{C}_R^{\vee},
\end{align}
where the $\dagger$ refers to the fact that we are considering unitary topological line defects, $\mathcal{B}$ is the modular tensor category of anyons of the bulk TQFT, and $\mathcal{C}_R^{\vee}$ is the category dual to $\mathcal{C}_R$.\footnote{Both $\mathcal{C}_L$ and $\mathcal{C}_R$ are $\mathcal{B}$-module categories because anyons of the bulk TQFT can be fused onto topological line operators on the boundary.} We do not need the fully general statement: it suffices to observe that, when $\mathcal{C}_L$ and $\mathcal{C}_R$ are invertible symmetries described by groups $G_L$ and $G_R$ (likewise, call $H$ the group describing the bulk anyons $\mathcal{B}$), then this formula reduces to the \emph{amalgamated product}, 
\begin{align}    \mathbf{Sym}^\dagger(\mathcal{Q})=(G_L\times G_R)/H.
\end{align}
The intuitive idea behind this formula is that $G_L\times G_R$ overcounts the lines of the full theory, because e.g.\ lines on the left-moving chiral algebra in the subgroup $H\subset G_L$ can be dragged into the bulk and transported to the right-moving chiral algebra where they become lines in the subgroup $H\subset G_R.$ Thus, the two $H$ subgroups of $G_L$ and $G_R$ must be identified, which is what the quotient achieves. The relative Deligne product is a precise implementation of this phenomenon in the non-invertible setting.

For the $SU(2)_1$ WZW model, we determined earlier that the full category of unitary topological lines on its $\widehat{\mathfrak{su}}(2)_1$ chiral algebra is described by the group $G_L=SU(2)_L$ (and likewise $G_R=SU(2)_R$). On the other hand, the $SU(2)_1$ Chern-Simons theory has an $H=\mathbb{Z}_2$ worth of anyons, and so we compute that the category of lines of the full theory is described by $SO(4)\cong (SU(2)_L\times SU(2)_R)/\mathbb{Z}_2$.

Finally, before moving on, we record how the topological lines which implement the $(SU(2)_L\times SU(2)_R)/\mathbb{Z}_2$ symmetry of $SU(2)_1$ act on the boundary conditions by parallel fusion. Calling $\mathcal{L}_{(g_L,g_R)}$ the topological line defect which implements the symmetry $(g_L,g_R)\in (SU(2)_L\times SU(2)_R)/\mathbb{Z}_2$, it is known that (see also \eqref{eq:action_on_bdy_states})
\begin{align}\label{eqn:SO(4)bdyact}
    \mathcal{L}_{(g_L,g_R)}\otimes \mathbf{B}_g \cong \mathbf{B}_{g_L g g_R^{-1}},
\end{align}
where $\mathcal{L}_{(g_L,g_R)}\otimes \mathbf{B}_g$ denotes the boundary obtained by fusing the line $\mathcal{L}_{(g_L,g_R)}$ onto $\mathbf{B}_g$.

\section{Boundaries and Lines of Non-Compact Boson from Flat-Gauging} \label{sec:NCBAbs}

In the previous section, we described the boundary conditions and topological line defects of the $SU(2)_1$ WZW model.
In this section, we will derive these same observables for the non-compact boson theory using the fact that the non-compact boson can be obtained from $SU(2)_1$ by flat-gauging its $U(1)$ winding symmetry. 

Our main goal in this section is to calculate the moduli space of conformal boundaries and topological lines using  abstract algebraic properties of gauging interfaces and topological defects. In Section \ref{sec:NCBLag}, we will reproduce this moduli space with concrete Lagrangian descriptions. In Section \ref{sec:explicitNCB}, we will carry out more detailed calculations.

As we have emphasized earlier, one upshot of these investigations is that, because our classification of the boundaries and lines of $SU(2)_1$ is complete, we will likewise be able to conclude that the boundaries and topological lines we construct here exhaust \emph{all} unitary conformal boundaries and topological lines of the non-compact boson theory. Interestingly, most of the topological lines we find were not known in the literature before (though some of them did appear as boundaries of the $c=2$ theory of two non-compact bosons \cite{Callan:1995em}). 

Although we specialize  to the non-compact boson, much of our exposition applies mutatis mutandis to any theory obtained by flat gauging the diagonal $U(1)$ subgroup of a model with a symmetry group $(SU(2)_L\times SU(2)_R)/\mathbb{Z}_2$ and anomaly coinciding with that of the $SU(2)_1$ WZW model.

\subsection{Gauging Interface} \label{sec:int_abs}

Denote the topological line defects of $SU(2)_1$ which implement the $U(1)$ winding symmetry of the theory as 
\begin{align} \label{eq:LW}
    \mathcal{L}_\alpha^W\equiv \mathcal{L}_{(e^{i\alpha\sigma_3/2},e^{i\alpha\sigma_3/2})}, \ \ \ \ \alpha\sim \alpha+2\pi,
\end{align}
where $\sigma_3 = \left(\begin{smallmatrix} 1 & 0 \\ 0 & -1\end{smallmatrix}\right)$ is a Pauli matrix. It is known that flat gauging the $U(1)$ winding symmetry of $SU(2)_1$ yields the non-compact boson \cite{Argurio:2024ewp,Argurio:2026txf,Jia:2026tfh}. We will give a careful Lagrangian derivation of this fact in Section \ref{subsec:Lagflatgauging}. For now, we take this for granted and explore the consequences.

A crucial ingredient which enters our constructions in the sequel is the notion of a \emph{gauging interface} between a theory and its orbifold \cite{Carqueville:2012dk,Bhardwaj:2017xup,Choi:2021kmx,Choi:2022zal,Choi:2023vgk,Diatlyk:2023fwf}. In the present situation, this is constructed by starting with the $SU(2)_1$ WZW model and flat gauging the $U(1)$ winding symmetry in, say, the left half of spacetime, taking Dirichlet boundary conditions for the gauge field at the location of the interface. (See Section \ref{subsec:Laggauginginterface} for an explicit Lagrangian construction, as well as Section \ref{sec:gauging_interface_U1} for its detailed properties.) We draw this as follows: 
\begin{align}
\begin{tikzpicture}[baseline={([yshift=-0.5ex]0,1)}]
  \fill[ncb] (-2,0) rectangle (0,2);
  \fill[su]  (0,0) rectangle (2,2);
  \draw[frame] (-2,0) rectangle (2,2);
  \draw[iface] (0,0) -- (0,2);
  \node at (-1,1) {$\mathrm{NCB}$};
  \node at ( 1,1) {$SU(2)_1$};
  \node[ifacecol, below=1pt] at (0,0) {$\mathcal{I}_{U(1)}$};
  \node[overlay, anchor=base west, inner sep=0pt] at ([xshift=7pt]2,0.2) {.};
\end{tikzpicture}
\end{align}

We will only need a couple of properties of this gauging interface in this section. First, in general, fusing a gauging interface with its orientation reversal gives a direct sum over the topological lines which are being gauged (see \cite{Choi:2023vgk,Diatlyk:2023fwf} for a more precise statement in terms of algebra objects in the category of topological lines). In the present setting, we have that 
\begin{align}\label{eqn:IastI}
    \begin{tikzpicture}[baseline={([yshift=-0.5ex]0,1)}]
  \fill[su]  (-1.8,0) rectangle (0,2);
  \fill[ncb] (0,0) rectangle (1.6,2);
  \fill[su]  (1.6,0) rectangle (3.4,2);
  \draw[frame] (-1.8,0) rectangle (3.4,2);
  \draw[iface] (0,0) -- (0,2);
  \draw[iface] (1.6,0) -- (1.6,2);
  \node at (-0.9,1) {$SU(2)_1$};
  \node at ( 0.8,1) {$\mathrm{NCB}$};
  \node at ( 2.5,1) {$SU(2)_1$};
  \draw[->, line width=0.6pt] (0.56,0.5) -- (1.04,0.5);
  \node[ifacecol, below=1pt] at (0,0)   {$\overline{\mathcal{I}}_{U(1)}$};
  \node[ifacecol, below=1pt] at (1.6,0) {$\mathcal{I}_{U(1)}$};
\end{tikzpicture}
\;=\frac{\mathcal{N}}{2\pi}\int_{[0,2\pi)}^{\oplus} d\alpha\;
\begin{tikzpicture}[baseline={([yshift=-0.5ex]0,1)}]
  \fill[su] (-1.8,0) rectangle (1.8,2);
  \draw[frame] (-1.8,0) rectangle (1.8,2);
  \draw[tline, midarrow] (0,0) -- (0,2);
  \node at (-0.9,1) {$SU(2)_1$};
  \node at ( 0.9,1) {$SU(2)_1$};
  \node[below=1pt] at (0,0) {$\mathcal{L}^{W}_{\alpha}$};
  \node[overlay, anchor=base west, inner sep=0pt] at ([xshift=7pt]2,0.2) {,};
\end{tikzpicture}
\end{align}
where $\mathcal{N}$ is a (formally divergent) normalization constant which we will specify in Section \ref{sec:explicitNCB}. 

Let us denote the topological line defects which implement the $\mathbb{R}$ momentum symmetry $\Phi\to \Phi+\xi$ of the non-compact boson using $\mathcal{M}_\xi$. If we fuse the gauging interfaces in the other order, we instead obtain a direct sum over the topological line defects of the dual/quantum symmetry \cite{Vafa:1989ih}, which in this case is the $\mathbb{Z}$ subgroup of the $\mathbb{R}$ momentum symmetry generated by $\mathcal{M}_{2\pi}$. In pictures:
\begin{align}\label{eqn:IIast}
    \begin{tikzpicture}[baseline={([yshift=-0.5ex]0,1)}]
  \fill[ncb] (-1.8,0) rectangle (0,2);
  \fill[su]  (0,0) rectangle (1.6,2);
  \fill[ncb] (1.6,0) rectangle (3.4,2);
  \draw[frame] (-1.8,0) rectangle (3.4,2);
  \draw[iface] (0,0) -- (0,2);
  \draw[iface] (1.6,0) -- (1.6,2);
  \node at (-0.9,1) {$\mathrm{NCB}$};
  \node at ( 0.8,1) {$SU(2)_1$};
  \node at ( 2.5,1) {$\mathrm{NCB}$};
  \draw[->, line width=0.6pt] (0.56,0.5) -- (1.04,0.5);
  \node[ifacecol, below=1pt] at (0,0)   {$\mathcal{I}_{U(1)}$};
  \node[ifacecol, below=1pt] at (1.6,0) {$\overline{\mathcal{I}}_{U(1)}$};
\end{tikzpicture}
\;=\bigoplus_{n\in\mathbb{Z}}\,
\begin{tikzpicture}[baseline={([yshift=-0.5ex]0,1)}]
  \fill[ncb] (-1.8,0) rectangle (1.8,2);
  \draw[frame] (-1.8,0) rectangle (1.8,2);
  \draw[tline, midarrow] (0,0) -- (0,2);
  \node at (-0.9,1) {$\mathrm{NCB}$};
  \node at ( 0.9,1) {$\mathrm{NCB}$};
  \node[below=1pt] at (0,0) {$\mathcal{M}_{2\pi n}$};
  \node[overlay, anchor=base west, inner sep=0pt] at ([xshift=7pt]2,0.2) {.};
\end{tikzpicture}
\end{align}

One other useful feature we will exploit is that the topological lines which implement the symmetry that is being gauged are able to terminate topologically on these gauging interfaces. Concretely, this means that each of the $U(1)$ winding lines in $SU(2)_1$, and each of the $\mathbb{Z}$ momentum lines in the non-compact boson, admit a unique (up to scaling) topological point junction on $\mathcal{I}_{U(1)}$. Thus, we can draw pictures like this one:
\begin{align}
    \begin{tikzpicture}[baseline={([yshift=-0.5ex]0,1.3)}]
  \fill[ncb] (-2,0) rectangle (0,2.6);
  \fill[su]  (0,0) rectangle (2,2.6);
  \draw[frame] (-2,0) rectangle (2,2.6);
  \draw[tline, midarrow] (0,1.75) -- (-2,1.75);
  \draw[tline, midarrow] (0,0.85) -- (2,0.85);
  \draw[iface] (0,0) -- (0,2.6);
  \node at (-1,0.85) {$\mathrm{NCB}$};
  \node at ( 1,1.75) {$SU(2)_1$};
  \node[above=2pt] at (-1,1.75) {$\mathcal{M}_{2\pi n}$};
  \node[below=2pt] at ( 1,0.85) {$\mathcal{L}^{W}_{\alpha}$};
  \node[ifacecol, below=1pt] at (0,0) {$\mathcal{I}_{U(1)}$};
  \node[overlay, anchor=base west, inner sep=0pt] at ([xshift=7pt]2,0.2) {,};
\end{tikzpicture}
\end{align}
for any $\alpha \sim \alpha + 2\pi$ and $n\in\mathbb{Z}.$

\subsection{Conformal Boundaries}\label{subsec:NCBconformalbdy}

As in Section \ref{subsec:CBbdy}, the $SU(2)_1$ WZW model is known to admit a family of unitary conformal boundary conditions which form an $SU(2)$ group manifold \cite{Gaberdiel:2001zq,Callan:1994ub}. We will use gauging interfaces to determine the ``quantum'' boundary conditions of the non-compact boson theory which descend from this $SU(2)$ family. Because we are starting with the complete conformal moduli space of unitary boundary conditions of $SU(2)_1$, we will recover the complete conformal moduli space of unitary boundary conditions of the non-compact boson in this way. 

A similar determination of the full conformal manifold of unitary boundary conditions of the non-compact boson was made in \cite[Section 4.6]{Gannon:2026ttf}. We will reproduce the same answer obtained there using very different methods. We also comment that the boundary conditions we find here are equivalent to boundary conditions studied in \cite{Callan:1993mw,Callan:1994ub,Polchinski:1994my}, though these references did not discuss the question of completeness.

Every boundary condition of the non-compact boson can be reached by fusing the gauging interface $\mathcal{I}_{U(1)}$ described in the previous subsection onto a boundary condition $\mathbf{B}_g$ of $SU(2)_1$,
\begin{align}\label{eqn:GaugingIntFuseBdy}
\begin{tikzpicture}[baseline={([yshift=-0.5ex]0,1)}]
  \fill[ncb] (-2,0) rectangle (0,2);
  \fill[su]  (0,0) rectangle (2,2);
  \path[bstrip] (2,0) rectangle (2.15,2);
  \draw[frame] (2,0) -- (-2,0) -- (-2,2) -- (2,2);
  \draw[tline] (2,0) -- (2,2);
  \draw[iface] (0,0) -- (0,2);
  \node at (-1,1) {$\mathrm{NCB}$};
  \node at ( 1,1) {$SU(2)_1$};
  \draw[->, line width=0.6pt] (0.7,0.5) -- (1.3,0.5);
  \node[ifacecol, below=1pt] at (0,0) {$\mathcal{I}_{U(1)}$};
  \node[below=1pt] at (2,0) {$\mathbf{B}_{g}$};
\end{tikzpicture}
\;=\;
\begin{tikzpicture}[baseline={([yshift=-0.5ex]0,1)}]
  \fill[ncb] (-2,0) rectangle (0,2);
  \path[bstrip] (0,0) rectangle (0.15,2);
  \draw[frame] (0,0) -- (-2,0) -- (-2,2) -- (0,2);
  \draw[tline] (0,0) -- (0,2);
  \node at (-1,1) {$\mathrm{NCB}$};
  \node[below=1pt, inner xsep=0pt] (b) at (0,0) {$\mathcal{I}_{U(1)}\otimes\mathbf{B}_{g}$};
  \node[overlay, anchor=base west, inner sep=0pt] at (b.base east) {.};
\end{tikzpicture}
\end{align}
In fact, an analogous statement is true in general: any boundary condition of a theory obtained by orbifolding can be reached by fusing the corresponding gauging interface onto some boundary of the original theory. However we must take two subtleties into account.

The first subtlety is that some of these boundaries are equivalent to each other. 
Indeed, recall that the $U(1)$ winding symmetry lines $\mathcal{L}_\alpha^W$ are able to terminate on the gauging interface $\mathcal{I}_{U(1)}$. Using this fact, we can construct a topological point interface between $\mathcal{I}_{U(1)}\otimes \mathbf{B}_g$ and $\mathcal{I}_{U(1)}\otimes \mathbf{B}_{g'}$ if the two boundaries $\mathbf{B}_g$ and $\mathbf{B}_{g'}$ are related by the $U(1)$ winding symmetry, 
\begin{align} \label{eq:xalpha}
    \begin{tikzpicture}[baseline={([yshift=-0.5ex]0,1)}]
  \fill[ncb] (-2,0) rectangle (0,2);
  \fill[su]  (0,0) rectangle (2,2);
  \path[bstrip] (2,0) rectangle (2.15,2);
  \draw[frame] (2,0) -- (-2,0) -- (-2,2) -- (2,2);
  \draw[tline, midarrow] (2,1) -- (0,1);
  \draw[tline] (2,0) -- (2,2);
  \draw[iface] (0,0) -- (0,2);
  \node at (-1,1.55) {$\mathrm{NCB}$};
  \node at ( 1,1.55) {$SU(2)_1$};
  \node[below=2pt] at (1,1) {$\mathcal{L}^{W}_{\alpha}$};
  \node[ifacecol, below=1pt] at (0,0) {$\mathcal{I}_{U(1)}$};
  \node[anchor=west, inner xsep=1.5pt] at (2.15,1.5) {$\mathbf{B}_{g'}$};
  \node[anchor=west, inner xsep=1.5pt] at (2.15,0.5) {$\mathbf{B}_{g}$};
\end{tikzpicture}
\;=\;
\begin{tikzpicture}[baseline={([yshift=-0.5ex]0,1)}]
  \fill[ncb] (-2,0) rectangle (0,2);
  \path[bstrip] (0,0) rectangle (0.15,2);
  \draw[frame] (0,0) -- (-2,0) -- (-2,2) -- (0,2);
  \draw[tline] (0,0) -- (0,2);
  \fill (0,1) circle (2.2pt);
  \node at (-1,1.55) {$\mathrm{NCB}$};
  \node[left=3pt] at (0,1) {$x_{\alpha}$};
  \node[anchor=west, inner xsep=1.5pt] at (0.15,1.5) {$\mathcal{I}_{U(1)}\otimes\mathbf{B}_{g'}$};
  \node[anchor=west, inner xsep=1.5pt] at (0.15,0.5) {$\mathcal{I}_{U(1)}\otimes\mathbf{B}_{g}$};
\end{tikzpicture}
\end{align}
where we have assumed that $ \mathbf{B}_g\cong \mathcal{L}^W_\alpha\otimes\mathbf{B}_{g'}$ for some $\alpha \in U(1)$. Furthermore, it is straightforward to see that this topological point interface $x_\alpha$ is invertible, with inverse given (up to a possible constant of proportionality) by $x_{-\alpha}$.
Indeed, one can readily see that $x_{-\alpha}\otimes x_\alpha \propto \mathds{1}.$ An invertible topological point interface between two boundaries is essentially an equivalence of those boundaries, and hence we can conclude that 
\begin{align}\label{eqn:equivalentbdys}
    \mathcal{I}_{U(1)}\otimes \mathbf{B}_g\cong \mathcal{I}_{U(1)}\otimes \mathbf{B}_{g'} \ \ \ \text{ if }  \mathbf{B}_g\cong \mathcal{L}_\alpha^W\otimes\mathbf{B}_{g'},
\end{align}
for some $\alpha\in U(1)$.

The second subtlety is that $\mathcal{I}_{U(1)}\otimes\mathbf{B}_g$ may not be simple, i.e.\ it may decompose into a direct sum of other boundary conditions. Indeed, in the situation that a boundary $\mathbf{B}_g$ is invariant under the $U(1)$ winding symmetry, $\mathcal{L}_\alpha^W\otimes \mathbf{B}_g\cong \mathbf{B}_g$, it follows that the topological point interface $x_\alpha$ constructed above is actually a topological point operator on $\mathcal{I}_{U(1)}\otimes \mathbf{B}_g$, i.e.\ a topological point interface from this boundary to itself. The existence of a non-identity topological point operator on a boundary signals that the boundary is not simple \cite{Choi:2023xjw}, i.e.\ that it decomposes into a direct sum of other boundary conditions, 
\begin{align}\label{eqn:decompbdysimples}
    \mathcal{I}_{U(1)}\otimes \mathbf{B}_g \cong n_1 B_1\oplus n_2 B_2\oplus \cdots,
\end{align}
where each $B_i$ is simple and occurs with multiplicity $n_i\in\mathbb{Z}_{\geq 0}$ in $\mathcal{I}_{U(1)}\otimes \mathbf{B}_g$.  This  decomposition tells us that the algebra of topological point operators on $\mathcal{I}_{U(1)}\otimes \mathbf{B}_g$ is
\begin{align}\label{eq:homIB}
    \mathrm{Hom}(\mathcal{I}_{U(1)}\otimes\mathbf{B}_g,\mathcal{I}_{U(1)}\otimes\mathbf{B}_g)\cong \bigoplus_{i}M_{n_i}(\mathbb{C}),
\end{align}
where $M_n(\mathbb{C})$ is the algebra of $n\times n$ matrices with complex entries.

Now, it is clear that, if $\mathbf{B}_g$ is invariant under the $U(1)$ winding symmetry, then the algebra of topological point operators on $\mathcal{I}_{U(1)}\otimes \mathbf{B}_g$, which is generated by the $x_\alpha$, is isomorphic to the group algebra of $U(1)$,\footnote{More precisely, we need the \emph{group von Neumann algebra of $U(1)$}, and when we write $\bigoplus$ we really mean the ``bounded product'', but we will be cavalier about these distinctions below.}  
\begin{align}\label{HomIBg}
    \mathrm{Hom}(\mathcal{I}_{U(1)}\otimes \mathbf{B}_g,\mathcal{I}_{U(1)}\otimes \mathbf{B}_g)\cong \mathbb{C}[U(1)]\cong \bigoplus_{Q\in\mathbb{Z}}M_1(\mathbb{C}). 
\end{align}
Here, we use the fact that the group algebra $\mathbb{C}[G]$ for a compact group $G$ can be decomposed as
\ie
\mathbb{C}[G]\cong \bigoplus_{R\in\text{Irr}(G)}M_{d_R}(\mathbb{C})
\fe
where $R$ labels inequivalent irreducible complex representations of $G$ and $d_R$ is the dimension of the representation $R$. Since $U(1)$ representations are all one-dimensional and are labeled by their integer charges $Q$, the group algebra $\mathbb{C}[U(1)]$ decomposes into a countably infinite direct sum of $1\times 1$ matrix algebras.
Matching \eqref{eq:homIB} and \eqref{HomIBg} onto each other, we learn that $\mathcal{I}_{U(1)}\otimes\mathbf{B}_g$ dissolves/fractionalizes into $\mathbb{Z}$-many simple summands, each occuring with multiplicity $1$.  Each simple summand can be obtain by projecting $\mathcal{I}_{U(1)}\otimes \mathbf{B}_g$ to the corresponding charge $Q$ representation of $\mathbb{C}[U(1)]$ as
\begin{align} \label{eq:bdy_proj}
    \left[\mathcal{I}_{U(1)}\otimes\mathbf{B}_{g}\right]^{Q}
=\frac{1}{2\pi}\int_0^{2\pi} d\alpha\, e^{-i\alpha Q}\;
\begin{tikzpicture}[baseline={([yshift=-0.5ex]0,1)}]
  \fill[ncb] (-2,0) rectangle (0,2);
  \path[bstrip] (0,0) rectangle (0.15,2);
  \draw[frame] (0,0) -- (-2,0) -- (-2,2) -- (0,2);
  \draw[tline] (0,0) -- (0,2);
  \fill (0,1) circle (2.2pt);
  \node at (-1,1.55) {$\mathrm{NCB}$};
  \node[left=3pt] at (0,1) {$x_{\alpha}$};
  \node[anchor=west, inner xsep=1.5pt] at (0.15,0.5) {$\mathcal{I}_{U(1)}\otimes\mathbf{B}_{g}$};
  \node[overlay, anchor=base west, inner sep=0pt] at ([xshift=7pt]2,0.2) {.};
\end{tikzpicture}
\end{align}

Now, let us unpack what this means for the non-compact boson theory. We use $\mathcal{B}_{SU(2)_1}$ to denote the category of unitary conformal boundary conditions of the $SU(2)_1$ WZW model, and $\mathrm{Irr}(\mathcal{B}_{SU(2)_1})$ to denote the set of simple boundaries in $\mathcal{B}_{SU(2)_1}$. Following the discussion in Section \ref{subsec:CBbdy}, the simple boundary conditions can be parametrized as
\begin{equation}
    \mathrm{Irr}(\mathcal{B}_{SU(2)_1}) = \{ \mathbf{B}_{a,b} \mid a, b \in \mathbb{C} , |a|^2 + |b|^2 = 1\} \,,
\end{equation}
where we write $\mathbf{B}_{a,b}\equiv \mathbf{B}_g$ for $g$ the $SU(2)$ matrix given by
\begin{align} \label{eq:gab}
    g=\left(\begin{array}{cc} a & b \\ -\bar b & \bar a\end{array}\right), \ \ \ \ a,b\in\mathbb{C}\,, \ |a|^2+|b|^2=1\,.
\end{align}
Equation \eqref{eqn:SO(4)bdyact} shows that the $U(1)$ winding symmetry \eqref{eq:LW} acts on the simple boundary conditions according to
\begin{align}\label{eqn:fusewindingbdy}
\begin{split}
    \mathcal{L}^W_{\alpha} \otimes \mathbf{B}_{a,b} = \mathbf{B}_{a,be^{i\alpha}} \,.
\end{split}
\end{align}
There are therefore two cases to consider.
\begin{enumerate}
\item If $|a|<1$, then $|b|>0$ and hence the boundary $\mathbf{B}_{a,b}$ breaks the $U(1)$ winding symmetry. In fact, any two boundaries $\mathbf{B}_{a,b}$ and $\mathbf{B}_{a,b'}$ with the same value of $a$ are related by fusing a $U(1)$ winding topological line defect, $\mathbf{B}_{a,b}\cong \mathcal{L}_\alpha^W\otimes \mathbf{B}_{a,b'}$, for some $\alpha \in U(1)$. Using \eqref{eqn:equivalentbdys}, we learn that the boundary of the non-compact boson CFT obtained by fusing the gauging interface onto $\mathbf{B}_{a,b}$ depends only on $a$, and we label it as
\begin{align}\label{eqn:Nadef}
    \mathbf{N}_a\equiv \mathcal{I}_{U(1)}\otimes \mathbf{B}_{a,b}, \ \ \ \ a\in \mathbb{C}, \ |a|<1.
\end{align}
As we will demonstrate in Section \ref{sec:explicitNCB}, $\mathbf{N}_0$ is the Neumann boundary condition of the non-compact boson. The $\mathbf{N}_a$ form a two-parameter family which we refer to as \emph{generalized Neumann} boundary conditions.
\item If $|a|=1$, then $b=0$ and $\mathbf{B}_{a,b}=\mathbf{B}_{a,0}$ is invariant under the $U(1)$ winding symmetry. Thus, the boundary obtained by fusing the gauging interface onto $\mathbf{B}_{a,0}$ dissolves into an infinite direct sum of elementary boundary conditions  
\begin{equation}
\mathbf{D}_{\Phi_0}\equiv \left[\mathcal{I}_{U(1)}\otimes \mathbf{B}_{e^{i\Phi_0},0}\right]^Q\,,\quad Q=\left(\frac{1}{2\pi}\Phi_0\text{ mod }1\right)-\frac{1}{2\pi}\Phi_0.
\end{equation}
In later sections, we will show that $\mathbf{D}_{\Phi_0}$ is a Dirichlet boundary condition of the non-compact boson which sets $\Phi\vert_{\mathrm{bdy}}=\Phi_0$,  and that their superposition gives\footnote{The fact that a discrete superposition of Dirichlet boundary conditions arises as a limit of the generalized Neumann boundary conditions was first anticipated in \cite{Callan:1993mw}.}
\begin{align}\label{eqn:latticeofdirichlet}
    \mathcal{I}_{U(1)}\otimes \mathbf{B}_{a,0}\cong \bigoplus_{Q\in\mathbb{Z}}\mathbf{D}_{\arg(a)-2\pi Q}.
\end{align}
\end{enumerate}

Summarizing, the space of unitary conformal boundary conditions of the non-compact boson theory which follows from the $U(1)$ flat gauging procedure is a disk, parameterized by $a$.
The boundary conditions are simple inside the disk, and become an infinite direct sum of Dirichlet boundary conditions on the boundary of the disk, 
\begin{align}\label{eqn:Na_dissolve}
    \mathbf{N}_a \xrightarrow{|a|\to 1}\bigoplus_{Q\in\mathbb{Z}}\mathbf{D}_{\mathrm{arg}(a)-2\pi Q}.
\end{align}

Let $\mathcal{B}_{NCB}$ be the category of unitary conformal boundary conditions of the non-compact boson, and $\mathrm{Irr}(\mathcal{B}_{\NCB} )$ be the set of simple boundaries.
From the above, we have that
\begin{equation}\label{eq:NCB_BC_space}
    \mathrm{Irr}(\mathcal{B}_{\NCB} ) = \mathring{D}^2 \cup \mathbb{R} \,,
\end{equation}
where $\mathring{D}^2$ is the interior of a unit disk representing generalized Neumann boundary conditions, and $\mathbb{R}$ corresponds to the set of Dirichlet boundary conditions.
See Figure \ref{fig:ncb_bdy} for an illustration.

\begin{figure}
    \begin{center}
        \begin{tikzpicture}[baseline={([yshift=-0.5ex]0,1.2)}]
  \fill[su]  (-2.6,0) -- (-2.2,0) -- (0,1.2) -- (0,2.4) -- (-2.6,2.4) -- cycle;
  \fill[ncb] (-2.2,0) -- (0,0) -- (0,1.2) -- cycle;
  \path[bstrip] (0,0) rectangle (0.15,2.4);
  \draw[frame] (0,0) -- (-2.6,0) -- (-2.6,2.4) -- (0,2.4);
  \draw[iface] (0,1.2) -- (-2.2,0);
  \draw[tline] (0,0) -- (0,2.4);
  \fill (0,1.2) circle (2.2pt);
  \node at (-1.45,1.75) {$SU(2)_1$};
  \node at (-0.7,0.3)   {$\mathrm{NCB}$};
  \node[above left=1pt] at (0,1.2) {$x$};
  \path (0,1.2) -- (-2.2,0) node[pos=0.6, anchor=south east, ifacecol, inner sep=2pt] {$\overline{\mathcal{I}}_{U(1)}$};
  \node[anchor=west, inner xsep=1.5pt] at (0.15,1.8) {$\mathbf{B}_{a,b}$};
  \node[anchor=west, inner xsep=1.5pt] at (0.15,0.6) {$\mathbf{D}_{\Phi_0}$};
\end{tikzpicture}
$\quad\longleftrightarrow\quad$
\begin{tikzpicture}[baseline={([yshift=-0.5ex]0,1.2)}]
  \fill[ncb] (-2.6,0) -- (0,0) -- (0,1.2) -- (-2.2,2.4) -- (-2.6,2.4) -- cycle;
  \fill[su]  (0,1.2) -- (0,2.4) -- (-2.2,2.4) -- cycle;
  \path[bstrip] (0,0) rectangle (0.15,2.4);
  \draw[frame] (0,0) -- (-2.6,0) -- (-2.6,2.4) -- (0,2.4);
  \draw[iface] (0,1.2) -- (-2.2,2.4);
  \draw[tline] (0,0) -- (0,2.4);
  \fill (0,1.2) circle (2.2pt);
  \node at (-0.7,2.12)  {$SU(2)_1$};
  \node at (-1.45,0.65) {$\mathrm{NCB}$};
  \node[below left=1pt] at (0,1.2) {$y$};
  \path (0,1.2) -- (-2.2,2.4) node[pos=0.6, anchor=north east, ifacecol, inner sep=2pt] {$\mathcal{I}_{U(1)}$};
  \node[anchor=west, inner xsep=1.5pt] at (0.15,1.8) {$\mathbf{B}_{a,b}$};
  \node[anchor=west, inner xsep=1.5pt] at (0.15,0.6) {$\mathbf{D}_{\Phi_0}$};
\end{tikzpicture}
        \caption{The proof of the first equality of Equation \eqref{eqn:frobrep}. There is a one-to-one correspondence between topological junctions $x\in\mathrm{Hom}(\overline{\mathcal{I}}_{U(1)}\otimes \mathbf{D}_{\Phi_0},\mathbf{B}_{a,b})$ and junctions $y\in \mathrm{Hom}(\mathbf{D}_{\Phi_0},\mathcal{I}_{U(1)}\otimes \mathbf{B}_{a,b})$. The correspondence is obtained by bending the gauging interface upwards.}\label{fig:frobrep}
    \end{center}
\end{figure}
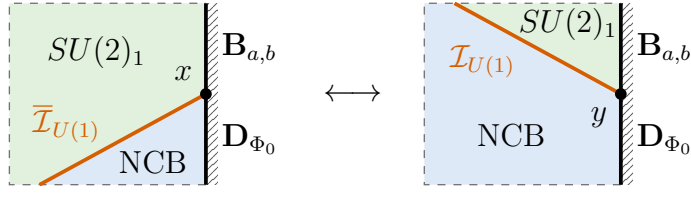

Before moving on, we comment that it is straightforward to determine what happens when one fuses the (orientation reversal of the) gauging interface onto boundaries of the non-compact boson, which will reproduce boundaries of $SU(2)_1$. For example, we can easily calculate that
\begin{align}\label{eqn:I*N}
    \overline{\mathcal{I}}_{U(1)}\otimes \mathbf{N}_a\cong \overline{\mathcal{I}}_{U(1)}\otimes \mathcal{I}_{U(1)}\otimes \mathbf{B}_{a,b} \cong \frac{\mathcal{N}}{2\pi}\int_{[0,2\pi)}d\alpha ~\mathcal{L}_\alpha^W\otimes \mathbf{B}_{a,b}\cong  \frac{\mathcal{N}}{2\pi}\int_{[0,2\pi)}d\alpha~\mathbf{B}_{a,be^{i\alpha}},
\end{align}
where we have made use of \eqref{eqn:IastI} and \eqref{eqn:fusewindingbdy}. The action on Dirichlet boundary conditions can be determined using that (cf.\ Figure \ref{fig:frobrep}) 
\begin{align}\label{eqn:frobrep}
    \dim(\mathrm{Hom}(\overline{\mathcal{I}}_{U(1)}\otimes \mathbf{D}_{\Phi_0},\mathbf{B}_{a,b}))=\dim(\mathrm{Hom}(\mathbf{D}_{\Phi_0},\mathcal{I}_{U(1)}\otimes \mathbf{B}_{a,b}))=\delta_{e^{i\Phi_0},a},
\end{align}
where the second equality was obtained by plugging in \eqref{eqn:latticeofdirichlet}.
This in particular implies that 
\begin{align}\label{eqn:I*D}
    \overline{\mathcal{I}}_{U(1)}\otimes \mathbf{D}_{\Phi_0}\cong \mathbf{B}_{e^{i\Phi_0},0}.
\end{align}
Being able to go back and forth between the boundaries of the non-compact boson and the boundaries of $SU(2)_1$ will be useful for the computations we carry out below.

\subsection{Topological Lines}\label{subsec:abstract_top_lines}

We now move on to discuss the unitary topological line defects of the non-compact boson theory. 
Again, we obtain these by applying the $U(1)$ flat gauging to the $SU(2)_1$ WZW model, whose unitary topological lines are all known as in Section \ref{subsec:SU(2)1lines}.

\subsubsection{Moduli Space}\label{subsubsec:modulispacelines}

Let us first abstractly describe the moduli space of topological line defects of the non-compact boson theory.

Analogously to boundaries, every unitary topological line defect of the non-compact boson theory can be reached by sandwiching a topological line defect of the $SU(2)_1$ WZW model with a gauging interface:
\begin{align} \label{eq:sandwich}
    \begin{tikzpicture}[baseline={([yshift=-0.5ex]0,1)}]
  \fill[ncb] (-1.8,0) rectangle (0,2);
  \fill[su]  (0,0) rectangle (3.2,2);
  \fill[ncb] (3.2,0) rectangle (5,2);
  \draw[frame] (-1.8,0) rectangle (5,2);
  \draw[iface] (0,0) -- (0,2);
  \draw[tline, midarrow] (1.6,0) -- (1.6,2);
  \draw[iface] (3.2,0) -- (3.2,2);
  \node at (-0.9,1.45) {$\mathrm{NCB}$};
  \node at ( 0.8,1.45) {$SU(2)_1$};
  \node at ( 2.4,1.45) {$SU(2)_1$};
  \node at ( 4.1,1.45) {$\mathrm{NCB}$};
  \draw[->, line width=0.6pt] (0.56,0.5) -- (1.04,0.5);
  \draw[->, line width=0.6pt] (2.64,0.5) -- (2.16,0.5);
  \node[ifacecol, below=1pt] at (0,0)   {$\mathcal{I}_{U(1)}$};
  \node[below=1pt]           at (1.6,0) {$\mathcal{L}_{(g_L,g_R)}$};
  \node[ifacecol, below=1pt] at (3.2,0) {$\overline{\mathcal{I}}_{U(1)}$};
\end{tikzpicture}
\;\equiv\;
\begin{tikzpicture}[baseline={([yshift=-0.5ex]0,1)}]
  \fill[ncb] (-1.8,0) rectangle (1.8,2);
  \draw[frame] (-1.8,0) rectangle (1.8,2);
  \draw[tline, midarrow] (0,0) -- (0,2);
  \node at (-0.9,1.45) {$\mathrm{NCB}$};
  \node at ( 0.9,1.45) {$\mathrm{NCB}$};
  \node[below=1pt] at (0,0) {$\mathcal{S}(g_L,g_R)$};
  \node[overlay, anchor=base west, inner sep=0pt] at ([xshift=3pt]1.8,0) {.};
\end{tikzpicture}
\end{align}
To save ink, we have defined 
\begin{align}
    \mathcal{S}(g_L,g_R)\equiv \mathcal{I}_{U(1)}\otimes \mathcal{L}_{(g_L,g_R)}\otimes \overline{\mathcal{I}}_{U(1)}.
\end{align}As in our discussion of conformal boundary conditions in Section \ref{subsec:NCBconformalbdy}, there are two subtleties we must take into account. 

The first subtlety is that two distinct topological line defects of $SU(2)_1$ may lead to the same line in the non-compact boson after sandwiching with gauging interfaces. More specifically, one can see that
\begin{align}
    \mathcal{S}(g_L,g_R)\cong \mathcal{S}(g_L',g_R') \ \ \  \text{ if } \mathcal{L}_\alpha^W\otimes \mathcal{L}_{(g_L,g_R)}\otimes \mathcal{L}_\beta^W \cong \mathcal{L}_{(g_L',g_R')},
\end{align}
for some $\alpha,\beta\in U(1)$. Indeed, in this situation, one can construct a topological point interface $x_{\alpha,\beta}$ between $\mathcal{S}(g_L,g_R)$ and $\mathcal{S}(g_L',g_R')$, 
\begin{align}\label{eqn:xalphabeta}
    \begin{tikzpicture}[baseline={([yshift=-0.5ex]0,1.2)}]
  \fill[ncb] (-1.8,0) rectangle (0,2.4);
  \fill[su]  (0,0) rectangle (3.2,2.4);
  \fill[ncb] (3.2,0) rectangle (5,2.4);
  \draw[frame] (-1.8,0) rectangle (5,2.4);
  \draw[tline, midarrow] (0,1.2) -- (1.6,1.2);
  \draw[tline, midarrow] (3.2,1.2) -- (1.6,1.2);
  \draw[tline, midarrow] (1.6,0) -- (1.6,1.2);
  \draw[tline, midarrow] (1.6,1.2) -- (1.6,2.4);
  \draw[iface] (0,0) -- (0,2.4);
  \draw[iface] (3.2,0) -- (3.2,2.4);
  \node at (-0.9,1.9) {$\mathrm{NCB}$};
  \node at ( 0.8,1.9) {$SU(2)_1$};
  \node at ( 2.4,1.9) {$SU(2)_1$};
  \node at ( 4.1,1.9) {$\mathrm{NCB}$};
  \node[below=2pt] at (0.8,1.2) {$\mathcal{L}^{W}_{\alpha}$};
  \node[below=2pt] at (2.4,1.2) {$\mathcal{L}^{W}_{\beta}$};
  \node[ifacecol, below=1pt] at (0,0)   {$\mathcal{I}_{U(1)}$};
  \node[below=1pt]           at (1.6,0) {$\mathcal{L}_{(g_L,g_R)}$};
  \node[above=1pt]           at (1.6,2.4) {$\mathcal{L}_{(g'_L,g'_R)}$};
  \node[ifacecol, below=1pt] at (3.2,0) {$\overline{\mathcal{I}}_{U(1)}$};
\end{tikzpicture}
\;\equiv\;
\begin{tikzpicture}[baseline={([yshift=-0.5ex]0,1.2)}]
  \fill[ncb] (-1.8,0) rectangle (1.8,2.4);
  \draw[frame] (-1.8,0) rectangle (1.8,2.4);
  \draw[tline, midarrow] (0,0) -- (0,1.2);
  \draw[tline, midarrow] (0,1.2) -- (0,2.4);
  \fill (0,1.2) circle (2.2pt);
  \node at (-0.9,1.9) {$\mathrm{NCB}$};
  \node at ( 0.9,1.9) {$\mathrm{NCB}$};
  \node[left=3pt] at (0,1.2) {$x_{\alpha,\beta}$};
  \node[below=1pt] at (0,0)   {$\mathcal{S}(g_L,g_R)$};
  \node[above=1pt] at (0,2.4) {$\mathcal{S}(g'_L,g'_R)$};
\end{tikzpicture}
\end{align}
which can easily seen to be invertible. In particular, this means that the sandwich line $\mathcal{S}(g_L,g_R)$ depends only on the double coset $U(1)(g_L,g_R)U(1)$ of $(g_L,g_R)\in SO(4)$.\footnote{Similar discussions in the case of ordinary gauging of finite symmetries are in \cite{Hsin:2024aqb,Hsin:2025ria}.}

The second subtlety is that $\mathcal{S}(g_L,g_R)$ may not be simple. If $\mathcal{L}_\alpha^W\otimes \mathcal{L}_{(g_L,g_R)}\otimes \mathcal{L}_\beta^W\cong \mathcal{L}_{(g_L,g_R)}$, then the topological point interface $x_{\alpha,\beta}$ is in fact a topological point operator on $\mathcal{S}(g_L,g_R)$, which signals that $\mathcal{S}(g_L,g_R)$ decomposes non-trivially into a direct sum of simple lines. Parametrizing group elements of $SO(4)$ as 
\begin{equation}\label{eqn:SO(4)elements}
    g_L = \begin{pmatrix}
        a_L & b_L \\ -\overline{b}_L & \overline{a}_L
    \end{pmatrix} \,, \quad g_R = \begin{pmatrix}
        a_R & b_R \\ -\overline{b}_R & \overline a_R
    \end{pmatrix} \,, \quad |a_L|^2 + |b_L|^2 = |a_R|^2 + |b_R|^2 = 1  \,,
\end{equation}
the algebra of topological point operators on $\mathcal{S}(g_L,g_R)$ is straightforwardly seen to be $\mathbb{C}[H^{(g_L,g_R)}]$, where 
\begin{equation} \label{eq:u1_Hg}
    H^{(g_L , g_R)} = U(1) \cap (g_L , g_R) U(1) (g_L , g_R)^{-1} = \begin{dcases}
        U(1) ~~~& \text{if $a_L = a_R = 0$ or $b_L = b_R = 0$} \,,\\
        \{1\} ~~~& \text{otherwise}.
    \end{dcases}
\end{equation}
Thus, if $a_L=a_R=0$ or $b_L=b_R=0$, the same logic we employed in the case of conformal boundary conditions tells us that $\mathcal{S}(g_L,g_R)$ decomposes into $\mathbb{Z}$-many simple topological lines,
\begin{align}
    \mathcal{S}(g_L,g_R)\cong \bigoplus_{Q\in\mathbb{Z}} [\mathcal{S}(g_L,g_R)]^Q, \ \ \text{if }a_L=a_R=0 \text{ or }b_L=b_R=0,
\end{align}
which we have denoted
$[\mathcal{S}(g_L,g_R)]^Q$, with $Q\in\mathbb{Z}$.

Putting this all together, we learn that the simple topological line defects in the non-compact boson theory are 
\begin{align}
   [\mathcal{S}(g_L,g_R)]^Q, \ \ \ \ [(g_L,g_R)]\in U(1)\backslash SO(4)/U(1), \ \ Q\in\mathrm{Rep}(H^{(g_L,g_R)}).
\end{align}
Let us unpack this moduli space more explicitly. To begin, we compute the double coset space
\begin{equation} \label{eq:u1_double_coset}
    U(1) \backslash SO(4) / U(1) \,,
\end{equation}
where the $U(1) \subset SO(4)$ subgroup is generated by elements of the form $\left( e^{i\alpha \sigma_3 /2} , e^{i\alpha \sigma_3 /2} \right)$ \eqref{eq:LW}.
The left and right actions of $U(1)$ on $SO(4)$ imply the following identifications on the group elements in \eqref{eqn:SO(4)elements}:
\begin{equation} \label{eq:u1_double_coset_identifications}
    (a_L , b_L ; a_R , b_R) \sim (a_L e^{i(\alpha + \alpha')/2} , b_L e^{i(\alpha - \alpha')/2} ;a_R e^{i(\alpha + \alpha')/2}, b_R e^{i(\alpha - \alpha')/2}) \,,
\end{equation}
where $\alpha$ and $ \alpha'$ are arbitrary.\footnote{The identification $[(a_L , b_L ) , (a_R , b_R)] = [(-a_L , -b_L ) , (-a_R , -b_R)]$ coming from the $\mathbb{Z}_2$ quotient in $SO(4) \cong [SU(2)_L \times SU(2)_R]/\mathbb{Z}_2$ is subsumed by \eqref{eq:u1_double_coset_identifications}.}
It is convenient to parametrize the double coset space by associating, to any element $(g_L,g_R)\in SO(4)$,  the following quantities
\begin{align}\label{eqn:doublecosetcoordinates}
    x_L\equiv |a_L|, \ \ \ x_R \equiv |a_R|, \ \ \ \theta_a\equiv \arg(a_L\overline{a_R}), \ \ \ \theta_b \equiv \arg(b_L\overline{b_R}).
\end{align}
We always take $\theta_a , \theta_b \in [0,2\pi)$, and when e.g.\ $a_L\overline{a_R}=0$, we adopt the convention that $\theta_a=0$.

It is not hard to convince oneself that the quantities in \eqref{eqn:doublecosetcoordinates} are invariant along every double coset. Thus, if $(g_L,g_R)$ and $(g_L',g_R')$ have different invariants $(x_L,x_R,\theta_a,\theta_b)$ and $ (x_L',x_R',\theta_a',\theta'_b)$, then they belong to different double cosets. Furthermore, we can construct a distinguished representative which realizes each $(x_L,x_R,\theta_a,\theta_b)$ as 
\begin{align}\label{eqn:doublecosetrep}
    \mathbf{g}(x_L,x_R,\vec\theta)\equiv\big(g(x_Le^{i\theta_a/2},\theta_b/2),g(x_Re^{-i\theta_a/2},-\theta_b/2)\big),
\end{align}
where we have defined 
\begin{align}\label{eqn:gapsi}
    g(a,\psi) = \left(\begin{array}{cc} a & \sqrt{1-|a|^2}e^{i\psi} \\ 
    -\sqrt{1-|a|^2}e^{-i\psi} & \bar a\end{array}\right).
\end{align}
Thus, we will coordinatize the double-coset space by $(x_L,x_R,\theta_a,\theta_b)$.\footnote{If we shift, say, $\theta_a \to \theta_a + 2\pi$, then the representative $\mathbf{g}(x_L,x_R,\vec\theta)$ in \eqref{eqn:doublecosetrep} changes to another group element which is in the same double coset as $\mathbf{g}(x_L,x_R,\vec\theta)$.}

Let us understand its global structure. The coordinates $(x_L,x_R)$ trace out a square, 
\begin{align}
    (x_L,x_R)\in [0,1]\times [0,1].
\end{align}
The angles $(\theta_a,\theta_b)$ can be thought of as defining a (singular) torus $T^2=S^1_a\times S^1_b$ fibration over this square. The reason we use the word ``singular'' is that the torus fibration degenerates over certain loci in the base. Specifically, at the boundary of the square,
\begin{align}
\begin{split}
    x_L\to 1 \text{ or } x_R\to 1 \implies S^1_b \text{ degenerates to a point,} \\
    x_L\to 0 \text{ or } x_R\to 0 \implies S^1_a \text{ degenerates to a point.}
\end{split}
\end{align}
In particular, at the special points $(x_L,x_R)=(0,1)$ and $(1,0)$ in the base, both circles in the $T^2$ fiber degenerate to a point.

It turns out that a $T^2$ fibration over $[0,1]^2$ with this structure is topologically the same as $S^4$, so that in particular 
\begin{align}
    U(1)\backslash SO(4)/U(1)\cong S^4.
\end{align}
Let us argue for this by exhibiting a fibration with the same structure for $S^4$ in its usual presentation. We model the sphere as the subspace of $\mathbb{C}_u\times \mathbb{C}_v\times\mathbb{R}_t$ with $|u|^2+|v|^2+t^2=1$. Consider the map  $\pi : S^4 \to \mathbb{R}^2$ given by 
\begin{align}
    \pi(u,v,t)=(s,t)=(|u|,t).
\end{align}
The image is $\pi(S^4)=\{(s,t)\mid 0\leq s\leq \sqrt{1-t^2}\}$, which one can think of as the portion of the unit disk which resides in the upper half-plane (topologically the same as a square $[0,1]\times [0,1]$). We will think of $\pi(S^4)$ as the base of our fibration, and the preimage $\pi^{-1}(s,t)$ of a point $(s,t)$ in the base as the fiber over that point. The preimage over each point $(s,t)$ in the interior of the base is $T^2=S^1_u\times S^1_v$, consisting of the phase of $u$ and the phase of $v$. The points of $S^4$ which map to the interval on the boundary defined by $s=0$ all have $u=0$, and hence the circle $S^1_u$ in the fiber degenerates to a point. The points which map to the interval on the boundary defined by $s=\sqrt{1-t^2}$ all have $v=0$, and hence the circle $S^1_v$ in the fiber degenerates to a point. Over the two points $(s,t)=(0,\pm 1)$ in the base, both circles in the fiber degenerate to a point. Thus, we see that the structure of the singular fibration defined by $\pi$ is identical to the one uncovered in the previous paragraph.

We note that, in the coordinates $(x_L,x_R,\theta_a,\theta_b)$, it is precisely the two circles defined by $(x_L,x_R)=(1,1)$ and $(x_L,x_R)=(0,0)$ where the group $H^{(g_L,g_R)}$ \eqref{eq:u1_Hg} enhances from the trivial group to $U(1)$. At these points, if we are interested in simple topological line defects, we must keep track of an additional integer $Q\in\mathbb{Z}$ corresponding to an irreducible representation of $H^{(g_L,g_R)}$. At e.g.\ $(x_L,x_R)=(1,1)$, we think of this integer $Q\in\mathbb{Z}$ and circle $S^1_a$ traced out by $\theta_a$ in the fiber as combining to form a copy of the real line $\mathbb{R}_\xi$, with $\xi\equiv \theta_a-2\pi Q$. Similar comments apply at $(x_L,x_R)=(0,0)$.

\begin{table}
\begin{center}
    \begin{tabular}{c|c|c}
    topological line & base coordinate & fiber coordinate \\\midrule 
    $\mathcal{X}^{\vec\theta}_{(x_L,x_R)}$ & $(x_L,x_R)\in (0,1)^2$ & $\vec\theta\in S^1_a\times S^1_b$ \\
    $\mathcal{D}^L_a$ ($a\equiv x_Le^{i\theta_a}$) & $(x_L,x_R)\in (0,1)\times \{1\}$ & $\vec\theta \in S^1_a\times\mathrm{pt}$ \\
    $\mathcal{D}_a^R$ ($a\equiv x_Re^{-i\theta_a}$) & $(x_L,x_R)\in \{1\}\times (0,1)$ & $\vec\theta \in S^1_a\times \mathrm{pt}$ \\
    $\widetilde{\mathcal{D}}_b^L$ $\big(b\equiv \sqrt{(1-x_L^2)}e^{i\theta_b}\big)$ & $(x_L,x_R)\in (0,1)\times \{0\}$ & $\vec\theta\in \mathrm{pt}\times S^1_b$ \\
    $\widetilde{\mathcal{D}}^R_b$ $\big( b\equiv \sqrt{(1-x_R^2)}e^{-i\theta_b} \big)$ & $(x_L,x_R)\in \{0\}\times (0,1)$ & $\vec\theta \in \mathrm{pt}\times S^1_b$ \\
    $\mathcal{D}^L$ & $(x_L,x_R)=(0,1)$ & $\vec\theta\in \mathrm{pt}\times\mathrm{pt}$ \\
    $\mathcal{D}^R$ & $(x_L,x_R)=(1,0)$ & $\vec\theta \in \mathrm{pt}\times \mathrm{pt}$ \\
    $\mathcal{M}_\xi$ ($\xi = \theta_a-2\pi Q$) & $(x_L,x_R)=(1,1)$ & $\vec\theta\in S^1_a\times\mathrm{pt}$ \\
    $\mathcal{M}_\xi^C$ ($\xi =  \theta_b + 2\pi Q$) & $(x_L,x_R)=(0,0)$ & $\vec\theta\in \mathrm{pt}\times S^1_b$
    \end{tabular}
    \caption{The simple topological lines of the non-compact boson. The first column provides the notation we use for labeling the lines. The second and third columns give its coordinates in the moduli space. The line with coordinates $(x_L,x_R,\theta_a,\theta_b)$ is either equal to, or occurs inside of, the line obtained by sandwiching the line $\mathcal{L}_{\mathbf{g}(x_L,x_R,\vec\theta)}$ of $SU(2)_1$, with $\mathbf{g}(x_L,x_R,\vec\theta)$ defined in \eqref{eqn:doublecosetrep}, by a $U(1)$ gauging interface.  }\label{tab:simplelinesNCB}
\end{center}
\end{table}

\begin{figure}
    \begin{center}
        \input{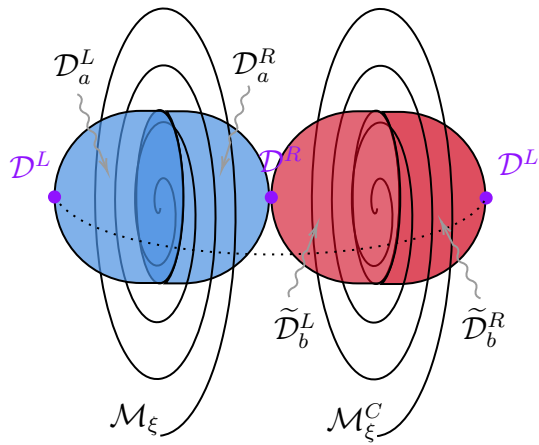}
        \caption{A more detailed depiction of some of the lines of the non-compact boson. Moving from left to right corresponds to traversing the boundary of the square in Figure \ref{fig:stackyS4}. The point on the far left is periodically identified with the point on the far right, as depicted by the dashed line (both correspond to $\mathcal{D}^L$). Although it appears that the points $\mathcal{D}^L$ and $\mathcal{D}^R$ are singular, the space is in fact smooth at these points when they are viewed inside of the entire moduli space.  }\label{fig:boundarysquare}
    \end{center}
\end{figure}

We name the simple topological lines of the non-compact boson as in Table \ref{tab:simplelinesNCB}. A visual depiction of the moduli space of lines is given in Figure \ref{fig:stackyS4} and Figure \ref{fig:boundarysquare}. By definition, the generic line $\mathcal{X}^{\vec\theta}_{(x_L,x_R)}$ is obtained by sandwiching the following $SO(4)$ line in the $SU(2)_1$ WZW model with a half-space gauging interface $\mathcal{I}_{U(1)}$, 
\begin{align}\label{eqn:Xsandwich}
    \mathcal{X}^{\vec\theta}_{(x_L,x_R)}\equiv \mathcal{S}\big(\mathbf{g}(x_L,x_R,\vec\theta)\big)=\mathcal{I}_{U(1)}\otimes \mathcal{L}_{\mathbf{g}(x_L,x_R,\vec\theta)}\otimes \overline{\mathcal{I}}_{U(1)},
\end{align}
and similar comments apply to the remaining lines. In various formulae that follow, we will make use of the following limiting behaviors of these topological lines,
\begin{align}\label{eqn:limitingidNCBlines}
\begin{split}
    & \hspace{.3in}\mathcal{X}^{\vec\theta}_{(x_L,x_R)}\xrightarrow{x_R\to 1}\mathcal{D}^L_{x_Le^{i\theta_a}}, \ \  \mathcal{X}^{\vec\theta}_{(x_L,x_R)}\xrightarrow{x_L\to 1}\mathcal{D}^R_{x_Re^{-i\theta_a}}, \\
    &\mathcal{X}^{\vec\theta}_{(x_L,x_R)}\xrightarrow{x_R\to 0 }\widetilde{\mathcal{D}}^L_{\sqrt{(1-x_L^2)}e^{i\theta_b}}, \ \ \mathcal{X}^{\vec\theta}_{(x_L,x_R)}\xrightarrow{x_L\to 0}\widetilde{\mathcal{D}}^R_{\sqrt{(1-x_R^2)}e^{-i\theta_b}} \\
    &\hspace{.3in}\lim_{(x_L,x_R)\to (0,1)}\mathcal{X}^{\vec\theta}_{(x_L,x_R)}=\lim_{a\to 0}\mathcal{D}^L_a=\lim_{b\to 0}\widetilde{\mathcal{D}}^R_b=\mathcal{D}^L, \\ 
    &\hspace{.3in}\lim_{(x_L,x_R)\to (1,0)}\mathcal{X}^{\vec\theta}_{(x_L,x_R)}=\lim_{a\to 0}\mathcal{D}^R_a=\lim_{b\to 0}\widetilde{\mathcal{D}}^L_b=\mathcal{D}^R \\
   & \lim_{(x_L,x_R)\to (1,1)}\mathcal{X}^{\vec\theta}_{(x_L,x_R)}=\lim_{a\to e^{i\theta_a}}\mathcal{D}^L_a=\lim_{a\to e^{-i\theta_a}}\mathcal{D}^R_a=\bigoplus_{Q\in\mathbb{Z}}\mathcal{M}_{\theta_a-2\pi Q}, \\ 
   & \lim_{(x_L,x_R)\to (0,0)}\mathcal{X}^{\vec\theta}_{(x_L,x_R)}=\lim_{b\to e^{i\theta_b}}\widetilde{\mathcal{D}}^L_b=\lim_{b\to e^{-i\theta_b}}\widetilde{\mathcal{D}}^R_b=\bigoplus_{Q\in\mathbb{Z}}\mathcal{M}^C_{\theta_b + 2\pi Q}.
\end{split}
\end{align}

We can summarize the moduli space of topological lines of the non-compact boson as follows. If one starts with a generic simple topological line (i.e.\ a topological line which is not $\mathcal{M}_\xi$ or $\mathcal{M}_\xi^C$) and explores the topological lines which are continuously connected to it, then one traces out the smooth space $S^4\cong U(1)\backslash SO(4)/U(1)$. The generic line one reaches in this way is simple, however precisely at the two $S^1$ loci defined by $(x_L,x_R)=(1,1)$ and $(x_L,x_R)=(0,0)$, each topological line dissolves into an infinite direct sum as in the last two lines of \eqref{eqn:limitingidNCBlines}. We will see that the lines $\mathcal{M}_\xi$ can be identified with the shift symmetry of the non-compact boson, and that the lines $\mathcal{M}_\xi^C$ can be identified with a product of a shift symmetry and charge conjugation.

Before moving on, we make one more comment on the structure of this moduli space. Analogously to our discussion of the $SU(2)_1$ WZW model in Section \ref{sec:SU(2)1}, there is a 3D TQFT presentation of the non-compact boson. Specifically, one can engineer the 2D non-compact boson by placing $\mathbb{R}$ Chern-Simons theory on an interval with a $\widehat{\mathfrak{u}}(1)$ Kac-Moody chiral algebra boundary condition imposed at one end of the interval, and its complex conjugate imposed at the other end of the interval. (See e.g.\ Figure \ref{fig:su(2)1sandwich} for the analogous picture of the $SU(2)_1$ WZW model.) 

Now, the lines which live on the left-moving $\widehat{\mathfrak{u}}(1)$ chiral algebra boundary condition give rise to ``purely left-moving'' symmetries of the non-compact boson theory, in much the same way that $SU(2)_L\subset (SU(2)_L\times SU(2)_R)/\mathbb{Z}_2$ can be thought of as the left-moving symmetries of the $SU(2)_1$ WZW model. Similarly, the lines living on the other end of the interval will give rise to purely right-moving symmetries of the non-compact boson.\footnote{The techniques developed in \cite{Gannon:2026ttf} then imply that the full category of symmetries of the non-compact boson theory can be recovered as the relative Deligne product $\mathcal{C}_L\boxtimes_{\mathbb{R}}\mathcal{C}_R^{\mathrm{op}}$ over the category of anyons of the bulk $\mathbb{R}$ Chern-Simons theory, but we do not pursue this perspective further here.}  Where do the subcategories $\mathcal{C}_L$ and $\mathcal{C}_R$ of  purely left-moving and purely right-moving topological line defects live in the full moduli space of Figure \ref{fig:stackyS4}?

Our claim is that 
\begin{align}\label{eqn:leftrightmovingNCBlines}
\begin{split}
    \mathrm{Irr}(\mathcal{C}_L)&=\{\mathcal{D}^L_a \mid a\in\mathbb{C}, \ |a|<1\} \cup \{\mathcal{M}_\xi\mid \xi\in\mathbb{R}\}, \\ \mathrm{Irr}(\mathcal{C}_R)&=\{\mathcal{D}^R_a\mid a\in\mathbb{C}, \ |a|<1\} \cup \{\mathcal{M}_\xi\mid \xi\in\mathbb{R}\},
\end{split}
\end{align}
where we are identifying $\mathcal{D}^L_0=\mathcal{D}^L$ and $\mathcal{D}^R_0=\mathcal{D}^R$, as in \eqref{eqn:limitingidNCBlines}.  As a consistency check, $\mathrm{Irr}(\mathcal{C}_L)$ and $\mathrm{Irr}(\mathcal{C}_R)$ indeed are identical to $\mathrm{Irr}(\mathcal{B}_{\NCB} )$, the space of unitary boundary conditions of the non-compact boson theory summarized in \eqref{eq:NCB_BC_space}, as one expects by the folding trick. Note that the $\mathbb{R}$-momentum lines $\mathcal{M}_\xi$ of the non-compact boson occur both inside of $\mathcal{C}_L$ and also $\mathcal{C}_R$. The reason for this is that the $\mathcal{M}_\xi$ are the $\mathbb{R}$ worth of \emph{bulk} Wilson lines of $\mathbb{R}$ Chern-Simons theory, and lines of the bulk TQFT can be pushed either onto the left-moving chiral algebra or onto the right-moving chiral algebra. This is analogous to the fact that in the $SU(2)_1$ WZW model, $SU(2)_L$ and $SU(2)_R$ both contain the $\mathbb{Z}_2$ Verlinde line of the theory (i.e.\ the non-trivial central element of $SO(4)$): this is precisely because the $\mathbb{Z}_2$ Verlinde line is the fundamental Wilson line of the bulk $SU(2)_1$ Chern-Simons theory.

One way to see the validity of \eqref{eqn:leftrightmovingNCBlines} is to observe that the lines in e.g.\ $\mathcal{C}^L$ come from double cosets $[(g_L,g_R)]\in U(1)\backslash SO(4)/U(1)$ which admit a representative $(g_L,g_R)$ with $g_R=\mathds{1}$. We will give more evidence in support of \eqref{eqn:leftrightmovingNCBlines} as we go.

\subsubsection{Fusion Algebra and \texorpdfstring{$\mathbb{R}$}{R} Tambara-Yamagami Symmetry} \label{sec:fusion_TY}

Let us now discuss the fusion rules obeyed by the topological line defects of the non-compact boson theory.

To warm-up, we start by discussing all the invertible symmetries, which are $\mathcal{M}_{\xi}$ and $\mathcal{M}^C_{\xi}$.
These invertible topological line operators form a group $\mathbb{R} \rtimes \mathbb{Z}_2$, with the fusion rules given by 
\begin{align}
\begin{split}
    \mathcal{M}_{\xi} \otimes \mathcal{M}_{\xi'} &\cong \mathcal{M}_{\xi+\xi'} \,, \quad  \mathcal{M}^C_{\xi} \otimes \mathcal{M}^C_{\xi'} \cong \mathcal{M}_{\xi-\xi'} \,,\\
    \mathcal{M}_{\xi} \otimes \mathcal{M}^C_{\xi'} &\cong \mathcal{M}^C_{\xi+\xi'} \,, \quad \mathcal{M}^C_{\xi} \otimes \mathcal{M}_{\xi'} \cong \mathcal{M}^C_{\xi-\xi'} \,.
\end{split}
\end{align}
They correspond to the familiar target space $\mathbb{R}$ momentum symmetry of the non-compact boson $\Phi \to \Phi + \xi$,  the $\mathbb{Z}_2$ charge conjugation symmetry $\Phi \to - \Phi$, and combinations thereof. (We will more explicitly verify this in Section \ref{subsec:explicit_con_lines}.)

Next, let us consider the fusion of $\mathcal{X}_{(x_L,x_R)}^{\vec\theta}$ with $\mathcal{X}_{(x_L',x_R')}^{\vec\theta'}$. This case is fairly straightforward using the fact that both lines are obtained by sandwiching $SO(4)$ lines of the $SU(2)_1$ WZW model with $U(1)$ gauging interfaces, as in \eqref{eqn:Xsandwich}: 
\begin{align}\label{eqn:XXfusion}
\begin{split}
    &\mathcal{X}_{(x_L,x_R)}^{\vec\theta}\otimes \mathcal{X}_{(x_L',x_R')}^{\vec\theta'}   \\
    &\hspace{.2in}\cong \mathcal{I}_{U(1)}\otimes \mathcal{L}_{\mathbf{g}(x_L,x_R,\vec\theta)}\otimes \overline{\mathcal{I}}_{U(1)} \otimes \mathcal{I}_{U(1)}\otimes \mathcal{L}_{\mathbf{g}(x_L',x_R',\vec\theta')}\otimes \overline{\mathcal{I}}_{U(1)} \\
    &\hspace{.2in}\cong\frac{\mathcal{N}}{2\pi} \int_{[0,2\pi)}^{\oplus}d\alpha~\mathcal{I}_{U(1)}\otimes \mathcal{L}_{\mathbf{g}(x_L,x_R,\vec\theta)}\otimes \mathcal{L}_{(e^{i\alpha \sigma_3/2},e^{i\alpha\sigma_3/2})}\otimes \mathcal{L}_{\mathbf{g}(x_L',x_R',\vec\theta')}\otimes \overline{\mathcal{I}}_{U(1)} \\
    &\hspace{.2in}\cong\frac{\mathcal{N}}{2\pi} \int_{[0,2\pi)}^{\oplus}d\alpha~ \mathcal{X}^{\vec\theta''(\alpha)}_{(x_L''(\alpha),x_R''(\alpha))}.
\end{split}
\end{align}
In the last line, $(x_L''(\alpha),x_R''(\alpha),\theta''_a(\alpha),\theta_b''(\alpha))$ are the invariants (see \eqref{eqn:doublecosetcoordinates}) of the double coset of the $SO(4)$ element $\mathbf{g}(x_L,x_R,\vec\theta)\cdot (e^{i\alpha\sigma_3/2},e^{i\alpha\sigma_3/2})\cdot \mathbf{g}(x_L',x_R',\vec\theta')$.  Doing this calculation explicitly, one finds 
\begin{align}
\begin{split}
    & \ \ \ \ \ x_L''(\alpha)=\left|z_L(\alpha)\right|, \ \ \ \ \ x_R''(\alpha) = \left|z_R(\alpha) \right|, \\
    &\theta_a''(\alpha)=\theta_a+\theta_a'+\mathrm{Arg}\left[z_L(\alpha) \right]-\mathrm{Arg}\left[z_R(\alpha)   \right], \\
    &\theta''_b(\alpha)=\theta_a+\theta_b'+\mathrm{Arg}\left[w_L(\alpha)\right]-\mathrm{Arg}\left[w_R(\alpha)\right],
\end{split}
\end{align}
where we have defined the following quantities:
\begin{align}
\begin{split}
   \Delta&\equiv\frac12(\theta_a+\theta_a'-\theta_b+\theta_b'), \\
   z_L(\alpha)&\equiv x_L x_L'-\sqrt{(1-x_L^2)(1-x_L^{'2})}e^{-i(\alpha+\Delta)}, \\
   z_R(\alpha)&\equiv x_Rx_R'-\sqrt{(1-x_R^2)(1-x_R^{'2})}e^{-i(\alpha-\Delta)}, \\
   w_L(\alpha) &= x_L\sqrt{1-x_L'^2}
    +x_L'\sqrt{1-x_L^2}e^{-i(\alpha+\Delta)},\\
    w_R(\alpha) &= x_R\sqrt{1-x_R'^2}
    +x_R'\sqrt{1-x_R^2}e^{-i(\alpha-\Delta)}.
\end{split}
\end{align}
When $(x_L''(\alpha),x_R''(\alpha))$ are outside of $(0,1)\times (0,1)$, we identify $\mathcal{X}^{\vec\theta''(\alpha)}_{(x_L''(\alpha),x_R''(\alpha))}$ with its various limits appearing in \eqref{eqn:limitingidNCBlines}. In other words, even though it naively appears that the fusion of two $\mathcal{X}$ lines only ever involves a direct integral of other $\mathcal{X}$ lines, the other types of line operators ($\mathcal{D}^L_a$, $\mathcal{D}^R_a$, etc.) can also appear.

By the same token, \eqref{eqn:XXfusion} actually covers every fusion rule of the form $A\otimes B$ with $A,B\notin \{\mathcal{M}_\xi\mid \xi\in\mathbb{R}\}\cup \{\mathcal{M}_\xi^C\mid \xi\in\mathbb{R}\}$. For example, by specializing $x_R=x_R'=1$, we can recover the fusion rules 
\begin{align}
    \mathcal{D}^L_a\otimes \mathcal{D}^L_{a'}\cong \frac{\mathcal{N}}{2\pi}\int_{[0,2\pi)}^{\oplus}d\alpha ~\mathcal{D}^L_{aa'-\sqrt{(1-|a|^2)(1-|a'|^2)}e^{i\alpha}},
\end{align}
which agrees precisely with the independent computation in \cite[Claim 4.8]{Gannon:2026ttf} of the topological line operators supported on the $\widehat{\mathfrak{u}}(1)$ Kac-Moody algebra. This gives further evidence that the $\mathcal{D}^L_a$ should be thought of as purely left-moving symmetries of the non-compact boson as in \eqref{eqn:leftrightmovingNCBlines}. 

Another interesting special case of \eqref{eqn:XXfusion} is
\begin{align}
    \mathcal{D}^{L,R}\otimes \mathcal{D}^{L,R}\cong \frac{\mathcal{N}}{2\pi}\int_{\mathbb{R}}^{\oplus}d\xi ~\mathcal{M}_\xi, \ \ \ \ \ \ \mathcal{D}^{L,R}\otimes \mathcal{D}^{R,L}\cong \frac{\mathcal{N}}{2\pi}\int_{\mathbb{R}}^{\oplus}d\xi ~\mathcal{M}_\xi^C.
\end{align}
In particular, both $\{\mathcal{M}_\xi\mid \xi\in\mathbb{R}\}\cup \{\mathcal{D}^L\}$ and $\{\mathcal{M}_\xi\mid \xi\in\mathbb{R}\}\cup \{\mathcal{D}^R\}$ formally have the fusion rules of a Tambara-Yamagami category associated to the group $\mathbb{R}$.\footnote{Such categories were recently formalized in \cite{Marin-Salvador:2025stc}.} The duality lines $\mathcal{D}^L$ and $\mathcal{D}^R$ (which are exchanged by the charge conjugation symmetry $\mathcal{M}_0^C$) can be thought of as obtained from the fact that the non-compact boson is self-dual under flat gauging its $\mathbb{R}$-symmetry.

The most subtle fusion rules are those in which one line is invertible and the other is non-invertible. The reason is that the invertible lines $\mathcal{M}_\xi$ and $\mathcal{M}_\xi^C$ are not obtained as sandwiches of $SO(4)$ lines, so we cannot use the same manipulations as in \eqref{eqn:XXfusion} to determine their fusion. But the second best thing is true, which is that the quantity e.g.\ $\bigoplus_{Q\in\mathbb{Z}}\mathcal{M}_{\theta_a - 2\pi Q}$ is a sandwich, so we can determine its fusion rule with every non-invertible line. Indeed, using \eqref{eqn:limitingidNCBlines}, we can take the limit $(x_L,x_R)\to (1,1)$ of \eqref{eqn:XXfusion}, and find that 
\begin{align}
    \bigoplus_{Q\in\mathbb{Z}}\mathcal{M}_{\theta_a - 2\pi Q}\otimes \mathcal{X}^{\vec\theta'}_{(x_L',x_R')}\cong \frac{\mathcal{N}}{2\pi}\int_{[0,2\pi)}^{\oplus}d\alpha ~ \mathcal{X}_{(x_L',x_R')}^{(\theta_a+\theta_a',\theta_a+\theta'_b)} = \mathcal{N}\mathcal{X}_{(x_L',x_R')}^{(\theta_a+\theta_a',\theta_a+\theta'_b)}.
\end{align}
Since only a single topological line $\mathcal{X}_{(x_L',x_R')}^{(\theta_a+\theta_a',\theta_a+\theta'_b)}$ appears on the right-hand side (up to the factor of $\mathcal{N}$), this means that each line $\mathcal{M}_{\theta_a - 2\pi Q}$ must individually fuse with $\mathcal{X}^{\vec\theta'}_{(x_L',x_R')}$ to give $\mathcal{X}_{(x_L',x_R')}^{(\theta_a+\theta_a',\theta_a+\theta'_b)}$. In particular, we learn that 
\begin{align}
    \mathcal{M}_\xi \otimes \mathcal{X}^{\vec\theta}_{(x_L,x_R)}\cong \mathcal{X}^{(\xi+\theta_a,\xi+\theta_b)}_{(x_L,x_R)},
\end{align}
where there is no factor of $\mathcal{N}$ because $\mathcal{M}_\xi$ is invertible. Proceeding in the same way, we can calculate the rest of the fusion rules involving the invertibles:
\begin{align}\label{eqn:MXfusion}
\begin{split}
    \mathcal{M}_\xi\otimes\mathcal{X}^{(\theta_a,\theta_b)}_{(x_L,x_R)}
    &\cong \mathcal{X}^{(\theta_a+\xi,\ \theta_b+\xi)}_{(x_L,\,x_R)}\,,
    \qquad\qquad\quad\ \,
    \mathcal{X}^{(\theta_a,\theta_b)}_{(x_L,x_R)}\otimes\mathcal{M}_\xi
    \cong\mathcal{X}^{(\theta_a+\xi,\ \theta_b-\xi)}_{(x_L,\,x_R)}\,,\\
    \mathcal{M}^C_\xi\otimes\mathcal{X}^{(\theta_a,\theta_b)}_{(x_L,x_R)}
    &\cong\mathcal{X}^{(\xi-\theta_b,\ \xi-\theta_a)}
    _{\big(\sqrt{1-x_L^2},\ \sqrt{1-x_R^2}\big)}\,,
    \qquad
    \mathcal{X}^{(\theta_a,\theta_b)}_{(x_L,x_R)}\otimes\mathcal{M}^C_\xi
    \cong\mathcal{X}^{(\theta_b-\xi,\ \theta_a+\xi)}
    _{\big(\sqrt{1-x_L^2},\ \sqrt{1-x_R^2}\big)}\,.
\end{split}
\end{align}
Again, by taking various limits of these equations, we can obtain the fusion rules of $\mathcal{M}_\xi$, $\mathcal{M}_\xi^C$ with $\mathcal{D}_a^L$, $\mathcal{D}_a^R$, etc. This concludes the calculation of the fusion rules of the topological line operators of the non-compact boson.

\subsubsection{Action on Boundaries}

Topological line defects of a 2D CFT act on its boundary conditions by parallel fusion. For the non-compact boson theory, it is natural to ask how the topological line defects  act on the boundary conditions we found in Section \ref{subsec:NCBconformalbdy}.

The most straightforward case to analyze is the action of the generic topological line defects $\mathcal{X}^{\vec\theta}_{(x_L,x_R)}$ on the generalized Neumann boundary conditions $\mathbf{N}_a$, which we recall are labeled by a complex number $a\in\mathbb{C}$ satisfying $|a|<1$. Using the fact that $\mathcal{X}_{(x_L,x_R)}^{\vec\theta}$ can be obtained by sandwiching gauging interfaces onto a topological line of $SU(2)_1$ \eqref{eqn:Xsandwich}, and the fact that $\mathbf{N}_a$ is recovered by fusing a gauging interface onto a boundary of $SU(2)_1$ \eqref{eqn:Nadef}, we can calculate that
\begin{align}
\begin{split}
    \mathcal{X}^{\vec\theta}_{(x_L,x_R)}\otimes \mathbf{N}_a &\cong \mathcal{I}_{U(1)}\otimes \mathcal{L}_{\mathbf{g}(x_L,x_R,\vec\theta)}\otimes\overline{\mathcal{I}}_{U(1)} \otimes \mathcal{I}_{U(1)}\otimes \mathbf{B}_{a,b} \\
    &\cong \frac{\mathcal{N}}{2\pi}\int_{[0,2\pi)}^{\oplus} d\alpha~\mathcal{I}_{U(1)}\otimes \mathcal{L}_{\mathbf{g}(x_L,x_r,\vec\theta)}\otimes \mathcal{L}_\alpha^W\otimes \mathbf{B}_{a,b} \\
    &\cong\frac{\mathcal{N}}{2\pi}\int_{[0,2\pi)}^{\oplus}d\alpha~\mathbf{N}_{a'(\alpha)},
\end{split}
\end{align}
where we have defined the quantities
\begin{align}
\begin{split}
    a'(\alpha)&\equiv C+\sqrt{1-|a|^2}\,e^{\frac{i}{2}(\theta_a+\theta_b)}
    \Big(x_L\sqrt{1-x_R^2}\,e^{i\alpha}
    -x_R\sqrt{1-x_L^2}\,e^{-i\alpha}\Big), \\
    &\hspace{.3in}C\equiv a\,x_Lx_R\,e^{i\theta_a}
    +\bar a\,\sqrt{(1-x_L^2)(1-x_R^2)}\,e^{i\theta_b},
\end{split}
\end{align}
which is the top left entry of the $SU(2)$ matrix $g\big(x_Le^{i\theta_a/2},\tfrac{\theta_b}{2}\big)\;
    g(a,\alpha)\;
    g\big(x_Re^{-i\theta_a/2},-\tfrac{\theta_b}{2}\big)^{-1}$. The function $a'(\alpha)$ defines an ellipse centered at the point $C$. This equation also covers the action of the lines $\mathcal{D}^L_a$, $\mathcal{D}^R_a$, $\widetilde{\mathcal{D}}^L_b$, $\widetilde{\mathcal{D}}^R_b$, $\mathcal{D}^L$, and $\mathcal{D}^R$ on the generalized Neumann boundary conditions using the limiting identifications in \eqref{eqn:limitingidNCBlines}. We also emphasize that, if $|a'(\alpha)|=1$ for some $\alpha$, then one should understand $\mathbf{N}_{a'(\alpha)}$ as an infinite direct sum of Dirichlet boundary conditions as in \eqref{eqn:Na_dissolve}.

    The action of $\mathcal{X}_{(x_L,x_R)}^{\vec\theta}$ on Dirichlet boundary conditions of the non-compact boson can be calculated similarly. In particular, we find that 
    \begin{align}
    \begin{split}
        \mathcal{X}^{\vec\theta}_{(x_L,x_R)}\otimes \mathbf{D}_{\Phi_0}  &\cong \mathcal{I}_{U(1)}\otimes \mathcal{L}_{\mathbf{g}(x_L,x_R,\vec\theta)}\otimes \overline{\mathcal{I}}_{U(1)}\otimes \mathbf{D}_{\Phi_0} \cong \mathcal{I}_{U(1)}\otimes \mathcal{L}_{\mathbf{g}(x_L,x_R,\vec\theta)}\otimes \mathbf{B}_{e^{i\Phi_0},0} \\
        &\cong \mathbf{N}_a, \ \ \ \ \ \ a=e^{i(\theta_a+\Phi_0)}x_Lx_R+e^{i(\theta_b-\Phi_0)}\sqrt{(1-x_L^2)(1-x_R^2)},
    \end{split}
    \end{align}
where we have used \eqref{eqn:I*D}.

We can constrain the action of $\mathcal{M}_\xi$ on generalized Neumann boundary conditions by noting, using \eqref{eqn:I*N}, that
\begin{align}
    \bigoplus_{Q\in\mathbb{Z}}\mathcal{M}_{\xi-2\pi Q}\otimes \mathbf{N}_a \cong \mathcal{I}_{U(1)}\otimes \mathcal{L}_{(e^{i\xi \sigma_3},1)}\otimes \overline{\mathcal{I}}_{U(1)} \otimes \mathbf{N}_a \cong \mathcal{N}\cdot \mathbf{N}_{ae^{i\xi}}.
\end{align}
Since $\mathcal{M}_\xi$ is an invertible topological line operator, this is enough to conclude that 
\begin{align}
    \mathcal{M}_\xi\otimes \mathbf{N}_a\cong \mathbf{N}_{ae^{i\xi}}.
\end{align}
Nearly identical calculations yield that 
\begin{align}
    \mathcal{M}_\xi\otimes \mathbf{D}_{\Phi_0}\cong \mathbf{D}_{\Phi_0+\xi}, \ \ \ \ \mathcal{M}_\xi^C\otimes \mathbf{N}_a\cong \mathbf{N}_{e^{i\xi}\bar a}, \ \ \ \ \mathcal{M}_\xi^C\otimes \mathbf{D}_{\Phi_0}\cong \mathbf{D}_{\xi-\Phi_0}.
\end{align}

\section{Lagrangian Descriptions}\label{sec:NCBLag}

In this section, we provide Lagrangian descriptions for the various constructions which were carried out abstractly in Section \ref{sec:NCBAbs}. These descriptions ultimately owe their existence to the fact that the flat gauging of a $U(1)$ symmetry in a theory $\mathcal{T}$ can be implemented by coupling $\mathcal{T}$ to $U(1)$ BF theory. We review this first.

\subsection{Non-Compact Boson from Flat Gauging}\label{subsec:Lagflatgauging}

We start with the description of $SU(2)_1$ as a compact boson with radius chosen so that the theory is self-dual under T-duality,
\begin{align} \label{eq:cb_action}
    S_{\CB}=\frac{1}{4\pi}\int d\phi \wedge \star d\phi,
\end{align}
where $\phi \sim \phi+2\pi$ is a compact boson field.
This can be coupled to $U(1)$ BF theory through its winding\footnote{Due to $T$-duality, we may also flat-gauge the $U(1)$ momentum symmetry to obtain a non-compact boson theory.} symmetry, and we have the (Euclidean) action
\begin{equation} \label{eq:NCB1}
   S_{\mathrm{CB}\text{-}\mathrm{BF}} =\frac{1}{4\pi}\int  d\phi \wedge \star d\phi -i\int  J_w\wedge a - \frac{i}{2\pi}\int \Phi da \,, \ \ \ \ J_w = \frac{1}{2\pi}d\phi.
\end{equation}
Here, $a$ is a $U(1)$ gauge field, and it couples to the Noether current $J_{w}$ of the winding symmetry. $\Phi$ is a non-compact scalar which plays the role of ``B'' in BF theory.
In the last term of the action, the non-compact boson $\Phi$ acts as a Lagrange multiplier which forces the gauge field $a$ to be flat, $da = 0$, justifying the name ``flat gauging''. 

We refer to the theory $S_{\mathrm{BF}}=-\frac{i}{2\pi}\int \Phi da$ with $\Phi$ a non-compact scalar  as  $U(1)$ BF theory, as it is the direct $U(1)$ analog of BF theory for non-abelian Lie groups, where the scalar field $B$ is non-compact and takes values in the dual Lie algebra. This  should be distinguished from the theory $S'_{\mathrm{BF}}=-\frac{iN}{2\pi}\int Bda$ with $B$ a compact scalar field, $B\sim B+2\pi$ and $N\in\mathbb{Z}$, which is equivalent to $\mathbb{Z}_N$ gauge theory \cite{Kapustin:2014gua}.

Flat gauging is closely related to the more standard notion of gauging, which we might refer to as ``kinetic gauging''. Indeed, suppose we deform \eqref{eq:NCB1} by a mass term for $\Phi$, 
\begin{align}
    \frac{1}{4\pi}\int  d\phi \wedge \star d\phi -i\int  J_w\wedge a - \frac{i}{2\pi}\int \Phi da +\frac{e^2}{8\pi^2}\int \Phi^2 \,\mathrm{vol} \,.
\end{align}
(Here, $\mathrm{vol}$ is the volume form.)
This allows the gauge field $a$ to fluctuate away from flatness. When we integrate out $\Phi$, we are left with the action 
\begin{align}
    \frac{1}{4\pi}\int  d\phi \wedge \star d\phi -i\int  J_w\wedge a +\frac{1}{2e^2}\int f\wedge \star f,
    \label{eq:maxwell_coupling}
\end{align}
see e.g.\ \cite[Section 4.2]{Witten:1992xu}. The resulting action describes the standard coupling of $U(1)$ gauge theory to the compact boson through its $U(1)$ winding symmetry. Thus, flat gauging can be viewed as the $e^2\to 0$ limit of kinetic gauging. When we flat gauge the $U(1)$ winding symmetry of the compact boson, the theory remains conformal with the same central charge. By contrast, if we perform kinetic gauging, then the theory is actually gapped out. Indeed, in two dimensions, the electric coupling $e^2$ is relevant and flows to $e^2\to\infty$ in the infrared. The Maxwell term in \eqref{eq:maxwell_coupling} is then suppressed and the equation of motion for $a$ imposes $d\phi=0$, leaving a trivial theory at long distances.

A proper definition of the term
\begin{align}
    S_w[\phi,a]\equiv \text{``}-\frac{i}{2\pi}\int d\phi \wedge a\text{''}
\end{align}
is subtle. One common way to proceed is to cover spacetime in patches and define transition functions for the fields whenever those patches overlap (see \cite{Cordova:2019jnf} for details). We will be able to avoid having to do this below by simply appealing to two properties that $S_w$ enjoys once it is properly defined. First, whenever $b$ is a globally defined one-form, we have that 
\begin{align}
    S_w[\phi,a+b]-S_w[\phi,a]=-\frac{i}{2\pi}\int d\phi \wedge b,
\end{align}
where this is now a standard integral because both $d\phi$ and $b$ are globally well-defined. Similarly, whenever $\phi$ has a real lift $\tilde\phi$ (which exists if and only if $\phi$ winds trivially around all one-cycles), one has that 
\begin{align}
    S_w[\phi,a]=\frac{i}{2\pi}\int \tilde\phi da~~\mathrm{mod}~2\pi i,
\end{align}
which is again a standard integral because both $\tilde\phi$ and $da$ are globally well-defined.

To see that \eqref{eq:NCB1} describes a standard non-compact boson, we can integrate out $\phi$ and $a$. For simplicity, let us carry this out on a closed Riemann surface $\Sigma$ of genus $g$. It will be convenient to parametrize the most general gauge field configuration as follows. For each $n\in\mathbb{Z}$, fix a reference configuration $a^{(n)}$ whose flux is given by 
\begin{align}\label{eqn:fluxsector}
    \int_\Sigma da^{(n)}=2\pi n.
\end{align}
Choose one-cycles $A_I$ and $B_I$ of $\Sigma$ with $I=1,\dots, g$ and intersection numbers given by 
\begin{align}
    A_I\cdot B_J=\delta_{IJ}, \ \ \ \ A_I\cdot A_J=B_I\cdot B_J=0.
\end{align}
Pick also a basis $\omega_A^I$ and $\omega_B^I$ of real harmonic one-forms normalized so that their periods are 
\begin{align}
    \oint_{A_I}\omega_A^J=\oint_{B_I}\omega_B^{J}=\delta_{IJ}, \ \ \ \ \oint_{A_I}\omega_B^J=\oint_{B_I}\omega_A^J=0.
\end{align}
Then every other gauge field configuration with flux given by \eqref{eqn:fluxsector} can be obtained as 
\begin{align}
    a=a^{(n)}+\sum_{I=1}^{g}(h^A_I\omega_A^I+h_I^B\omega_B^I)+d\lambda + \star d\chi,
\end{align}
where $h^{A,B}_I\in \mathbb{R}/2\pi \mathbb{Z}$ are the holonomies of $a-a^{(n)}-\star d\chi$ around the one-cycles, and $\lambda,\chi$ are globally-defined real-valued functions. This is a kind of Hodge decomposition of $a$. This decomposition implies that the path integral over $a$ can be expressed as 
\begin{align}
    \int \frac{\mathcal{D}a}{\mathrm{Vol}(\mathrm{gauge})}\propto \sum_{n\in\mathbb{Z}}\prod_{I=1}^{g}\int_{0}^{2\pi} dh^A_I\int_0^{2\pi} dh_I^B\int \mathcal{D}\chi,
\end{align}
where we have not needed to path integrate over $\lambda$ because it parametrizes gauge equivalent configurations. 

Let us perform the path integral over the holonomies first. The relevant contribution is 
\begin{align}
    H\equiv \prod_{I=1}^{g}\left(\int_{0}^{2\pi}dh^A_I\int_0^{2\pi}dh_I^B ~ \exp\left(\frac{i}{2\pi}h_I^A\int_{\Sigma}d\phi \wedge \omega^I_A+\frac{i}{2\pi}h_I^B\int_\Sigma d\phi\wedge \omega_B^I\right)\right).
\end{align}
Using the standard formula that, for closed one-forms $\eta,\xi$,
\begin{align}
    \int_\Sigma \eta \wedge \xi=\sum_{I=1}^g\left[ \oint_{A_I}\eta \oint_{B_I}\xi -\oint_{B_I}\eta\oint_{A_I}\xi \right],
\end{align}
we learn that the holonomy contribution is given by
\begin{align}
H=\prod_{I=1}^g \left(\int_{0}^{2\pi}dh_I^A\int_0^{2\pi}dh_I^B  \exp\left(  -ih_I^A W_I^B+ih_I^BW_I^A\right)  \right)\propto \prod_{I=1}^g\delta_{W_I^A,0}\delta_{W_I^B,0},
\end{align}
where $W_I^A$ and $W_I^B$ are the winding numbers of $\phi$ around the A and B cycles of $\Sigma$. Thus, the path integral over the holonomies imposes that $\phi$ has trivial winding, and hence admits a lift to an $\mathbb{R}$-valued field, which we abusively also denote $\phi$.

Next, we can perform the path integral over $\chi$. This contribution (after integrating by parts several times and using that the Laplacian is $\Delta=-\star d\star d$) is 
\begin{align}
    \int \mathcal{D}\chi \exp\left(\frac{i}{2\pi}\int_\Sigma  \chi \Delta (\phi-\Phi) \, \mathrm{vol} \right).
\end{align}
Thus, $\chi$ is a Lagrange multiplier which sets $\Delta \phi = \Delta\Phi$ in the path integral. In particular, since $\phi$ has trivial winding and the only globally defined function $f$ on a closed Riemann surface with $\Delta f=0$ is a constant, this implies that $\phi=\Phi+c$ for some constant $c$.

Finally, the sum over flux sectors \eqref{eqn:fluxsector} requires that $c=0~\mathrm{mod}~2\pi$, and hence that $\phi=\Phi~\mathrm{mod}~2\pi$. Thus, the action reduces (up to overall $\Phi$-independent constants) to 
\begin{align}
    S_{\NCB}=\frac{1}{4\pi}\int d\Phi\wedge \star d\Phi
\end{align}
which is of course the action of a non-compact boson.

\subsection{Gauging Interface}\label{subsec:Laggauginginterface}

Since gauging interfaces play a crucial role in constructing observables of a theory obtained by orbifolding, it will be useful to describe how such interfaces are modeled in the Lagrangian approach.

A gauging interface between a theory $\mathcal{T}$ and one of its orbifolds is obtained by gauging $\mathcal{T}$ in half of spacetime, and choosing suitable boundary conditions for the gauge fields at the location of the interface. 
We take spacetime to be a closed Riemann surface $\Sigma$, and we choose a circle $C$ (where we will ultimately place the gauging interface) which divides $\Sigma$ into two pieces, $\Sigma_L$ and $\Sigma_R$. Locally near $C$, we let $\tau$ be the coordinate along $C$ and $x$ a coordinate normal to $C$. The interface lives at $x=0$, and we assume the metric near $C$ is $d\tau^2+dx^2$.

We model the gauging interface between $SU(2)_1$ on the right and the non-compact boson on the left with the following Lagrangian,
\begin{align}
    S_{\mathcal{I}_{U(1)}}=\frac{1}{4\pi}\int_{\Sigma_L}d\phi_L \wedge\star d\phi_L + \int_{\Sigma_R} \left(\frac{1}{4\pi }d\phi_R\wedge \star d\phi_R - \frac i{2\pi} d\phi_R\wedge a_R-\frac{i}{2\pi}\Phi_R da_R\right),
\end{align}
where $\phi_{L}$ (resp.\ $\phi_R$) is a compact boson defined on $\Sigma_L$ (resp.\ $\Sigma_R$), $\Phi_R$ is a non-compact scalar defined on $\Sigma_R$, and $a_R$ is a $U(1)$ gauge field defined on $\Sigma_R$.

To check the consistent choices of boundary conditions which can be imposed at $x=0$, the location of the interface, we calculate the variation of the action, which has a contribution from the bulk ($\delta S_\Sigma$) and from the interface ($\delta S_C$):
\begin{align}\label{eqn:NCB1gaugintvariation}
\begin{split}
    &\delta S_{\mathcal{I}_{U(1)}}=\delta S_{\Sigma}+\delta S_{C}, \\
    &\hspace{.3in}\delta S_{C}=\frac{1}{2\pi}\int_C \left(-\star d\phi_L~\delta\phi_L+\star d\phi_R ~ \delta\phi_R    -ia_{R}~\delta\phi_R-i\Phi_R~\delta a_{R}\right).
\end{split}
\end{align}
We obtain the gauging interface by imposing  Dirichlet boundary conditions for $a$ and imposing that the compact scalars are continuous across $C$ with continuous first derivative normal to the boundary,
\begin{align}
 a_{R}\vert_{C}=0, \ \ \ \ \ \ \ \phi_L\vert_{C}=\phi_R\vert_{C}, \ \ \ \ \ \ \ (\star d\phi_L)\vert_{C}=(\star d\phi_R)\vert_{C}.
\end{align}
The boson $\Phi$ is unconstrained at $x=0$. We henceforth drop the subscripts L and R.

In order to preserve the Dirichlet boundary condition for the gauge field $a$, we insist that gauge transformations $a\to a+d\lambda$ are constant at the interface, $d\lambda\vert_{C} =0$, so that $\lambda\vert_{C}=\lambda_0$. 
 We furthermore require the gauge transformations to be trivial on the boundary, that is, $\lambda_0 \in 2\pi \mathbb{Z}$.
(See, for instance, Section 2.2 of \cite{Gaiotto:2008sa}.)

Transformations with $\lambda_0 \notin 2\pi \mathbb{Z}$, on the other hand, are not gauge transformations. They change the action by
\begin{align}\label{eqn:gaugevariationaction}
 \delta_\lambda S_{\mathcal{I}_{U(1)}} = \frac{i}{2\pi}\lambda_0 \int_Cd\phi = i\lambda_0W~\mathrm{mod}~2\pi i,
\end{align}
where $W\in\mathbb{Z}$ is the winding number of $\phi$ around $C$. Suppose we fuse a $U(1)$ winding topological line defect $\exp\left(\frac{i}{2\pi}\alpha\int_Cd\phi\right)$ onto the gauging interface from the left. One of the consequences of \eqref{eqn:gaugevariationaction} is that the contribution to the path integral of such a winding defect can be canceled by performing $a\to a +d\lambda$ with $\lambda$ approaching $\lambda|_C = \lambda_0 = -\alpha$ at the location of the interface. We interpret this as saying that the gauging interface absorbs $U(1)$ winding topological line defects.

It is illuminating to see what the Lagrangian description of this gauging interface is if we use the conventional description of the non-compact boson. Recall that we can reach this description by integrating out $a$ and $\phi$ in \eqref{eq:NCB1}. If we perform this integration in the presence of the gauging interface, the same manipulations go through with minor modifications, and we obtain 
\begin{align} \label{eq:lag_interface}
\begin{gathered}
    S_{\mathcal{I}_{U(1)}}=\frac{1}{4\pi} \int_{\Sigma_L} d\phi \wedge\star d\phi + \frac{1}{4\pi}\int_{\Sigma_R} d\Phi\wedge \star d\Phi, \\
    \phi\vert_{C} = \Phi\vert_{C} ~\mathrm{mod}~2\pi, \ \ \star d \phi\vert_C=\star d \Phi\vert_C.
\end{gathered}
\end{align}
That is, the gauging interface simply glues the compact field $\phi$ to the non-compact field $\Phi$ at the interface modulo $2\pi$. 

This description also makes it clearer that the interface absorbs the topological line which implements the $U(1)$ winding symmetry of $\phi$: if one brings this line $\exp\left(i \frac{\alpha}{2\pi}\int d\phi\right)$ close to the interface, one can replace $\phi$ with $\Phi$ thanks to the gluing condition across the interface. But $\Phi$ is non-compact so it cannot wind, and hence $\exp\left(i\frac{\alpha}{2\pi}\int d\Phi\right)=1$ at the interface.

\subsection{Conformal Boundaries}\label{subsec:NCBbdyLag}

Next, let us describe how to obtain boundary conditions of the non-compact boson from boundary conditions of $SU(2)_1$.

 The boundary conditions which are most natural to describe in the compact boson description of $SU(2)_1$ are the Dirichlet and Neumann boundary conditions, and we can ask what becomes of them after gauging the $U(1)$ winding symmetry. Actually, in order to see the full moduli space of boundaries of the non-compact boson, it will be useful to consider certain deformations of the Neumann boundary conditions of $SU(2)_1$ by exactly marginal boundary vertex operators \cite{Callan:1993mw,Callan:1994ub,Polchinski:1994my}, as we describe below.

We would like to impose consistent boundary conditions on the action in \eqref{eq:NCB1}. In the absence of additional boundary terms in the action, the variation of it is obtained from \eqref{eqn:NCB1gaugintvariation} by simply setting $\Sigma_L$ to be empty. We drop the subscript $R$ on $\Sigma_{R}$ as well as on all the fields, and $C = \partial \Sigma $:
\begin{align}\label{eq:ncb_bdy_variation}
     \delta S^\partial_{\CB\text{-}\mathrm{BF}}=\delta S_{\Sigma}+\delta S_{\partial\Sigma}, \ \ \ \ \delta S_{\partial\Sigma}=\frac{1}{2\pi}\int_{\partial \Sigma}\left(\star d\phi \delta\phi-ia \delta\phi -i \Phi \delta a\right).
\end{align}
We would like to ask how the Dirichlet and (deformed) Neumann boundary conditions of $\phi$ are completed into consistent boundary conditions of the flat-gauged theory, \eqref{eq:NCB1}. It turns out that Dirichlet and (deformed) Neumann will complete differently, owing to the fact that they behave differently with respect to the $U(1)$ winding symmetry of $\phi$.

Let us begin with the deformed Neumann boundary conditions of the compact boson, which take the form 
\begin{align} \label{eq:cb_lag_bdy}
    S_{\CB}^\partial=\frac{1}{4\pi}\int_{\Sigma}d\phi\wedge\star d\phi +  \int_{\partial\Sigma}\left((\mu e^{i\phi}+\bar\mu e^{-i\phi})ds-\frac{i}{2\pi} \eta d\phi\right), \ \ \ \ \ \ \star d\phi\vert_{\partial\Sigma}=0, 
\end{align}
where the boundary theta angle is $2\pi$-periodic, $\eta\sim \eta+2\pi$, and $\mu$ is a complex number. Also, $ds$ is the line element on $\partial\Sigma$. Importantly, these boundary conditions break the $U(1)$ winding symmetry, which acts by shifting the boundary theta angle $\eta\to \eta+\alpha$. 
In the abstract approach, this means that they all combine into a single boundary condition of the gauged theory, which can be obtained by fusing the gauging interface onto any one of them. Literally implementing this in terms of Lagrangians using the formulation of the previous subsection suggests that we should take 
\begin{align}
\begin{split}
&S^\partial_{\CB\text{-}\mathrm{BF}}=\int_{\Sigma}\left(\frac{1}{4\pi}d\phi\wedge\star d\phi -\frac{i}{2\pi} d\phi\wedge a-\frac{i}{2\pi}\Phi da \right)+  \int_{\partial\Sigma}\left((\mu e^{i\phi}+\bar\mu e^{-i\phi})ds-\frac{i}{2\pi} \eta d\phi\right), \\
&\hspace{2in} (\star d \phi)\vert_{\partial\Sigma}=0, \ \ \ a\vert_{\partial \Sigma}=0\,.
\end{split}
\end{align}
Naively, it would seem that we still have a dependence on the boundary theta angle $\eta$. However, one can note that a transformation $a\to a+d\lambda$ with $\lambda\vert_{\partial X}=\lambda_0$ constant shifts the boundary theta angle, $\eta\to\eta+\lambda_0$, and it develops that  boundary conditions of the compact boson with the same value of $\mu$ all get identified after coupling to BF theory. 

We can confirm this by integrating out $a$ and $\phi$. As before, this identifies $\phi=\Phi~\mathrm{mod}~2\pi$, which leads to 
\begin{align}\label{eqn:GNLag}
    S_{\NCB}^{\partial}=\frac{1}{4\pi}\int_{\Sigma} d\Phi\wedge\star d\Phi+  \int_{\partial\Sigma}(\mu e^{i\phi}+\bar\mu e^{-i\phi})ds, \ \ \ \ \ (\star d \Phi)\vert_{\partial\Sigma}=0. 
\end{align}
Here, we have dropped a term $-\int_{\partial\Sigma}\frac{i}{2\pi} \eta d\Phi$ from the action because it is a total derivative when $\Phi$ is a non-compact boson. The upshot is that the generalized Neumann boundary conditions of the non-compact boson admit \eqref{eqn:GNLag} as a Lagrangian description. This Lagrangian appeared in e.g. \cite{Callan:1994ub}.

Next, let us discuss how the Dirichlet boundary conditions of the compact boson behave after coupling to BF theory. Before gauging, we have
\begin{align}\label{eqn:DirichletCB}
S_{\CB}^\partial=\frac{1}{4\pi}\int_{\Sigma} d\phi\wedge \star d\phi, \ \ \ \ \ \ \phi\vert_{\partial\Sigma}=\phi_0.
\end{align}
Unlike Neumann boundary conditions, Dirichlet boundary conditions for $\phi$ are symmetric with respect to the $U(1)$ winding symmetry. The abstract approach from Section \ref{subsec:NCBconformalbdy} then predicts that, if we fuse a gauging interface onto a Dirichlet boundary condition, we will get a non-simple boundary condition of the gauged theory. Another way to say this is that the boundary condition 
\begin{align}
    S_{\CB\text{-}\mathrm{BF}}^\partial=\int_{\Sigma}\left(\frac{1}{4\pi}d\phi\wedge\star d\phi-\frac{i}{2\pi}d\phi\wedge a-\frac{i}{2\pi}\Phi da\right), \ \ \ \ \phi\vert_{\partial\Sigma}=\phi_0, \ \ \ a\vert_{\partial\Sigma}=0,
\end{align}
supports topological point operators, namely 
\begin{align}
    x_\alpha\equiv \exp\left(i\frac{\alpha}{2\pi}(\Phi-\phi_0)\right), \ \ \ \ \alpha\in U(1).
\end{align}
The operator $x_\alpha$ is topological due to the bulk equation of motion for $a$, which imposes $d(\Phi-\phi)=0$.
Thus, in order to get a simple boundary condition, we must evidently choose different boundary conditions for the BF sector. 

Before we explain this, it will be useful to give a slightly different Lagrangian description of the Dirichlet boundary conditions of the compact boson. Rather than using \eqref{eqn:DirichletCB}, we implement $\phi\vert_{\partial\Sigma}=\phi_0$ using a compact Lagrange multiplier field  $\widetilde{\phi}\sim \widetilde{\phi}+2\pi$ which lives entirely on the boundary \cite{Choi:2025ebk}. In more detail, we take our action before flat gauging the $U(1)$ symmetry to be
\begin{align}
    S_{\CB}^\partial=\frac{1}{4\pi}\int_{\Sigma}d\phi\wedge\star d\phi+ \frac{i}{2\pi}\int_{\partial\Sigma}(\phi-\phi_0)d\widetilde{\phi}.
\end{align}
Then, once we couple to $U(1)$ BF theory, we should take \emph{Neumann} boundary conditions for the gauge sector. That is, we allow $a$ and $\Phi$ to fluctuate on the boundary. In particular, we take
\begin{align}\label{eqn:gaugeddirichlet}
\begin{split}
&S_{\CB\text{-}\mathrm{BF}}^\partial=\int_{\Sigma}\left(\frac{1}{4\pi}d\phi\wedge\star d\phi -\frac{i}{2\pi}d\phi\wedge a - \frac{i}{2\pi}\Phi da \right) \\
&\hspace{1in}+\frac{i}{2\pi}\int_{\partial\Sigma}(\phi-\phi_0)d\widetilde{\phi}+ i\left(\frac{\phi_0}{2\pi}-Q\right)\int_{\partial\Sigma}a.
\end{split}
\end{align}
We have deformed the action after gauging by the addition of a boundary Wilson line for $a$ with charge $\frac{\phi_0}{2\pi}-Q$. The reason for the improper quantization of the charge of this Wilson line is gauge invariance. Specifically, if we postulate that the fields transform as
\begin{align}
    \phi\to\phi, \ \ \ \Phi\to\Phi, \ \ \ a\to a+d\lambda, \ \ \ \widetilde{\phi} \to \widetilde{\phi}+\lambda|_{\partial \Sigma},
\end{align}
then one readily sees that the term $\frac{i}{2\pi}\phi_0\int_{\partial\Sigma}a$ is needed to cancel the variation of the term $-\frac{i}{2\pi}\phi_0\int_{\partial\Sigma}d\widetilde{\phi}$. 

The action \eqref{eqn:gaugeddirichlet} reproduces the prediction of the abstract approach in Section \ref{subsec:NCBconformalbdy} that each Dirichlet boundary condition of the compact boson splits, after flat gauging the $U(1)$ winding symmetry, into $\mathbb{Z}$-many boundary conditions. Here, we see that these $\mathbb{Z}$-many boundaries of the compact boson are parametrized roughly by the charge of the Wilson line that we have the freedom to dress the boundary by.

\subsection{Topological Lines}\label{subsec:Lag_lines}

Finally, we turn to topological line defects of the non-compact boson. We once again work on a closed Riemann surface $\Sigma$ which is divided into two pieces $\Sigma_L$ and $\Sigma_R$ by a separating cycle $C$. Topological defects are wrapped on $C$.

To begin with, we would like to obtain a Lagrangian description of the $SO(4)$ symmetry lines in the $SU(2)_1$ WZW model. These are somewhat difficult to describe in the compact boson variables for $SU(2)_1$ because the vertex operators which serve as currents for the symmetry do not admit a straightforward Lagrangian description.\footnote{By contrast, the Lagrangian description of $SU(2)_1$ as a sigma model into the group manifold $SU(2)$ manifests the full symmetry group more easily.}

We will get around this by constructing $SO(4)$ topological defects as continuous deformations of the $\mathbb{Z}_2$ T-duality topological defect of $SU(2)_1$, rather than as continuous deformations of the trivial line. As it turns out, it is easier to describe the dimension-1 operators living on the world-line of this defect using the elementary fields of the Lagrangian than it is to describe the ordinary Noether currents of the theory.

To this end, let us begin by considering the T-duality symmetry of the $SU(2)_1$ WZW-model \cite{Harvey:2017rko,Choi:2021kmx,Shao:2025qvf},
\begin{align}\label{eqn:TdualitydefectLag}
\begin{split}
   S_{\CB}^{T\text{-duality}}&= \frac{1}{4\pi}\int_{\Sigma_L}d\phi_L\wedge \star d\phi_L+\frac{1}{4\pi}\int_{\Sigma_R}d\phi_R\wedge \star d\phi_R- \frac{i}{2\pi}\int_{C}\phi_L d\phi_R .
\end{split}
\end{align}
The precise definition of the last term is subtle, and we refer readers to \cite{Shao:2025qvf} for further details.
This is a particular $\mathbb{Z}_2$ symmetry in $SO(4)$. We can then deform this defect by exactly marginal operators which live on its worldline:
\begin{align}\label{eqn:TdualitydefectLag_deform}
\begin{split}
   &\Delta S = \int_C \left(\mu_+ e^{i(\phi_L+\phi_R)} +\mu_-e^{i(\phi_L-\phi_R)}+\mathrm{c.c.}\right)ds +\frac{i}{2\pi}\int_C(\eta_Ld\phi_L+\eta_R d\phi_R),
\end{split}
\end{align}
where $ds$ is the line element on $C$, and $\mu_\pm\in\mathbb{C}$ and $\eta_{L,R}\in\mathbb{R}/2\pi\mathbb{Z}$. This leads to a 6-dimensional moduli space of lines, which is of course the dimension of $SO(4)$.

The intuition for why this Lagrangian presentation works is that the compact boson $\phi_L$ is related to $\phi_R$ by T-duality. Hence, on the defect $C$, we have access to both the compact boson and its T-dual field at the same time, and therefore can construct a larger number of local operators than we would otherwise be able to construct with just the compact boson field in a single duality frame.

Now, we couple to $U(1)$ BF theory. Using the intuition from the abstract approach of the previous section that we are simply sandwiching $SO(4)$ lines with the gauging interface, we are led to the following Lagrangian:
\begin{align}
\begin{split}
    &S^{\mathrm{TDL}}_{\CB\text{-}\mathrm{BF}}=\int_{\Sigma_L}\left(\frac{1}{4\pi}d\phi_L\wedge\star d\phi_L -\frac{i}{2\pi}d\phi_L\wedge a_L -\frac{i}{2\pi}\Phi_L da_L \right)+(L\leftrightarrow R) -\frac{i}{2\pi}\int_C \phi_L d\phi_R \\
    &\hspace{.5in}+\int_C \left(\mu_+ e^{i(\phi_L+\phi_R)} +\mu_-e^{i(\phi_L-\phi_R)}+\mathrm{c.c.}\right)ds +\frac{i}{2\pi}\int_C(\eta_Ld\phi_L+\eta_R d\phi_R),
\end{split}
\end{align}
where we also take Dirichlet boundary conditions for the gauge fields at the location of the defect:
\begin{align}
    a_L\vert_C=a_R\vert_C=0.
\end{align}
We can then integrate out $a_L$, $a_R$, $\phi_L$, and $\phi_R$ just as we did in Section \ref{subsec:Laggauginginterface} to obtain that 
\begin{align}
\begin{split}
    &S^{\mathcal{X}}_{\NCB}=\frac{1}{4\pi}\int_{\Sigma_L}d\Phi_L\wedge\star d\Phi_L + \frac{1}{4\pi}\int_{\Sigma_R}d\Phi_R\wedge \star d\Phi_R  \\
    &\hspace{.5in}-\frac{i}{2\pi} \int_C \Phi_L d\Phi_R+\int_C(\mu_+e^{i(\Phi_L+\Phi_R)}+\mu_-e^{i(\Phi_L-\Phi_R)}+\mathrm{c.c.})ds.
\end{split}
\end{align}
We have dropped a contribution to the action of the form 
\begin{align}
    \frac{i}{2\pi}\int_C (\eta_L d\Phi_L+\eta_R d\Phi_R)
\end{align}
because it is a total derivative when $\Phi_{L,R}$ are non-compact and hence vanishes. One should think of $\mu_+,\mu_-\in\mathbb{C}$ as coordinates of the moduli space of topological lines we discovered in Section \ref{subsubsec:modulispacelines} in a neighborhood of the $\mathbb{R}$ Tambara-Yamagami duality line $\mathcal{D}^R$. A special case of this Lagrangian, after folding and performing a change of variables, appears in \cite{Callan:1995em} who studied boundary conditions of the $c=2$ theory of two non-compact bosons.

\section[Boundary States, Defect Operators, and Interval/Twisted Hilbert Spaces]
{Boundary States, Defect Operators, and Interval/\protect\\Twisted Hilbert Spaces}\label{sec:explicitNCB}

In this section, we will provide descriptions of the conformal boundary conditions and the topological line defects of the non-compact boson theory in abstract CFT language.

For the conformal boundaries, we will derive the corresponding conformal boundary states and the interval Hilbert spaces, and check that  open/closed duality is satisfied.
For the topological lines, we will explain how they act on the Hilbert space of the non-compact boson theory, and derive the spectrum of twisted Hilbert spaces (equivalently, the spectrum of defect operators attached to the topological lines). 

Throughout, we will invoke various known facts about the $SU(2)_1$ WZW model and the non-compact boson theory.
They are reviewed in Appendices \ref{app:su2} and \ref{app:ncb}, where the notations are also set.

\subsection{Gauging Interface} \label{sec:gauging_interface_U1}

We first lay out some more detailed properties of the $U(1)$ flat-gauging interface $\mathcal{I}_{U(1)}$, which was discussed in Sections \ref{sec:int_abs} and \ref{subsec:Laggauginginterface}.

In particular, take as our spacetime manifold a cylinder, where the spatial direction is a circle.
On a portion of the cylinder before some given (Euclidean) time, we put the seed theory $SU(2)_1$, and after that, we put the non-compact boson theory, such that the two theories are separated by the gauging interface $\mathcal{I}_{U(1)}$ supported at a fixed time slice.

This defines a map from the Hilbert space of $SU(2)_1$ to that of the non-compact boson theory (both quantized on a circle).
More generally, we will consider a family of maps from the Hilbert spaces of $SU(2)_1$ twisted by a winding symmetry defect $\mathcal{L}_{\alpha}^W$, to the untwisted Hilbert space of the non-compact boson. These are pictorially given by
\begin{align}\label{eq:U1_interface}
\widehat{\mathcal{I}}^{\alpha}_{U(1)} =\;
\begin{tikzpicture}[baseline={([yshift=-0.5ex]0,1.4)}]
  \fill[su] (-1.5,0) arc[start angle=180, end angle=360, x radius=1.5, y radius=0.35]
        -- (1.5,1.4) arc[start angle=0, end angle=-180, x radius=1.5, y radius=0.35] -- cycle;
  \fill[ncb] (-1.5,1.4) arc[start angle=180, end angle=360, x radius=1.5, y radius=0.35]
        -- (1.5,2.8) arc[start angle=0, end angle=-180, x radius=1.5, y radius=0.35] -- cycle;
  \draw[silh] (-1.5,0) -- (-1.5,2.8)  (1.5,0) -- (1.5,2.8);
  \draw[frame] (0,2.8) ellipse[x radius=1.5, y radius=0.35];
  \draw[frame] (-1.5,0) arc[start angle=180, end angle=360, x radius=1.5, y radius=0.35];
  \draw[tline, midarrow] (0,-0.35) -- (0,1.05);
  \draw[iface] (-1.5,1.4) arc[start angle=180, end angle=360, x radius=1.5, y radius=0.35];
  \node at (0.78,1.8)  {$\mathrm{NCB}$};
  \node at (0.82,0.4)  {$SU(2)_1$};
  \node[ifacecol, anchor=north] at (-0.75,1.8) {$\mathcal{I}_{U(1)}$};
  \node[below=1pt] at (-1+.6,.6) {$\mathcal{L}^{W}_{\alpha}$};
\end{tikzpicture}
\;\;:\;\mathcal{H}^{\alpha}_{SU(2)_1}\to\mathcal{H}_{\mathrm{NCB}}\,.
\end{align}
Here, the hat on $\widehat{\mathcal{I}}_{U(1)}^\alpha$ is to indicate that it is an operator acting on a Hilbert space.
See Appendices \ref{app:su2} and \ref{app:ncb} for reviews of various Hilbert spaces which enter \eqref{eq:U1_interface}.
In this picture, the winding symmetry line $\mathcal{L}_\alpha^W$ ends topologically on the interface $\mathcal{I}_{U(1)}$ (see Section \ref{sec:int_abs}).
When $\alpha = 0$, we will also write $\widehat{\mathcal{I}}_{U(1)}^{\alpha = 0} \equiv \widehat{\mathcal{I}}_{U(1)}$.

The map $\widehat{\mathcal{I}}_{U(1)}^\alpha: \mathcal{H}_{SU(2)_1}^\alpha \rightarrow \mathcal{H}_{\NCB}$ can be uniquely fixed by imposing the following two conditions, up to an overall normalization.

First, it projects out every state in $\mathcal{H}_{SU(2)_1}^\alpha$ whose winding charge is nonzero.
Such states are not gauge-invariant once we flat-gauge the $U(1)$ winding symmetry to obtain the non-compact boson theory.
(A similar discussion in the case of discrete gauging can be found, for instance, in \cite{Fukusumi:2021zme}.)

Second, it commutes with the left- and right-moving $\widehat{\mathfrak{u}}(1)$ current algebras.
To be more precise, from the Lagrangian description of the gauging interface \eqref{eq:lag_interface}, it follows that the $\widehat{\mathfrak{u}}(1)$ currents $\partial \phi$, $\bar{\partial} \phi$ of $SU(2)_1$  (in the compact boson description of the theory) map to the $\widehat{\mathfrak{u}}(1)$ currents $\partial \Phi$, $\bar{\partial} \Phi$ of the non-compact boson theory across the interface.
In particular, the interface commutes with the Virasoro algebra, and is topological.

With these requirements, we find that
\begin{equation} \label{eq:u1_interface_ops}
    \widehat{\mathcal{I}}_{U(1)}^\alpha | n,w ;\alpha \rangle_{SU(2)_1} = \delta_{w,0} \int_{-\infty}^\infty dp \, \delta \left( p - \left(n+\frac{\alpha}{2\pi} \right) \right)\ket{p}_{\NCB} \,.
\end{equation}
Here, $\ket{n,w;\alpha}_{SU(2)_1} \in \mathcal{H}_{SU(2)_1}^\alpha$ and $\ket{p}_{\NCB} \in \mathcal{H}_{\NCB}$ are primary states with respect to the $\widehat{\mathfrak{u}}(1)$ current algebras of the two theories.
See Appendices \ref{app:su2} and \ref{app:ncb} for more details.
The choice of the overall normalization of the maps \eqref{eq:u1_interface_ops} will be explained momentarily.

Reversing the orientation of the gauging interface, we obtain the adjoint maps,
\begin{align}\label{eq:U1_interface_adj}
\left(\widehat{\mathcal{I}}^{\alpha}_{U(1)}\right)^{\dagger} =\;
\begin{tikzpicture}[baseline={([yshift=-0.5ex]0,1.4)}]
  \fill[ncb] (-1.5,0) arc[start angle=180, end angle=360, x radius=1.5, y radius=0.35]
        -- (1.5,1.4) arc[start angle=0, end angle=-180, x radius=1.5, y radius=0.35] -- cycle;
  \fill[su] (-1.5,1.4) arc[start angle=180, end angle=360, x radius=1.5, y radius=0.35]
        -- (1.5,2.8) arc[start angle=0, end angle=-180, x radius=1.5, y radius=0.35] -- cycle;
  \draw[silh] (-1.5,0) -- (-1.5,2.8)  (1.5,0) -- (1.5,2.8);
  \draw[frame] (0,2.8) ellipse[x radius=1.5, y radius=0.35];
  \draw[frame] (-1.5,0) arc[start angle=180, end angle=360, x radius=1.5, y radius=0.35];
  \draw[tline, midarrow] (0,1.05) -- (0,2.45);
  \draw[iface] (-1.5,1.4) arc[start angle=180, end angle=360, x radius=1.5, y radius=0.35];
  \node at (0.82,1.8)  {$SU(2)_1$};
  \node at (0.78,0.4)  {$\mathrm{NCB}$};
  \node[left=2pt] at (0,1.85) {$\mathcal{L}^{W}_{\alpha}$};
  \node[ifacecol, anchor=north] at (-0.75,1.0) {$\overline{\mathcal{I}}_{U(1)}$};
\end{tikzpicture}
\;\;:\;\mathcal{H}_{\mathrm{NCB}}\to\mathcal{H}^{\alpha}_{SU(2)_1}\,.
\end{align}
They act on states as
\begin{equation} \label{eq:u1_interface_dual}
    \left(\widehat{\mathcal{I}}_{U(1)}^{\alpha}\right)^\dagger \ket{p}_{\NCB} = \sum_{n\in\mathbb{Z}} \delta \left( p - \left(n+\frac{\alpha}{2\pi} \right) \right)\ket{n,0;\alpha}_{SU(2)_1} \,,
\end{equation}
which follows from \eqref{eq:u1_interface_ops}.

Let us discuss further properties of the gauging interface maps \eqref{eq:u1_interface_ops} and \eqref{eq:u1_interface_dual}.
It is instructive to compute their compositions
\begin{align}
\begin{split}
    \widehat{\mathcal{I}}_{U(1)}^{\alpha} \circ \left(\widehat{\mathcal{I}}_{U(1)}^{\alpha}\right)^\dagger :\quad& \mathcal{H}_{\NCB} \rightarrow \mathcal{H}_{\NCB} \,,\\
    \left(\widehat{\mathcal{I}}_{U(1)}^{\alpha}\right)^\dagger \circ \widehat{\mathcal{I}}_{U(1)}^{\alpha} :\quad& \mathcal{H}^{\alpha}_{SU(2)_1} \rightarrow \mathcal{H}^{\alpha}_{SU(2)_1} \,.
\end{split}
\end{align} 
First, we have
\begin{align} \label{eq:u1_comp2}
\begin{split}
    \widehat{\mathcal{I}}_{U(1)}^{\alpha} \circ \left(\widehat{\mathcal{I}}_{U(1)}^{\alpha}\right)^\dagger \ket{p}_{\NCB} &= \sum_{n\in\mathbb{Z}} \delta \left(p - \left( n + \frac{\alpha}{2\pi} \right) \right) \ket{p}_{\NCB} \\
    &= \sum_{s\in\mathbb{Z}} \exp \left[2\pi i \left( p- \frac{\alpha}{2\pi} \right)s \right] \ket{p}_{\NCB} \,,
\end{split}
\end{align}
where we used that
\begin{equation} \label{eq:delta_fourier}
    \sum_{s\in\mathbb{Z}} e^{2\pi i x s } = \sum_{n\in\mathbb{Z}} \delta (x - n ) \,,
\end{equation}
which holds for any real $x$.
When $\alpha = 0$, \eqref{eq:u1_comp2} becomes
\begin{equation} \label{eq:IIdop}
    \widehat{\mathcal{I}}_{U(1)} \circ \widehat{\mathcal{I}}_{U(1)}^\dagger \ket{p}_{\NCB} = \sum_{n \in \mathbb{Z} } \widehat{\mathcal{M}}_{2\pi n} \ket{p}_{\NCB} \,,
\end{equation}
where $\widehat{\mathcal{M}}_{2\pi n}$ denotes the action of the $\mathbb{Z} \subset \mathbb{R}$ momentum symmetry---which is the dual symmetry we get from the $U(1)$ flat-gauging---on the Hilbert space of the non-compact boson theory (see \eqref{eq:Rmop} below).
Note that \eqref{eq:IIdop} is consistent with the fusion algebra given in \eqref{eqn:IIast}.
The overall normalization of $\widehat{\mathcal{I}}_{U(1)}$ is chosen such that in the fusion algebra, the  momentum symmetry lines $\mathcal{M}_{2\pi n}$ appear with multiplicity 1.

For $\alpha \neq 0$ mod $2\pi$, the overall normalization of $\widehat{\mathcal{I}}_{U(1)}^\alpha$ can be tuned by rescaling the topological junction between the interface $\mathcal{I}_{U(1)}$ and the topological line $\mathcal{L}_{\alpha}^W$, and we choose the normalization of junctions such that \eqref{eq:u1_interface_ops} holds.

Next, the composition in the opposite order gives
\begin{align} \label{eq:u1_comp}
\begin{split}
    \left(\widehat{\mathcal{I}}_{U(1)}^{\alpha}\right)^\dagger \circ \widehat{\mathcal{I}}_{U(1)}^{\alpha} \ket{n,w ;\alpha}_{SU(2)_1} &= \delta_{w,0} \sum_{n' \in \mathbb{Z}} \delta (n-n' ) \ket{n' , 0;\alpha}_{SU(2)_1} \\
    &= \mathcal{N} \delta_{w,0} \ket{n , 0;\alpha}_{SU(2)_1} \,,
\end{split}
\end{align}
where $\mathcal{N} = {}_{\NCB}\langle p | p \rangle_{\NCB} = \delta (0)$ is proportional to the volume of the target space of the non-compact boson (see \eqref{eq:R_volume}).
Formally, this is divergent and it reflects the fact that the interface $\mathcal{I}_{U(1)}$ has an infinite quantum dimension.

We may define the quantum dimension of the gauging interface $\mathcal{I}_{U(1)}$ to be the square root of the eigenvalue of the operator $\widehat{\mathcal{I}}_{U(1)} \circ \widehat{\mathcal{I}}_{U(1)}^\dagger$ acting on the vacuum state of the non-compact boson theory, or alternatively, to be the square root of the eigenvalue of $\widehat{\mathcal{I}}_{U(1)}^\dagger \circ \widehat{\mathcal{I}}_{U(1)}$ acting on the vacuum state of $SU(2)_1$.

In either case, from \eqref{eq:u1_comp2} or from \eqref{eq:u1_comp}, we would formally conclude that the quantum dimension of the gauging interface $\mathcal{I}_{U(1)}$ is $\mathcal{N}^{1/2}$.
We will soon show that the divergent factor of $\mathcal{N}$ is in fact crucial for the consistency of the maps $\widehat{\mathcal{I}}_{U(1)}^\alpha$.

Note that, when $\alpha = 0$,  \eqref{eq:u1_comp} can be rewritten as
\begin{equation}
    \widehat{\mathcal{I}}_{U(1)}^\dagger \circ \widehat{\mathcal{I}}_{U(1)} \ket{n,w}_{SU(2)_1} = \frac{\mathcal{N}}{2\pi} \int_0^{2\pi} d\alpha \,\widehat{\mathcal{L}}_\alpha^W \ket{n,w}_{SU(2)_1} \,.
\end{equation}
This is consistent with the abstract fusion algebra in \eqref{eqn:IastI}.

Let us perform another consistency check, which will help us better understand the presence of the divergent factor $\mathcal{N}$ in \eqref{eq:u1_comp}.
On a torus, we must have
\begin{equation} \label{eq:u1_consistency}
    \mathrm{Tr}_{\mathcal{H}^{\alpha}_{SU(2)_1}} \left( \left(\widehat{\mathcal{I}}_{U(1)}^{\alpha}\right)^\dagger \circ \widehat{\mathcal{I}}_{U(1)}^{\alpha} \, q^{L_0 - \frac{1}{24}} \bar{q}^{\bar{L}_0 - \frac{1}{24}} \right) \stackrel{!}{=} \mathrm{Tr}_{\mathcal{H}_{\NCB}} \left( \widehat{\mathcal{I}}_{U(1)}^{\alpha} \circ \left(\widehat{\mathcal{I}}_{U(1)}^{\alpha}\right)^\dagger q^{L_0 - \frac{1}{24}} \bar{q}^{\bar{L}_0 - \frac{1}{24}} \right) \,,
\end{equation}
since $\widehat{\mathcal{I}}_{U(1)}^{\alpha}$ and $\left(\widehat{\mathcal{I}}_{U(1)}^{\alpha}\right)^\dagger$ commute with the Virasoro generators.
Let us verify \eqref{eq:u1_consistency} using \eqref{eq:u1_comp2} and \eqref{eq:u1_comp}.
The left-hand side of \eqref{eq:u1_consistency} is
\begin{align}
\begin{split}
     &\mathrm{Tr}_{\mathcal{H}^{\alpha}_{SU(2)_1}} \left( \left(\widehat{\mathcal{I}}_{U(1)}^{\alpha}\right)^\dagger \circ \widehat{\mathcal{I}}_{U(1)}^{\alpha} \, q^{L_0 - \frac{1}{24}} \bar{q}^{\bar{L}_0 - \frac{1}{24}} \right) \\
     &\hspace{.6in}= \mathcal{N} \sum_{n\in\mathbb{Z}} |\chi^{\widehat{\mathfrak{u}}(1)}_{n+\frac{\alpha}{2\pi}} (q) |^2 = \frac{\mathcal{N}}{|\eta(q)|^2} \sum_{n\in\mathbb{Z}} (q\bar{q})^{\frac{1}{4}\left( n + \frac{\alpha}{2\pi} \right)^2} \,,
\end{split}
\end{align}
and the right-hand side of \eqref{eq:u1_consistency} is easily seen to match this,
\begin{align}
\begin{split}
    &\mathrm{Tr}_{\mathcal{H}_{\NCB}} \left( \widehat{\mathcal{I}}_{U(1)}^{\alpha} \circ \left(\widehat{\mathcal{I}}_{U(1)}^{\alpha}\right)^\dagger q^{L_0 - \frac{1}{24}} \bar{q}^{\bar{L}_0 - \frac{1}{24}} \right) \\
    &\hspace{.3in}= \int_{-\infty}^{\infty} dp \sum_{s\in\mathbb{Z}} \exp \left[ 2\pi i \left( p - \frac{\alpha}{2\pi} \right) s \right] {}_{\NCB}\langle p | p \rangle_{\NCB} \times |\chi^{\widehat{\mathfrak{u}}(1)}_p (q)|^2 \\
    &\hspace{.3in}= \frac{\mathcal{N}}{|\eta(q)|^2} \sum_{n\in\mathbb{Z}} (q\bar{q})^{\frac{1}{4}\left( n + \frac{\alpha}{2\pi} \right)^2} \,,
\end{split}
\end{align}
where we used \eqref{eq:delta_fourier}.\footnote{When we take a trace over the states in $\mathcal{H}_{\NCB}$, each $\widehat{\mathfrak{u}}(1)$ character \eqref{eq:u1_char} contributes with an additional factor of $\mathcal{N} = \delta (0) = {}_{\NCB}\langle p | p \rangle_{\NCB}$ since in the full theory the primary states $\ket{p}_{\NCB}$ are only $\delta$-function normalizable as in \eqref{eq:p_normalization}.\label{ncb_part}}

Physically, the overall factor of $\mathcal{N}$ comes from the constant mode of the non-compact boson in the path integral.
We find that the two sides of Equation \eqref{eq:u1_consistency} indeed agree with each other, when the divergent volume of the non-compact target space is properly kept track of.

Finally, for later purposes, let us also mention how the gauging interface maps act on the  Virasoro primary states of the two theories.
As in \eqref{eq:cb_vir_primary}, we denote the Virasoro primary states in the untwisted Hilbert space of the $SU(2)_1$ WZW model by 
\begin{align}
    \ket{J,\bar{J},m,\bar{m}}_{SU(2)_1}, \ \ \ J,\bar{J} \in \frac{1}{2}\mathbb{Z}_{\geq 0},\ J - \bar{J} \in \mathbb{Z}, \ \  m = -J , \dots , J, \ \  \bar{m} = -\bar{J}  , \dots , \bar{J}.
\end{align}
In the Hilbert space of the non-compact boson, as in \eqref{eq:ncb_vir_primaries}, we have the Virasoro primary states 
\begin{align}
    \ket{p}_{\NCB}, \ \ \ p\notin\mathbb{Z}, \hspace{.5in} \ket{p,\ell,\bar{\ell}}_{\NCB}, \ \ \ p \in \mathbb{Z}, \ \ell , \bar{\ell} \in \mathbb{Z}_{\geq 0}.
\end{align} 
The latter states $\ket{p,\ell,\bar\ell}$  are $\widehat{\mathfrak{u}}(1)$ descendants except when $\ell = \bar{\ell} = 0$.

The gauging interface acts on these states as
\begin{equation} \label{eq:u1_interface_vir}
    \widehat{\mathcal{I}}_{U(1)} \ket{J,\bar{J},m,\bar{m}}_{SU(2)_1} = \delta_{m,\bar{m}} \ket{2m, J-|m|, \bar{J} - |\bar{m}|}_{\NCB} \,,
\end{equation}
and the adjoint map is
\begin{align} \label{eq:u1_interface_vir_adj}
\begin{split}
    &\widehat{\mathcal{I}}_{U(1)}^\dagger \ket{p,\ell,\bar{\ell}}_{\NCB} = \mathcal{N} | \tfrac{|p|}{2} + \ell , \tfrac{|p|}{2} + \bar{\ell} , \tfrac{p}{2} , \tfrac{p}{2} \rangle_{SU(2)_1} \,,\quad p\in\mathbb{Z}\,, \\ 
    &\widehat{\mathcal{I}}_{U(1)}^\dagger \ket{p}_{\NCB} =0 \,,\quad p\notin\mathbb{Z}\,.
\end{split}
\end{align}

\subsection{Conformal Boundaries} \label{sec:bdyabst}

\subsubsection{Boundary States}\label{subsubsec:NCBbdyGN}

We begin by computing conformal boundary states \cite{Cardy:2004hm} of all the conformal boundary conditions of the non-compact boson theory that were discussed in the previous sections, using the description of the gauging interface given in Section \ref{sec:gauging_interface_U1}.
We will reproduce known expressions for the boundary states, and this serves as a consistency check.

\paragraph{Generalized Neumann.}

Recall that, by fusing the $U(1)$ gauging interface onto simple conformal boundary conditions of the $SU(2)_1$ WZW model, we obtain a two-parameter family of generalized Neumann boundary conditions of the non-compact boson theory, as in \eqref{eqn:Nadef}.

Let us denote the boundary state corresponding to a generalized Neumann boundary condition $\mathbf{N}_a$ as $\ket{\mathbf{N}_a}_{\NCB}$, where $a\in \mathbb{C}$, $|a|<1$.
Similarly, denote the boundary state corresponding to the boundary condition $\mathbf{B}_{a,b}$ of the $SU(2)_1$ theory as $\ket{\mathbf{B}_{a,b}}_{SU(2)_1}$.
The boundary states $\ket{\mathbf{B}_{a,b}}_{SU(2)_1}$ were originally computed in \cite{Gaberdiel:2001xm,Gaberdiel:2001zq}, and they are reviewed in Appendix \ref{app:su2}.

From \eqref{eqn:Nadef}, it follows that\footnote{The diagonal matrix elements $D_{mm}^J (a,b)$ are determined by $a$, and they do not depend on $b$, other than through the condition $|a|^2 + |b|^2 = 1$ as in \eqref{eq:gab}. This is consistent with the fact that the generalized Neumann boundary conditions $\mathbf{N}_a$ are parameterized by $a$.}
\begin{align} \label{eq:ncb_gen_neu}
\begin{split}
    \ket{\mathbf{N}_a}_{\NCB} &= \widehat{\mathcal{I}}_{U(1)} \ket{\mathbf{B}_{a,b}}_{SU(2)_1} \\
    &= 2^{-1/4} \sum_{J \in \frac{1}{2}\mathbb{Z}_{\geq 0}} \sum_{m = -J}^J D^J_{m m} (a,b) |2m, J - |m| \rrangle_{\NCB} \,.
\end{split}
\end{align}
In the final expression, $D_{mm}^J (a,b)$ are (diagonal) elements of the Wigner D-matrices, which are given explicitly in \eqref{eq:su2_matrix_rep}, and $|2m, J - |m| \rrangle_{\NCB}$ are Virasoro Ishibashi states of the non-compact boson theory.
More details on the Ishibashi states are given in Appendix \ref{app:ncb}.

The boundary states \eqref{eq:ncb_gen_neu} were first studied perturbatively in \cite{Callan:1993mw}.
The exact boundary states later appeared in \cite{Callan:1994ub} where they were obtained by taking the infinite radius limit of known boundary states of the compact boson theory.
Here, we reproduce the exact boundary states from  $U(1)$ flat-gauging.

\paragraph{Dirichlet.}

Let us denote the boundary state that corresponds to the Dirichlet boundary condition $\mathbf{D}_{\Phi_0}$ of the non-compact boson theory as $\ket{\mathbf{D}_{\Phi_0}}_{\NCB}$, where $\Phi_0 \in \mathbb{R}$.
As in \eqref{eqn:latticeofdirichlet}, if we fuse the $U(1)$ gauging interface to a boundary condition $\mathbf{B}_{a,0}$ of the $SU(2)_1$ theory, with $|a| = 1$, we obtain an infinite discrete superposition of Dirichlet boundary conditions $\mathbf{D}_{\mathrm{arg}(a) - 2\pi Q}$, with $Q \in \mathbb{Z}$.

We can project onto a simple summand $\mathbf{D}_{\mathrm{arg}(a) - 2\pi Q}$ with some fixed value of $Q$ as in \eqref{eq:bdy_proj}.
In particular, combined with \eqref{eq:xalpha}, the conformal boundary state $\ket{\mathbf{D}_{\mathrm{arg}(a) - 2\pi Q}}_{\NCB}$ is given pictorially by
\begin{align}\label{eq:dirichlet}
\ket{\mathbf{D}_{\mathrm{arg}(a)-2\pi Q}}_{\mathrm{NCB}}
=\frac{1}{2\pi}\int_0^{2\pi} d\alpha\, e^{-i\alpha Q}\;
\begin{tikzpicture}[baseline={([yshift=-0.5ex]0,0.75)}]
  \fill[su] (-1.5,0) arc[start angle=180, end angle=360, x radius=1.5, y radius=0.3]
        -- (1.5,1.1) arc[start angle=0, end angle=-180, x radius=1.5, y radius=0.3] -- cycle;
  \fill[ncb] (-1.5,1.1) arc[start angle=180, end angle=360, x radius=1.5, y radius=0.3]
        -- (1.5,2.2) arc[start angle=0, end angle=-180, x radius=1.5, y radius=0.3] -- cycle;
  \draw[silh] (-1.5,0) -- (-1.5,2.2)  (1.5,0) -- (1.5,2.2);
  \draw[frame] (0,2.2) ellipse[x radius=1.5, y radius=0.3];
  \draw[tline, midarrow] (0,-0.3) -- (0,0.8);
  \draw[tline] (-1.5,0) arc[start angle=180, end angle=360, x radius=1.5, y radius=0.3];
  \draw[iface] (-1.5,1.1) arc[start angle=180, end angle=360, x radius=1.5, y radius=0.3];
  \node at (0.78,1.42) {$\mathrm{NCB}$};
  \node at (0.82,0.3)  {$SU(2)_1$};
  \node[left=4pt] at (0,0.3) {$\mathcal{L}^{W}_{\alpha}$};
  \node[ifacecol, anchor=south] at (-0.72,0.93) {$\mathcal{I}_{U(1)}$};
  \node[below=1pt] at (0,-0.3) {$\ket{\mathbf{B}_{a,0};\alpha}_{SU(2)_1}$};
  \node[overlay, anchor=base west, inner sep=0pt] at ([xshift=4pt]1.5,0) {.};
\end{tikzpicture}
\end{align}
On the right-hand side, for each $\alpha$, we have a \emph{twisted-sector boundary state}\footnote{General discussions on twisted-sector boundary states can be found in \cite{Choi:2024tri}, and various examples appear, for instance, in \cite{Fukusumi:2021zme,Ebisu:2021acm,Gaberdiel:2002qa,Fuchs:1997kt,Fuchs:2000cm,Fuchs:2004dz,Collier:2021ngi}.} $\ket{\mathbf{B}_{a,0};\alpha}_{SU(2)_1}$, which is acted upon by $\widehat{\mathcal{I}}_{U(1)}^\alpha$ given in \eqref{eq:U1_interface} and \eqref{eq:u1_interface_ops}.  
In other words, 
\begin{align}
    \ket{\mathbf{D}_{\mathrm{arg}(a) - 2\pi Q}}_{\NCB} = \frac{1}{2\pi} \int_0^{2\pi} d\alpha \, e^{-i\alpha Q} \, \widehat{\mathcal{I}}_{U(1)}^\alpha \ket{\mathbf{B}_{a,0};\alpha}_{SU(2)_1}.
\end{align}

The explicit expressions for the twisted-sector boundary states $\ket{\mathbf{B}_{a,0};\alpha}_{SU(2)_1}$ are given in \eqref{eq:twisted_Dir}.
Using this and \eqref{eq:dirichlet}, we find that
\begin{align}\label{eqn:DbdystateNCB}
\begin{split}
    \ket{\mathbf{D}_{\Phi_0}}_{\NCB} &= 2^{-1/4}  \int_{-\infty}^\infty dp \, e^{ip \Phi_0} |p\rrangle^{\widehat{\mathfrak{u}}(1),-}_{\NCB} \,.
\end{split}
\end{align}
Here, $|p\rrangle^{\widehat{\mathfrak{u}}(1),-}_{\NCB}$ are $\widehat{\mathfrak{u}}(1)$ Ishibashi states \cite{Ishibashi:1988kg} of the non-compact boson theory with momentum $p$, which we review in Appendix \ref{app:ncb}.
This reproduces the Dirichlet boundary state of the non-compact boson theory.

\paragraph{$g$-functions.}

Let us discuss the $g$-functions \cite{Affleck:1991tk} of the boundary conditions above.
The $g$-function of a conformal boundary condition is defined to be the overlap between the corresponding boundary state and the vacuum state of the theory.

First, all the generalized Neumann boundary conditions have infinite $g$-functions proportional to the volume of the target space.
Specifically, from \eqref{eq:ncb_gen_neu}, we find that
\begin{equation}
    g (\mathbf{N}_a ) \equiv {}_{\NCB} \langle 0 | \mathbf{N}_a \rangle_{\NCB} = 2^{-1/4} {}_{\NCB} \langle 0 | 0 \rangle_{\NCB} = 2^{-1/4}  \mathcal{N} \,.
\end{equation}
The $g$-function of $\mathbf{N}_a$ is independent of the value of $a$.
In the case of the ordinary Neumann boundary condition (for which $a=0$), the infinite $g$-function proportional to $\mathcal{N}$ is familiar, and it comes about physically due to the zero mode of the non-compact boson which is not constrained on the boundary.

For the Dirichlet boundary conditions, from \eqref{eqn:DbdystateNCB}, we have
\begin{equation}
    g(\mathbf{D}_{\Phi_0} ) \equiv {}_{\NCB} \langle 0 | \mathbf{D}_{\Phi_0} \rangle_{\NCB} = 2^{-1/4} \,,
\end{equation}
independent of the value of $\Phi_0$.
Unlike the case of the Neumann boundary condition, setting the non-compact boson to have a fixed value $\Phi_0$ at the boundary constrains its zero mode, and relatedly the $g$-function is finite.

\subsubsection{Interval Hilbert Spaces and Annulus Partition Functions}\label{subsubsec:intervalannulus}

We now study partition functions of the non-compact boson on the annulus $S^1\times I$. Let us set the length of the interval $I$ to be 1, and the circumference of the circle $S^1$ to be $\delta >0$.
Throughout, we write
\begin{align}
    q = e^{-\pi \delta}, \ \ \ \ \tilde{q} = e^{-\frac{4\pi}{\delta}}.
\end{align}
We impose a generalized Neumann boundary condition on the right end of the interval, and either a Dirichlet or a generalized Neumann boundary condition on the left end.\footnote{Special cases where one imposes the same generalized Neumann boundary condition at the two ends of the interval were discussed in \cite{Gaberdiel:2001zq}.} 
The case that both ends of the interval are equipped with Dirichlet boundary conditions is a standard string theory calculation, and the corresponding annulus partition function is
\begin{align}
    \mathcal{Z}_{\NCB}^{\mathbf{D}_{\Phi_0}\mathbf{D}_{\Phi_1}}(\delta)=\frac{q^{(\Phi_1-\Phi_0)^2/4\pi^2}}{\eta(q)},
\end{align}
and so we do not consider it further.

As a consistency check on our framework, we will compute these partition functions in two ways. First, we will compute them in the ``open string channel'' by determining and tracing over the appropriate interval Hilbert space. Second, we will compute them in the ``closed string channel,'' where we simply take the overlap of the boundary states calculated in Section \ref{subsubsec:NCBbdyGN}. The agreement of these two calculations constitutes a check of open/closed duality for these boundaries (a.k.a. the Cardy condition) \cite{Cardy:1989ir}.

\paragraph{Open string channel.} Let us begin with the open string channel. We take time to run along the circle direction of the annulus.

First, we impose a generalized Neumann boundary condition $\mathbf{N}_a$ on the left-hand side of the interval, and a generalized Neumann boundary condition $\mathbf{N}_{a'}$ on the right-hand side of the interval. We will call the corresponding interval Hilbert space $\mathcal{H}^{\mathbf{N}_a\mathbf{N}_{a'}}_{\NCB}$.
Through the state-operator correspondence, states in $\mathcal{H}^{\mathbf{N}_a\mathbf{N}_{a'}}_{\NCB}$ correspond to boundary-condition-changing operators between $\mathbf{N}_a$ and $\mathbf{N}_{a'}$.

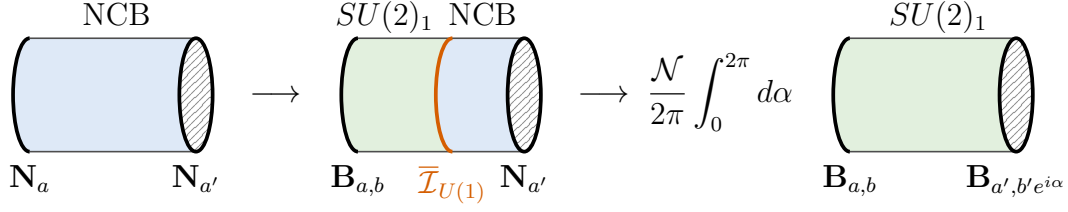
\begin{figure}
\begin{center}
    \begin{tikzpicture}[baseline={([yshift=-0.5ex]0,0)}]
  \fill[ncb] (2.2,0.75) -- (0,0.75) arc[start angle=90, end angle=270, x radius=0.22, y radius=0.75]
             -- (2.2,-0.75) arc[start angle=270, end angle=90, x radius=0.22, y radius=0.75] -- cycle;
  \draw[silh] (0,0.75) -- (2.2,0.75)  (0,-0.75) -- (2.2,-0.75);
  \draw[tline] (0,0.75) arc[start angle=90, end angle=270, x radius=0.22, y radius=0.75];
  \draw[endcap] (2.2,0) ellipse[x radius=0.22, y radius=0.75];
  \node[anchor=base] at (1.15,0.92) {$\mathrm{NCB}$};
  \node[below=1pt] at (0,-0.75)   {$\mathbf{N}_{a}$};
  \node[below=1pt] at (2.2,-0.75) {$\mathbf{N}_{a'}$};
\end{tikzpicture}
\;$\longrightarrow$\;
\begin{tikzpicture}[baseline={([yshift=-0.5ex]0,0)}]
  \fill[su]  (1.25,0.75) -- (0,0.75) arc[start angle=90, end angle=270, x radius=0.22, y radius=0.75]
             -- (1.25,-0.75) arc[start angle=270, end angle=90, x radius=0.22, y radius=0.75] -- cycle;
  \fill[ncb] (2.2,0.75) -- (1.25,0.75) arc[start angle=90, end angle=270, x radius=0.22, y radius=0.75]
             -- (2.2,-0.75) arc[start angle=270, end angle=90, x radius=0.22, y radius=0.75] -- cycle;
  \draw[silh] (0,0.75) -- (2.2,0.75)  (0,-0.75) -- (2.2,-0.75);
  \draw[tline] (0,0.75) arc[start angle=90, end angle=270, x radius=0.22, y radius=0.75];
  \draw[iface] (1.25,0.75) arc[start angle=90, end angle=270, x radius=0.22, y radius=0.75];
  \draw[endcap] (2.2,0) ellipse[x radius=0.22, y radius=0.75];
  \node[anchor=base] at (0.35,0.92) {$SU(2)_1$};
  \node[anchor=base] at (1.65,0.92) {$\mathrm{NCB}$};
  \node[below=1pt] at (0,-0.75)   {$\mathbf{B}_{a,b}$};
  \node[ifacecol, below=1pt] at (1.25,-0.75) {$\overline{\mathcal{I}}_{U(1)}$};
  \node[below=1pt] at (2.2,-0.75) {$\mathbf{N}_{a'}$};
\end{tikzpicture}
\;$\displaystyle\longrightarrow\;
\frac{\mathcal{N}}{2\pi}\int_0^{2\pi} d\alpha$\;
\begin{tikzpicture}[baseline={([yshift=-0.5ex]0,0)}]
  \fill[su] (2.2,0.75) -- (0,0.75) arc[start angle=90, end angle=270, x radius=0.22, y radius=0.75]
            -- (2.2,-0.75) arc[start angle=270, end angle=90, x radius=0.22, y radius=0.75] -- cycle;
  \draw[silh] (0,0.75) -- (2.2,0.75)  (0,-0.75) -- (2.2,-0.75);
  \draw[tline] (0,0.75) arc[start angle=90, end angle=270, x radius=0.22, y radius=0.75];
  \draw[endcap] (2.2,0) ellipse[x radius=0.22, y radius=0.75];
  \node[anchor=base] at (1.17,0.92) {$SU(2)_1$};
  \node[below=1pt] at (0,-0.75)   {$\mathbf{B}_{a,b}$};
  \node[below=1pt] at (2.2,-0.75) {$\mathbf{B}_{a',b'e^{i\alpha}}$};
\end{tikzpicture}
    \caption{The relationship between annulus partition functions of the non-compact boson and annulus partition functions of $SU(2)_1$. In the first step, we use the definition of the generalized Neumann boundary conditions in \eqref{eqn:Nadef} to peel a gauging interface off the left boundary. In the second step, we fuse the gauging interface onto the generalized Neumann boundary condition on the right using \eqref{eqn:I*N}.}\label{fig:openstringbdymanip}
\end{center}
\end{figure}

By performing the sequence of manipulations illustrated in Figure \ref{fig:openstringbdymanip}, we can determine the annulus partition function of the non-compact boson theory, which we denote by $\mathcal{Z}^{\mathbf{N}_a\mathbf{N}_{a'}}_{\mathrm{NCB}}(\delta)$, from the annulus partition functions of the $SU(2)_1$ WZW model.
In particular, we have
\begin{align} \label{eq:annulus_NCB_1}
\begin{split}
    \mathcal{Z}^{\mathbf{N}_a\mathbf{N}_{a'}}_{\mathrm{NCB}}(\delta) &= \frac{\mathcal{N}}{2\pi} \int_0^{2\pi} d\alpha \, \mathcal{Z}^{\mathbf{B}_{a,b}\mathbf{B}_{a',b' e^{i\alpha}}}_{SU(2)_1}(\delta) \\
    &= \frac{1}{\eta(q)}\frac{\mathcal{N}}{2\pi}\int_{0}^{2\pi} d\alpha \sum_{n\in\mathbb{Z}}q^{\left(n + \frac{\gamma_{a, a'}(\alpha)}{2\pi}\right)^2} \,.
\end{split}
\end{align}
Here, $\mathcal{Z}^{\mathbf{B}_{a,b}\mathbf{B}_{a',b' e^{i\alpha}}}_{SU(2)_1}(\delta)$ is the annulus partition function of the $SU(2)_1$ WZW model with the boundary conditions $\mathbf{B}_{a,b}$ and $\mathbf{B}_{a',b' e^{i\alpha}}$ imposed at the two ends along the interval direction (see \eqref{eq:annulus_CB}) and
\begin{align} \label{eq:chiaa}
    \gamma_{a,a'}(\alpha)= \cos^{-1}\left[\mathrm{Re}\left(a'\bar a+e^{-i\alpha}\sqrt{(1-|a|^2)(1-|a'|^2)}\right)\right] \,.
\end{align}
(We take the principal value of the inverse cosine function, and $0 \leq \gamma_{a,a'}(\alpha) \leq \pi$.)

To deduce the structure of the interval Hilbert space $\mathcal{H}^{\mathbf{N}_a\mathbf{N}_{a'}}_{\NCB}$ from the annulus partition function \eqref{eq:annulus_NCB_1}, we will perform the following manipulations.

First, we can replace $\gamma_{a,a'}(\alpha)$ \eqref{eq:chiaa} in the second line of \eqref{eq:annulus_NCB_1} by
\begin{equation} \label{eq:saa}
    s_{a,a'} (\alpha) = \begin{cases} \gamma_{a,a'}(\alpha) &~~~~~\text{for $0 \leq \alpha < \pi$} \,,\\
    2\pi - \gamma_{a,a'}(\alpha) &~~~~~\text{for $\pi < \alpha \leq 2\pi$} \,,
    \end{cases}
\end{equation}
by redefining the dummy variable $n$ for $\pi \leq \alpha < 2\pi$.
For generic values of $a$ and $a'$, the function $s_{a,a'} (\alpha)$ is discontinuous at $\alpha = \pi$, and for $a = a' = 0$, it evaluates to $s_{0,0} (\alpha) = \alpha$.

Second, we combine the integral over $\alpha \in [0,2\pi ]$ and the sum over $n \in \bZ$ into a single integral over a real number $p = n + \frac{\alpha}{2\pi} \in \bR$, such that $n = \lfloor p \rfloor$ and $\alpha = 2\pi ( p - \lfloor p \rfloor )$.

After these manipulations, \eqref{eq:annulus_NCB_1} becomes
\begin{align} \label{eq:annulus_NCB_2}
\begin{split}
    &\mathcal{Z}^{\mathbf{N}_a\mathbf{N}_{a'}}_{\mathrm{NCB}}(\delta) = \mathcal{N} \int_{-\infty}^\infty dp \, \frac{q^{ f_{a,a'}(p)^2  }}{\eta (q)} \,,\\
    & f_{a,a'}(p) = \lfloor p \rfloor + \frac{1}{2\pi} s_{a,a'} \big( 2\pi ( p - \lfloor p \rfloor ) \big) \,.
\end{split}
\end{align}
Let us now deduce the structure of the interval Hilbert space $\mathcal{H}^{\mathbf{N}_a\mathbf{N}_{a'}}_{\NCB}$ from this annulus partition function.

On the first line of \eqref{eq:annulus_NCB_2}, the integrand $\frac{q^{ f_{a,a'}(p)^2  }}{\eta (q)}$ on the right-hand side is the character of the $c=1$ Virasoro algebra representation with conformal weight $h = f_{a,a'}(p)^2$, if $f_{a,a'}(p) \notin \frac{1}{2}\mathbb{Z}$.
Depending on $a$ and $a'$, we may have $f_{a,a'}(p) \in \frac{1}{2}\mathbb{Z}$ for some isolated values of $p$, in which case the Virasoro representation with $h = f_{a,a'}(p)^2$ is degenerate. 

To simplify the problem, let us suppose $a \neq \pm a'$.
Then, we have $f_{a,a'}(p) \notin \frac{1}{2}\mathbb{Z}$ for all $p \in \mathbb{R}$, and the annulus partition function is given by an integral over $c=1$ Virasoro characters,
\begin{equation} \label{eq:aneqap}
    \mathcal{Z}^{\mathbf{N}_a\mathbf{N}_{a'}}_{\mathrm{NCB}}(\delta) = \mathcal{N} \int_{-\infty}^\infty dp \, \chi_{|f_{a,a'}(p)|}(q) \,, ~~~~\text{if $a \neq \pm a'$} \,,
\end{equation}
where $\chi_{|f_{a,a'}(p)|}(q)$ denotes the character of the $c=1$ Virasoro algebra representation with conformal weight $h = f_{a,a'}(p)^2$, as reviewed in \eqref{eq:vir_char}.

From \eqref{eq:aneqap}, we claim that
\begin{equation} \label{eq:NCB_interval_Hilb}
    \mathcal{H}^{\mathbf{N}_a\mathbf{N}_{a'}}_{\NCB}\cong \int^{\oplus}_{(-\infty, \infty)} dp \, \mathcal{V}_{|f_{a,a'}(p)|} \,, ~~~~\text{if $a \neq \pm a'$} \,.
\end{equation}
Here, $\mathcal{V}_{|f_{a,a'}(p)|}$ is the irreducible representation of the $c=1$ Virasoro algebra with $h = f_{a,a'}(p)^2$.\footnote{In the Hilbert space \eqref{eq:NCB_interval_Hilb}, states are delta-function normalized, and when we take the trace over \eqref{eq:NCB_interval_Hilb}, each representation $\mathcal{V}_{f_{a,a'}(p)}$ contributes by $\mathcal{N} \chi_{ f(p)}$, where the factor of $\mathcal{N} = \delta (0)$ comes from the formally divergent norm of the primary state. See Appendix \ref{app:ncb_spec} and footnote \ref{ncb_part} for similar subtleties in a more familiar situation.\label{ncb_part2}} 

\begin{figure}
    \centering 
    \input{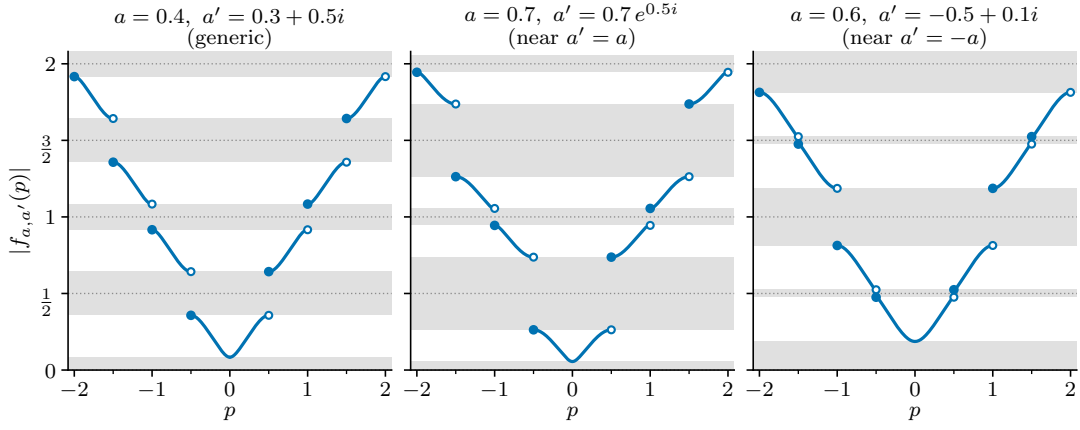}\caption{Plots of the open string spectrum between a pair $\mathbf{N}_a$, $\mathbf{N}_{a'}$ of generalized Neumann boundary conditions for various values of $a$ and $a'$.  Grey regions highlight gaps in the spectrum, which are around the locations where degenerate Virasoro representations would appear (dashed horizontal lines). When $a$ approaches $\pm a'$, some of these gaps close.}\label{fig:fplot}
\end{figure}

In \eqref{eq:NCB_interval_Hilb}, the function $f_{a,a'}(p)$ determines the spectrum of states in the interval Hilbert space $\mathcal{H}^{\mathbf{N}_a\mathbf{N}_{a'}}_{\NCB}$, or equivalently, the spectrum of boundary-condition-changing operators between two generalized Neumann boundary conditions.
Let us make a few observations about the properties of this function $f_{a,a'}(p)$.
First, for any $a$ and $a'$, it is a monotonically increasing function of $p$, and it satisfies $\lfloor p \rfloor \leq f_{a,a'}(p) \leq \lfloor p \rfloor + 1$.
Moreover, when $a \neq \pm a'$, the function $f_{a,a'}(p)$ has discontinuities at $p \in \frac{1}{2} \mathbb{Z}$.
These discontinuities are precisely such that in \eqref{eq:NCB_interval_Hilb}, the spectrum avoids all the degenerate representations of the $c=1$ Virasoro algebra. See Figure \ref{fig:fplot} for a plot.

When $a' = \pm a$, the interval Hilbert space is very similar, except that one must be more careful about what happens when $f_{a,a'}(p)\in\frac12\mathbb{Z}$. As one can imagine, the answer is simply to take
\begin{align}\label{eqn:HNaNpma}
    \mathcal{H}^{\mathbf{N}_a\mathbf{N}_{\pm a}}_{\NCB}\cong \int^{\oplus}_{(-\infty, \infty)} dp \, \widetilde{\mathcal{V}}_{|f_{a,\pm a}(p)|}, \ \ \ \ \ \ \ \ \widetilde{\mathcal{V}}_{\mu}\equiv \begin{cases}
        \mathcal{V}_\mu, & \mu\notin \frac12\mathbb{Z}, \\
        \bigoplus_{n\in\mathbb{Z}_{\geq 0}}\mathcal{V}_{\mu+n}, & \mu\in\frac12\mathbb{Z},
    \end{cases}
\end{align}
i.e.\ we understand $\widetilde{\mathcal{V}}_\mu$ to be an infinite direct sum of degenerate Virasoro representations when $\mu\in\frac12\mathbb{Z}$.

A special such case is when $a = a' = 0$, that is, when both of the boundary conditions are the ordinary Neumann boundary condition.
Then, we have $f_{0,0}(p) = p$, and the annulus partition function in \eqref{eq:annulus_NCB_2} is given by an integral of characters of the $\widehat{\mathfrak{u}}(1)$ current algebra (see \eqref{eq:u1_char}) that is preserved by the Neumann boundary condition.
This is consistent with the well-known spectrum of boundary vertex operators $e^{ip\Phi}$ on a Neumann boundary condition.

If we replace one of the generalized Neumann boundary conditions, say $\mathbf{N}_{a'}$, by a Dirichlet boundary condition $\mathbf{D}_{\Phi_0}$, the annulus partition function becomes
\begin{align}\label{eqn:openstringGND}
\begin{split}
    &\mathcal{Z}_{\NCB}^{\mathbf{N}_a\mathbf{D}_{\Phi_0}}(\delta) = \frac{1}{\eta(q)}\sum_{n\in\mathbb{Z}}q^{\left(n +  \frac{\gamma_{a,\Phi_0}}{2\pi} \right)^2} = \sum_{n\in\mathbb{Z}} \chi_{n +  \frac{\gamma_{a,\Phi_0}}{2\pi} }(q)  \,,\\
    & \ \ \ \ \ \ \ \gamma_{a,\Phi_0}=\cos^{-1} \left[\mathrm{Re}(ae^{-i\Phi_0})\right] \,,
\end{split}
\end{align}
where $\chi_{n +  \frac{\gamma_{a,\Phi_0}}{2\pi} }(q)$ denotes the character of the $c=1$ Virasoro algebra with conformal weight $h = \left(n +  \frac{\gamma_{a,\Phi_0}}{2\pi} \right)^2$, as reviewed in \eqref{eq:vir_char}.\footnote{In \eqref{eqn:openstringGND}, for any $|a| < 1$ and $\Phi_0 \in \bR$, we have $0< \gamma_{a,\Phi_0} < \pi$, and consequently, the virasoro representations at $h = \left(n +  \frac{\gamma_{a,\Phi_0}}{2\pi} \right)^2$ are all non-degenerate.}

From \eqref{eqn:openstringGND}, we deduce that the corresponding interval Hilbert space takes the form
\begin{align} \label{eq:GND_hilb}
    \mathcal{H}_{\NCB}^{\mathbf{N}_a\mathbf{D}_{\Phi_0}}\cong \bigoplus_{n\in\mathbb{Z}} \mathcal{V}_{n +  \frac{\gamma_{a,\Phi_0}}{2\pi} } \,,
\end{align}
where $\mathcal{V}_{n +  \frac{\gamma_{a,\Phi_0}}{2\pi} }$ is the irreducible $c=1$ Virasoro representation with conformal weight $h = \left(n +  \frac{\gamma_{a,\Phi_0}}{2\pi} \right)^2$, as in Appendix \ref{sec:su2_spec}.
The spectrum \eqref{eq:GND_hilb} is discrete. When $a= 0$, it is easy to see that $\gamma_{a,\Phi_0} = \frac{\pi}{2}$, and so we reproduce the known spectrum of boundary-condition-changing operators between the ordinary Neumann and Dirichlet boundary conditions.
(In particular, we see the familiar state with $h= \frac{1}{16}$.)

\paragraph{Closed string channel.} In order to compare these expressions to the closed string channel, we should take the modular S-transform. This can be facilitated by an application of Poisson resummation,
\begin{align}\label{eqn:Poissonresummation}
    \sum_{n\in\mathbb{Z}}q^{\left(n + \frac{\gamma}{2\pi} \right)^2}=\frac{1}{\sqrt{-2i\tau}}\sum_{k\in\mathbb{Z}}\tilde{q}^{k^2/4}e^{i k\gamma}, \ \ \ \ q=e^{2\pi i \tau}, \ \tilde{q}=e^{-2\pi i /\tau},
\end{align}
where, for us, $\tau = i\frac{\delta}{2}$.
Applying this to the annulus partition function \eqref{eq:annulus_NCB_1}, we find 
\begin{align}
\begin{split}
    \mathcal{Z}^{\mathbf{N}_a\mathbf{N}_{a'}}_{\mathrm{NCB}}(\delta) &= \frac{\mathcal{N}}{2\pi }\frac{1}{\sqrt{2}\eta(-1/\tau)}\sum_{m\in\mathbb{Z}}\tilde{q}^{m^2/4}\int_{0}^{2\pi} d\alpha \cos( m \gamma_{a,a'}(\alpha))\\
    &=\frac{\mathcal{N}}{2\pi \sqrt{2}\eta(-1/\tau)}\left(2\pi+2\sum_{m>0}\tilde{q}^{m^2/4}\int_{0}^{2\pi}d\alpha \cos( m \gamma_{a,a'}(\alpha))     \right)
\end{split}
\end{align}
where we have used the standard fact that $\eta(-1/\tau)=\sqrt{-i\tau}\eta(\tau)$. 

Now, let us focus on evaluating the integral over $\alpha$. Recall that the character of the spin-$J$ representation of $SU(2)$ is given by 
\begin{align}
    \mathrm{ch}_J[h]=\mathrm{Tr}_{R_J}[h]=\frac{\sin (2J+1)\theta}{\sin\theta}, \ \ \ \ \cos\theta\equiv \frac12\mathrm{Tr}[h],
\end{align}
where $\mathrm{Tr}[h]$ means the trace of $h$ in the fundamental representation, and $\mathrm{Tr}_{R_J}[h]$ denotes the trace of $h$ in the spin-$J$ representation. Application of basic trigonometric identities gives 
\begin{align}
    \frac12(\mathrm{ch}_{m/2}[h]-\mathrm{ch}_{(m-2)/2}[h])=\cos m\theta,
\end{align}
including for $m=1$ if we formally set $\mathrm{ch}_{-1/2}[h]=0$.
On the other hand, we note that 
\begin{align}
    \frac12\mathrm{Tr}[g(a)^{-1}U_\alpha g(a')U_{-\alpha}]=\frac12\mathrm{Tr}[g(a)^{-1}g(a',\alpha)]=\cos\gamma_{a,a'}(\alpha), 
\end{align}
where we  make use of the following definitions:
\begin{align}
    g(a,\psi )= \left(\begin{smallmatrix}a & e^{i\psi}\sqrt{1-|a|^2} \\ -e^{-i\psi}\sqrt{1-|a|^2} & \bar a\end{smallmatrix}\right), \ \ \ \ \  g(a)\equiv g(a,0), \ \ \ \ \ U_\alpha \equiv \left(\begin{smallmatrix} e^{i\alpha/2} & 0 \\ 0 & e^{-i\alpha/2}\end{smallmatrix}\right).
\end{align}  
Thus, it follows that we can rewrite 
\begin{align}
    \cos( m \gamma_{a,a'}(\alpha))= \frac12 (\mathrm{ch}_{m/2}[h]-\mathrm{ch}_{(m-2)/2}[h]), \ \ \ \ h=g(a)^{-1}g(a',\alpha).
\end{align}

Now, integrals over the characters can be expressed in terms of the Wigner D-matrices (see Equation \eqref{eq:su2_matrix_rep}),
\begin{align}
\begin{split}
   \int_0^{2\pi} d\alpha ~ \mathrm{ch}_J[h] &= \int_0^{2\pi} d\alpha \sum_{\mu,\nu,\rho,\sigma=-J}^J D_{\mu\nu}^J(g(a)^{-1})D_{\nu\rho}^J(U_\alpha)D_{\rho\sigma}^J(g(a'))D^J_{\sigma\mu}(U_{-\alpha}) \\
   &=\int_0^{2\pi} d\alpha\sum_{\mu,\rho=-J}^{J} D_{\mu\rho}^J(g(a)^{-1})e^{i\alpha\rho}D_{\rho\mu}^J(g(a'))e^{-i\alpha\mu} \\
   &=2\pi \sum_{\mu=-J}^J\overline{D^J_{\mu\mu}(g(a))}D^J_{\mu\mu}(g(a'))=: 2\pi \mathcal{D}_J(a,a').
\end{split}
\end{align}
Plugging this all in, we find that 
\begin{align}
\begin{split}
   Z^{\mathbf{N}_a\mathbf{N}_{a'}}_{\mathrm{NCB}}(\delta) &= \frac{\mathcal{N}}{\sqrt{2}\eta(-1/\tau)}\left(1+\sum_{m>0}\tilde{q}^{m^2/4}(\mathcal{D}_{m/2}(a,a')-\mathcal{D}_{(m-2)/2}(a,a'))\right) \\
    &= \frac{\mathcal{N}}{\sqrt{2}\eta(\tilde\tau)}\sum_{J\in\frac12\mathbb{Z}_{\geq 0}}\mathcal{D}_{J}(a,a')(\tilde{q}^{J^2}-\tilde{q}^{(J+1)^2} ).
\end{split}
\end{align}
This is exactly the answer one obtains if one takes the overlap of the boundary states calculated in Section \ref{subsubsec:NCBbdyGN}. So we find that the generalized Neumann boundary conditions of the non-compact boson satisfy open-closed duality, 
\begin{align}
    \mathrm{Tr}_{\mathcal{H}_{\NCB}^{\mathbf{N}_a\mathbf{N}_{a'}}}(e^{-\delta H_{\mathrm{open}}})=\langle \mathbf{N}_a|e^{-H_{\mathrm{closed}}/\delta}|\mathbf{N}_{a'}\rangle.
\end{align}

The case that one of the boundaries is Dirichlet can be treated similarly. For example, the overlap of boundary states \eqref{eq:ncb_gen_neu} and \eqref{eqn:DbdystateNCB} is given by
\begin{align}
\begin{split}
    \langle \mathbf{N}_a|e^{-H_{\mathrm{closed}}/\delta}|\mathbf{D}_{\Phi_0}\rangle &= \frac{1}{\sqrt{2}\eta(\tilde\tau)}\sum_{J\in\frac12\mathbb{Z}_{\geq 0}}\sum_{m=-J}^J(\tilde{q}^{J^2}-\tilde{q}^{(J+1)^2})\overline{D_{mm}^J(a,b)}e^{2 m i \Phi_0}\\
     &= \frac{1}{\sqrt{2}\eta(\tilde\tau)} \sum_{J\in\frac12\mathbb{Z}_{\geq0}}(\tilde q^{J^2}-\tilde{q}^{(J+1)^2})\mathrm{ch}_{J}[h], \ \ \ \ h=g(a)^{-1}g(e^{i\Phi_0}) .
\end{split}
\end{align}
After performing completely analogous manipulations to the Neumann-Neumann overlaps, one eventually finds 
\begin{align}
    \langle \mathbf{N}_a|e^{-H_{\mathrm{closed}}/\delta}|\mathbf{D}_{\Phi_0}\rangle=\frac{1}{\sqrt{2}\eta(\tilde\tau)}\sum_{m\in\mathbb{Z}}\tilde{q}^{m^2/4}e^{i m \gamma_{a,\Phi_0}}.
\end{align}
After applying Poisson resummation \eqref{eqn:Poissonresummation}, this can be rewritten as 
\begin{align}
     \langle \mathbf{N}_a|e^{-H_{\mathrm{closed}}/\delta}|\mathbf{D}_{\Phi_0}\rangle=\frac{1}{\eta(\tau)}\sum_{n\in\mathbb{Z}}q^{\left(n + \frac{\gamma_{a,\Phi_0}}{2\pi} \right)^2}=\mathrm{Tr}_{\mathcal{H}_{\NCB}^{\mathbf{N}_a\mathbf{D}_{\Phi_0}}}(e^{-\delta H_{\mathrm{open}}}).
\end{align}
which is precisely the answer obtained by tracing over the interval Hilbert space in \eqref{eqn:openstringGND}.

\subsection{Topological Lines}\label{subsec:explicit_con_lines}

\subsubsection{Action on States}

Here, we compute the action of all the topological lines in Table \ref{tab:simplelinesNCB} on the Hilbert space of the non-compact boson theory $\mathcal{H}_{\NCB}$, or equivalently, their action on local operators.
See Appendix \ref{app:ncb} for the review of the Hilbert space of the non-compact boson theory.

\paragraph{Non-invertible lines.}

Let us begin with the generic lines $\mathcal{X}^{\vec\theta}_{(x_L,x_R)}$. As an operator acting on the Hilbert space, we obtain 
\begin{align}
    \widehat{\mathcal{X}}^{\vec\theta}_{(x_L,x_R)} = \widehat{\mathcal{I}}_{U(1)}\circ \widehat{\mathcal{L}}_{\mathbf{g}(x_L,x_R,\vec\theta)}\circ \widehat{\mathcal{I}}_{U(1)}^\dagger,
\end{align}
where $\mathcal{L}_{\mathbf{g}(x_L,x_R,\vec\theta)}$ is the topological line operator of $SU(2)_1$ corresponding to the element $\mathbf{g}(x_L,x_R,\vec\theta)\in SO(4)$ defined in \eqref{eqn:doublecosetrep}.

Since the gauging interface map $\widehat{\mathcal{I}}_{U(1)}^\dagger$ annihilates every state $\ket{p}_{\NCB}$ with $p \notin \mathbb{Z}$ (see \eqref{eq:u1_interface_dual}), we immediately learn that the same is true for $\widehat{\mathcal{X}}^{\vec\theta}_{(x_L,x_R)}$.
That is,
\begin{equation}\label{eqn:vanishingXaction}
    \widehat{\mathcal{X}}^{\vec\theta}_{(x_L,x_R)}\ket{p}_{\NCB} = 0, \ \ \ \text{ for }p\not\in \mathbb{Z}.
\end{equation}
The action on other Virasoro primary states of the non-compact boson theory $\ket{p,\ell,\bar{\ell}}_{\NCB}$ \eqref{eq:ncb_vir_primaries} can be obtained by combining the action of the $U(1)$ gauging interface maps on Virasoro primary states in \eqref{eq:u1_interface_vir} and \eqref{eq:u1_interface_vir_adj}, with the action of the $SO(4)$ symmetry lines of the compact boson theory in \eqref{eq:so4_action}.
We get
\begin{align}\label{eqn:nonvanishingXaction}
\begin{split}
    &\widehat{\mathcal{X}}^{\vec\theta}_{(x_L,x_R)} \ket{p,\ell,\bar{\ell}}_{\NCB} \\
    &= {\cal N} \sum_{m = - J_{\mathrm{min}}}^{J_{\mathrm{min}}} D^{\frac{|p|}{2} + \ell}_{m, \frac{p}{2}} \big(g_L\big) D^{\frac{|p|}{2} + \bar{\ell}}_{\frac{p}{2},m} \big(g_R^{-1}\big) \ket{2m, \frac{|p|}{2} + \ell - |m| , \frac{|p|}{2} + \bar{\ell} - |m|}_{\NCB}, \ \text{for }p\in\mathbb{Z},
\end{split}
\end{align}
where $g_L\equiv g(x_Le^{i\theta_a/2},\theta_b/2)$, $g_R\equiv g(x_Re^{-i\theta_a/2},-\theta_b/2)$, and
 $J_{\mathrm{min}} \equiv \mathrm{min} \left( \frac{|p|}{2}  + \ell ,  \frac{|p|}{2}  + \bar{\ell} \right)$ (recall the definition of $g(a,\psi)$ from \eqref{eqn:gapsi}).
As written now, the right-hand side of \eqref{eqn:nonvanishingXaction} contains a divergent factor of $\mathcal{N} = \delta (0)$, and we will soon comment more on this.
 
As a quick consistency check, we can see that the conformal weights $h = (|p|/2 + \ell)^2$, $\bar{h} = (|p|/2 + \bar{\ell})^2$ are preserved.
Note that the operator carries an explicit factor of $\mathcal{N}$, and the topological lines $\mathcal{X}_{(x_L,x_R)}^{\vec\theta}$ all have infinite quantum dimensions.
We will comment more on this below.

The action of most of the other lines can be determined similarly, using the limiting behaviors from \eqref{eqn:limitingidNCBlines}. Indeed, it follows from \eqref{eqn:vanishingXaction} that
\begin{align}
\begin{split}
    &\mathcal{D}^L_a|p,\ell,\bar\ell\rangle_{\NCB}= \mathcal{D}^R_a|p,\ell,\bar\ell\rangle_{\NCB}=\widetilde{\mathcal{D}}^L_b|p,\ell,\bar\ell\rangle_{\NCB} = \widetilde{\mathcal{D}}^R_b|p,\ell,\bar\ell\rangle_{\NCB} \\
    &\hspace{.2in}=\mathcal{D}^L|p,\ell,\bar\ell\rangle_{\NCB}= \mathcal{D}^R|p,\ell,\bar\ell\rangle_{\NCB}=0, \ \  \text{for }p\notin\mathbb{Z}.
\end{split}
\end{align}
Also, by taking various limits of \eqref{eqn:nonvanishingXaction}, we find that, for $p\in\mathbb{Z}$,
\begin{align} \label{eq:D_ops}
\begin{split}
    \widehat{\mathcal{D}}_{a}^L|p,\ell,\bar\ell\rangle_{NCB} &= \mathcal{N}D^{\frac{|p|}{2}+\ell}_{\frac p2,\frac p2}(g(a,0))|p,\ell,\bar\ell\rangle_{\NCB} \\
    \widehat{\mathcal{D}}^R_a|p,\ell,\bar\ell\rangle_{NCB} &=\mathcal{N}D^{\frac{|p|}{2}+\bar\ell}_{\frac p2,\frac p2}(g(a,0)^{-1})|p,\ell,\bar\ell\rangle_{\NCB} \\
    {\widehat{\widetilde{\mathcal{D}}}}^L_b|p,\ell,\bar\ell\rangle_{NCB} &= \mathcal{N}(-1)^{\frac{|p|+p}{2}+\bar\ell}D_{-\frac p2,\frac p2}^{\frac{|p|}{2}+\ell}(g(x_L,\theta_b)) \ket{-p,\ell,\bar\ell}_{\NCB}, \ \ \ \ \ \ \ \ \  b=\sqrt{1-x_L^2}e^{i\theta_b} \\
    \widehat{\widetilde{\mathcal{D}}}^R_b|p,\ell,\bar \ell\rangle_{NCB} &= \mathcal{N}(-1)^{\frac{|p|+p}{2}+\ell}D^{\frac{|p|}{2}+\bar\ell}_{\frac p2,-\frac p2}(g(x_R,-\theta_b)^{-1})\ket{-p,\ell,\bar\ell}_{\NCB}, \ \ \ \ b=\sqrt{1-x_R^2}e^{-i\theta_b} \\
    \widehat{\mathcal{D}}^L|p,\ell,\bar\ell\rangle_{NCB} &= \mathcal{N}(-1)^{\ell}\delta_{p,0}|p,\ell,\bar\ell\rangle_{\NCB} \\
    \widehat{\mathcal{D}}^R|p,\ell,\bar\ell\rangle_{NCB} &= \mathcal{N}(-1)^{\bar\ell}\delta_{p,0}\ket{p,\ell,\bar\ell}_{\NCB},
\end{split}
\end{align}
where again, we are using the definition of $g(a,\psi)$ in \eqref{eqn:gapsi}. As one expects, these topological line operators do not mix different Virasoro primary states $|p,\ell,\bar\ell\rangle_{\NCB}$. Every $|p,\ell,\bar\ell\rangle_{\NCB}$ is an eigenstate with an eigenvalue proportional to $\mathcal{N}$.

Note that, consistent with $\mathcal{D}^L_a$ (resp.\  $\mathcal{D}^R_a$) being a purely left-moving (resp.\ purely right-moving) symmetry, its action on $|p,\ell,\bar\ell\rangle_{\NCB}$ is only sensitive to $\ell$ (resp.\ $\bar\ell$), and is insensitive to $\bar\ell$ (resp.\ $\ell$).

\paragraph{Infinite quantum dimensions.}

The action of the non-invertible topological line operators in \eqref{eqn:nonvanishingXaction} and \eqref{eq:D_ops} contains divergent factors proportional to $\cal N$, or to the volume of the target space of the non-compact boson.
If we define the quantum dimension of a line operator as its eigenvalue on the vacuum state of the theory, then formally all the non-invertible topological lines above have a divergent quantum dimension $\mathcal{N}$.\footnote{As we discussed in the paragraphs below \eqref{eq:u1_comp}, the quantum dimension of the gauging interface $\mathcal{I}_{U(1)}$ is formally $\mathcal{N}^{1/2}$. The fact that non-invertible lines of the non-compact boson theory all have quantum dimension $\mathcal{N}$ is consistent with the sandwich picture \eqref{eq:sandwich}, if we assume that the quantum dimensions multiply under a parallel fusion, which is the case for finite fusion categories.}

To better understand this, let us take $\widehat{\mathcal{D}}^L$ as an example. 
The operator is of the form
\begin{equation} \label{eq:DL}
    \widehat{\mathcal{D}}^L = \ket{0} \bra{0} + \cdots \,,
\end{equation}
where $\cdots$ represent its action on $\widehat{\mathfrak{u}}(1)$ descendants of the vacuum state $\ket{0}$. (We omitted the subscripts ``NCB'' in $\ket{0}$ and $\bra{0}$ for brevity.)

We can simplify further by restricting the theory to the zero mode of the non-compact boson.
Then, the problem is that of the quantum mechanics of a free particle (moving on a line), and we ignore $\cdots$ in \eqref{eq:DL}.

Formally, when $\widehat{\mathcal{D}}^L$ acts on the zero momentum state $\ket{0}$, the divergent norm $\langle 0 | 0 \rangle = \delta (0 ) = \mathcal{N}$ gives a factor of $\mathcal{N}$ in $\widehat{\mathcal{D}}^L \ket{0} = \mathcal{N} \ket{0}$.
This is analogous to the way we wrote the action of the non-invertible topological line operators in \eqref{eqn:nonvanishingXaction} and \eqref{eq:D_ops}.

A more careful statement is the following.
Let $\Psi_1 (p)$ and $\Psi_2 (p)$ be square-integrable and smooth wavefunctions (in momentum space), and let $\ket{\Psi_a} = \int dp \, \Psi_a (p) \ket{p}$, $a = 1,2$.
Then, the matrix element $\langle \Psi_1 | \widehat{\mathcal{D}}^L | \Psi_2 \rangle = \overline{\Psi_1 (0) } \Psi_2 (0)$ is finite and well-defined. 
Mathematically, $\widehat{\mathcal{D}}^L$ defines a sesquilinear form on the space of such square-integrable and smooth wavefunctions of the particle, which forms a dense subset of the Hilbert space of all square-integrable wavefunctions.

In the full non-compact boson theory, we expect that the non-invertible topological line operators in \eqref{eqn:nonvanishingXaction} and \eqref{eq:D_ops} similarly give rise to well-defined sesquilinear forms on a dense subset of the Hilbert space of the theory, but we will not attempt to fully specify the rigorous mathematical details.

\paragraph{Invertible lines.}

The remaining topological lines are the two families $\mathcal{M}_\xi$ and $\mathcal{M}_\xi^C$, both parameterized by $\xi\in\mathbb{R}$. Their analysis requires more care, since they cannot be obtained as the sandwich $\mathcal{S}(g_L,g_R)$ of a line in $SU(2)_1$. Rather, each $\mathcal{M}_\xi$ or $\mathcal{M}_\xi^C$ arises as a summand inside of some $\mathcal{S}(g_L,g_R)$. We of course anticipate that $\mathcal{M}_\xi$ corresponds to the shift symmetry of the non-compact boson, and that $\mathcal{M}_\xi^C$ corresponds to the composition of the shift symmetry with charge conjugation, but we will not assume this. Instead, we will consider the derivation of this as a consistency check on our general techniques.

Our calculations rest on a simple observation. If $\mathcal{L}$ is a topological line which decomposes into a direct sum $\mathcal{L}_1\oplus \mathcal{L}_2$, then the action of e.g.\ $\mathcal{L}_1$ on states in the $S^1$ Hilbert space is the same as the action of $\mathcal{L}$ with a topological point operator inserted on its world-line. Specifically, we should insert the point operator which is the identity morphism on $\mathcal{L}_1$ and the zero morphism on $\mathcal{L}_2$. The defining property of this point operator is that it is a \emph{central primitive idempotent}, i.e.\ it squares to itself, it commutes with every other point operator, and it cannot be expressed as a sum of other point operators with the same properties. 

In our non-compact boson example, the algebra of point operators on $\mathcal{S}(g_L,g_R)$ is always the group algebra of $U(1)$ when it is not trivial. The central primitive idempotents in a group algebra are well-known: they are in one-to-one correspondence with irreducible representations.

As an example, let us first discuss the lines $\mathcal{M}_\xi$. These are obtained as simple summands of the sandwich line,  
\begin{align}\label{eqn:sandwichMxi}
    \mathcal{S}\big(e^{i\xi\sigma_3},1\big) \cong \bigoplus_{Q\in\mathbb{Z}}\mathcal{M}_{\xi-2\pi Q}.
\end{align}
Recalling the definition of $x_{\alpha,\beta}$ in \eqref{eqn:xalphabeta}, it is straightforward to see that $x_{-\alpha,\alpha}$ defines a topological point operator on the line $\mathcal{S}(e^{i\xi \sigma_3},1)$ for any $\alpha\in U(1)$, and that these fuse according to $x_{-\alpha,\alpha}\otimes x_{-\beta,\beta}=x_{-\alpha-\beta,\alpha+\beta}$. One can then easily check that the point operator 
\begin{align}
    \frac{1}{2\pi}\int_0^{2\pi} d\alpha \, e^{-i\alpha Q}x_{-\alpha,\alpha}
\end{align}
is a central primitive idempotent for any $Q\in\mathbb{Z}$, so inserting it on the world-line of \eqref{eqn:sandwichMxi} projects onto simple summands.
Thus, we can determine the action of $\mathcal{M}_\xi$ on $S^1$ states as follows:
\begin{align}
\widehat{\mathcal{M}_{\xi}}
=\frac{1}{2\pi}\int_0^{2\pi} d\alpha\, e^{-i\alpha Q}\;
\begin{tikzpicture}[baseline={([yshift=-0.5ex]0,1.5)}]
  \fill[ncb] (-1.5,0) arc[start angle=180, end angle=360, x radius=1.5, y radius=0.18]
        -- (1.5,3.0) arc[start angle=0, end angle=-180, x radius=1.5, y radius=0.18] -- cycle;
  \draw[silh] (-1.5,0) -- (-1.5,3.0)  (1.5,0) -- (1.5,3.0);
  \draw[frame] (0,3.0) ellipse[x radius=1.5, y radius=0.18];
  \draw[frame] (-1.5,0) arc[start angle=180, end angle=360, x radius=1.5, y radius=0.18];
  \draw[tline, arcarrows] (-1.5,1.5) arc[start angle=180, end angle=360, x radius=1.5, y radius=0.18];
  \fill (0,1.32) circle (2.2pt);
  \node at (-0.75,2.2) {$\mathrm{NCB}$};
  \node[anchor=south west, inner sep=1pt] at (-.6,1.38) {$x_{-\alpha,\alpha}$};
  \node[anchor=north] at (0,1.2) {$\mathcal{S}\big(e^{i\theta_a\sigma_3},1\big)$};
\end{tikzpicture}
\;=\frac{1}{2\pi}\int_0^{2\pi} d\alpha\, e^{-i\alpha Q}\;
\begin{tikzpicture}[baseline={([yshift=-0.5ex]0,1.5)}]
  \fill[ncb] (-1.5,0) arc[start angle=180, end angle=360, x radius=1.5, y radius=0.18]
        -- (1.5,0.75) arc[start angle=0, end angle=-180, x radius=1.5, y radius=0.18] -- cycle;
  \fill[su]  (-1.5,0.75) arc[start angle=180, end angle=360, x radius=1.5, y radius=0.18]
        -- (1.5,2.25) arc[start angle=0, end angle=-180, x radius=1.5, y radius=0.18] -- cycle;
  \fill[ncb] (-1.5,2.25) arc[start angle=180, end angle=360, x radius=1.5, y radius=0.18]
        -- (1.5,3.0) arc[start angle=0, end angle=-180, x radius=1.5, y radius=0.18] -- cycle;
  \draw[silh] (-1.5,0) -- (-1.5,3.0)  (1.5,0) -- (1.5,3.0);
  \draw[frame] (0,3.0) ellipse[x radius=1.5, y radius=0.18];
  \draw[frame] (-1.5,0) arc[start angle=180, end angle=360, x radius=1.5, y radius=0.18];
  \draw[tline, midarrow] (0.55,0.5825) -- (0.55,1.3325);
  \draw[tline, midarrow] (0.55,1.3325) -- (0.55,2.0825);
  \draw[iface] (-1.5,0.75) arc[start angle=180, end angle=360, x radius=1.5, y radius=0.18];
  \draw[tline, arcarrowsat={0.22}{0.9}] (-1.5,1.5) arc[start angle=180, end angle=360, x radius=1.5, y radius=0.18];
  \draw[iface] (-1.5,2.25) arc[start angle=180, end angle=360, x radius=1.5, y radius=0.18];
  \node at (0.85,2.48)  {$\mathrm{NCB}$};
  \node at (-0.55,0.96) {$SU(2)_1$};
  \node at (0.85,0.25)  {$\mathrm{NCB}$};
  \node[ifacecol, anchor=south] at (-0.75,2.124) {$\mathcal{I}_{U(1)}$};
  \node[anchor=south east, inner sep=1pt] at (0.37,1.41) {$\mathcal{L}_{(e^{i\theta_a\sigma_3},1)}$};
  \node[anchor=west, inner sep=0.5pt] at (0.64,0.92) {$\mathcal{L}^{W}_{\alpha}$};
  \node[ifacecol, anchor=north] at (-0.75,0.564) {$\overline{\mathcal{I}}_{U(1)}$};
\end{tikzpicture}~~.
\end{align}
In equations, we have 
\begin{equation}
     \widehat{\mathcal{M}}_\xi= \frac1{2\pi}\int_0^{2\pi}d\alpha \, e^{-i\alpha Q} \, \widehat{\mathcal{I}}_{U(1)}^\alpha\circ \widehat{\mathcal{L}}_{(e^{i\theta_a\sigma_3},1)}^\alpha \circ \left( \widehat{\mathcal{I}}_{U(1)}^{\alpha} \right)^\dagger \,, \ \ \ \  \xi=\theta_a-2\pi Q \,, \ \ \theta_a \in [0, 2\pi) \,,
\end{equation}
where $\widehat{\mathcal{L}}_{(e^{i\theta_a \sigma_3},1)}^\alpha$ is a special case of \eqref{eq:twisted_symm}. Plugging in the explicit expressions, we find that
\begin{align} \label{eq:Rmop}
\begin{split}
    \widehat{\mathcal{M}}_{\xi} \ket{p}_{\NCB} &= \frac{1}{2\pi} \int_0^{2\pi} d\alpha \, e^{-i\alpha Q}\, \sum_{n\in\mathbb{Z}} \delta \left( p - \left(n + \frac{\alpha}{2\pi} \right) \right) e^{i\left(n + \frac{\alpha}{2\pi} \right)\theta_a} \ket{p}_{\NCB} \\
    &= \int_{-\infty}^\infty dp' e^{i\xi p'} \delta (p - p') \ket{p}_{\NCB}= e^{i\xi p} \ket{p}_{\NCB},
\end{split}
\end{align}
for all $p\in\mathbb{R}$.
The line operators $\widehat{\mathcal{M}}_{\xi}$ commute with the $\widehat{\mathfrak{u}}(1)$ current algebra of the theory and \eqref{eq:Rmop} determines their action on arbitrary states.
On the second line, we combined the infinite sum over $n$ and the integral over $\alpha$ into a single integral over $p' = n + \frac{\alpha}{2\pi}$.
As anticipated, we find that the topological lines $\mathcal{M}_{\xi}$ generate the $\mathbb{R}$ momentum symmetry of the non-compact boson.

Finally, let us discuss the last family of lines $\mathcal{M}_\xi^C$. Using an almost identical argument to the one we used to calculate $\widehat{\mathcal{M}}_\xi$, we find that
\begin{equation}
    \widehat{\mathcal{M}}^C_{\xi} = \frac{1}{2\pi} \int_0^{2\pi} d\alpha \, e^{-i\alpha Q} \, \widehat{\mathcal{I}}_{U(1)}^{2\pi - \alpha} \circ \widehat{\mathcal{L}}_{(e^{i\theta_b \sigma_3 }g_C,g_C)}^\alpha \circ \left( \widehat{\mathcal{I}}_{U(1)}^\alpha \right)^\dagger \,, \ \ \ \ \xi=\theta_b+2\pi Q\,, \ \ \theta_b\in [0,2\pi),
\end{equation}
where $g_C=\left(\begin{smallmatrix} 0 & 1 \\ -1 & 0 \end{smallmatrix}\right)$.
The operators $\widehat{\mathcal{L}}_{(e^{i\theta \sigma_3 }g_C,g_C)}^\alpha$ are given in \eqref{eq:twisted_symm_C}.
Plugging this in, we determine the action of $\widehat{\mathcal{M}}_\xi^C$ to be
\begin{align}
\begin{split}
    \widehat{\mathcal{M}}^C_{\xi} \ket{p}_{\NCB} &= \frac{1}{2\pi} \int_0^{2\pi} d\alpha \, e^{-i\alpha Q}\, \sum_{n\in\mathbb{Z}} \delta \left( p - \left(n + \frac{\alpha}{2\pi} \right) \right) e^{-i\left(n + \frac{\alpha}{2\pi} \right)\theta_b} \ket{-p}_{\NCB} \\
    &= \int_{-\infty}^\infty dp' e^{-i\xi p'} \delta (p - p') \ket{-p}_{\NCB} = e^{-i\xi p} \ket{-p}_{\NCB},
\end{split}
\end{align}
for all $p \in \mathbb{R}$. Moreover,
\begin{align}
\begin{split}
    \widehat{\mathcal{M}}^C_{\xi} \ket{p,\ell,\bar{\ell}}_{\NCB} = (-1)^{\ell + \bar{\ell}} e^{-i\xi p} \ket{-p , \ell , \bar{\ell}}_{
    \NCB} \,,
\end{split}
\end{align}
for $p \in \mathbb{Z}$ and $\ell, \bar{\ell} \in \mathbb{Z}_{\geq 0}$.
In particular, we note that $\mathcal{M}^C_{\xi = 0}$ corresponds to the topological line that generates the $\mathbb{Z}_2$ charge conjugation symmetry of the non-compact boson.

\paragraph{Remote detectability.}

Earlier, we advocated for a remote detectability condition in 2D conformal field theories. We used it in Section \ref{subsec:SU(2)1lines} to argue for the completeness of the known spectrum of topological lines in the $SU(2)_1$ WZW model. In turn, we argued that the topological line defects we found for the non-compact boson are complete because they came from these complete set of lines in $SU(2)_1$. 

We can also argue for the completeness of the topological lines in the non-compact boson theory more directly by invoking remote detectability. Indeed, from the action of topological line operators that we derived above, we find that only the vacuum state $\ket{0}_{\NCB}$ and its Virasoro descendants are transparent to every line in Table \ref{tab:simplelinesNCB}.
Therefore, invoking the remote detectability criterion, independent of the relation to the $SU(2)_1$ WZW model, we can again conclude that the set of (unitary) simple topological lines of the non-compact boson theory given in Table \ref{tab:simplelinesNCB} is complete.

On the other hand, if we take a proper tensor subcategory of the set of all lines, we expect there to be an operator in a nontrivial Virasoro representation which is charged under it.
For instance, take the proper subcategory of lines given by the momentum symmetry lines $\mathcal{M}_{\xi}$, $\xi \in \bR$, and one non-invertible line $\mathcal{D}^L$.
As we discussed in Section \ref{sec:fusion_TY}, they generate an $\mathbb{R}$ Tambara-Yamagami symmetry.
We find that infinitely many Virasoro primary states $\ket{0,\ell,\bar{\ell}}_{NCB}$ with $\ell \in 2\mathbb{Z}_{\geq 0}$ and $\bar{\ell} \in \mathbb{Z}_{\geq 0}$ are neutral under this $\mathbb{R}$ Tambara-Yamagami symmetry, which is consistent with our remote detectability criterion.

\subsubsection{Twisted Hilbert Spaces}

Finally, let us determine the twisted Hilbert spaces of these topological lines.
Equivalently, through the state-operator correspondence, we determine the spectrum of the defect operators attached at the endpoint of the topological lines.

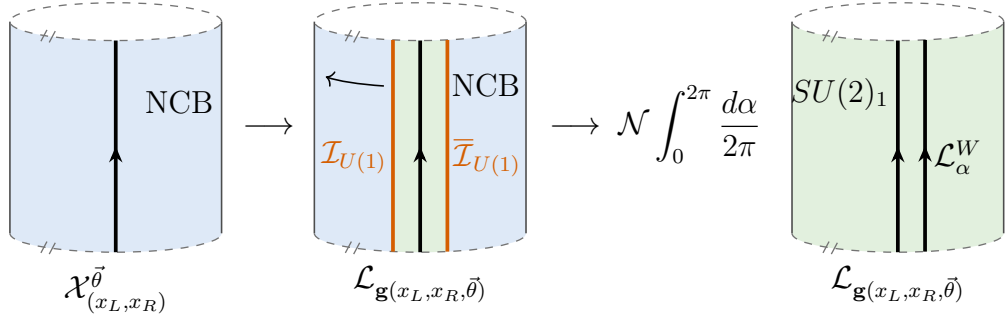
\begin{figure}
    \centering
    \begin{tikzpicture}[baseline={([yshift=-0.5ex]0,1.4)}]
  \fill[ncb] (-1.4,0) arc[start angle=180, end angle=360, x radius=1.4, y radius=0.25]
        -- (1.4,2.8) arc[start angle=0, end angle=-180, x radius=1.4, y radius=0.25] -- cycle;
  \draw[silh] (-1.4,0) -- (-1.4,2.8)  (1.4,0) -- (1.4,2.8);
  \draw[frame] (0,2.8) ellipse[x radius=1.4, y radius=0.25];
  \draw[frame] (-1.4,0) arc[start angle=180, end angle=360, x radius=1.4, y radius=0.25];
  \draw[black!55, line width=0.5pt] (-0.99,-0.259) -- (-0.91,-0.109);
  \draw[black!55, line width=0.5pt] (-0.88,-0.275) -- (-0.8,-0.125);
  \draw[black!55, line width=0.5pt] (-0.99,2.541) -- (-0.91,2.691);
  \draw[black!55, line width=0.5pt] (-0.88,2.525) -- (-0.8,2.675);
  \draw[tline, midarrow] (0,-0.25) -- (0,2.55);
  \node at (0.84,1.736) {$\mathrm{NCB}$};
  \node[below=1pt] at (0,-0.25) {$\mathcal{X}^{\vec{\theta}}_{(x_L,x_R)}$};
\end{tikzpicture}
$\;\longrightarrow\;$
\begin{tikzpicture}[baseline={([yshift=-0.5ex]0,1.4)}]
  \fill[ncb] (-1.4,0) arc[start angle=180, end angle=360, x radius=1.4, y radius=0.25]
        -- (1.4,2.8) arc[start angle=0, end angle=-180, x radius=1.4, y radius=0.25] -- cycle;
  \fill[su] (-0.36,-0.242) arc[start angle=255.099, end angle=284.901, x radius=1.4, y radius=0.25]
        -- (0.36,2.558) arc[start angle=284.901, end angle=255.099, x radius=1.4, y radius=0.25] -- cycle;
  \draw[silh] (-1.4,0) -- (-1.4,2.8)  (1.4,0) -- (1.4,2.8);
  \draw[frame] (0,2.8) ellipse[x radius=1.4, y radius=0.25];
  \draw[frame] (-1.4,0) arc[start angle=180, end angle=360, x radius=1.4, y radius=0.25];
  \draw[black!55, line width=0.5pt] (-0.99,-0.259) -- (-0.91,-0.109);
  \draw[black!55, line width=0.5pt] (-0.88,-0.275) -- (-0.8,-0.125);
  \draw[black!55, line width=0.5pt] (-0.99,2.541) -- (-0.91,2.691);
  \draw[black!55, line width=0.5pt] (-0.88,2.525) -- (-0.8,2.675);
  \draw[tline, midarrow] (0,-0.25) -- (0,2.55);
  \draw[iface] (-0.36,-0.242) -- (-0.36,2.558);
  \draw[iface] (0.36,-0.242) -- (0.36,2.558);
  \draw[->, line width=0.6pt] (-0.48,1.949) arc[start angle=249.949, end angle=206.766, x radius=1.4, y radius=0.25];
  \node at (0.868,1.96) {$\mathrm{NCB}$};
  \node[ifacecol, anchor=east, inner sep=1pt] at (-0.4,1.008) {$\mathcal{I}_{U(1)}$};
  \node[ifacecol, anchor=west, inner sep=1pt] at (0.4,1.008) {$\overline{\mathcal{I}}_{U(1)}$};
  \node[below=1pt] at (0,-0.25) {$\mathcal{L}_{\mathbf{g}(x_L,x_R,\vec{\theta})}$};
\end{tikzpicture}
$\displaystyle\;\longrightarrow\;
\mathcal{N}\int_0^{2\pi}\frac{d\alpha}{2\pi}\;$
\begin{tikzpicture}[baseline={([yshift=-0.5ex]0,1.4)}]
  \fill[su] (-1.4,0) arc[start angle=180, end angle=360, x radius=1.4, y radius=0.25]
        -- (1.4,2.8) arc[start angle=0, end angle=-180, x radius=1.4, y radius=0.25] -- cycle;
  \draw[silh] (-1.4,0) -- (-1.4,2.8)  (1.4,0) -- (1.4,2.8);
  \draw[frame] (0,2.8) ellipse[x radius=1.4, y radius=0.25];
  \draw[frame] (-1.4,0) arc[start angle=180, end angle=360, x radius=1.4, y radius=0.25];
  \draw[black!55, line width=0.5pt] (-0.99,-0.259) -- (-0.91,-0.109);
  \draw[black!55, line width=0.5pt] (-0.88,-0.275) -- (-0.8,-0.125);
  \draw[black!55, line width=0.5pt] (-0.99,2.541) -- (-0.91,2.691);
  \draw[black!55, line width=0.5pt] (-0.88,2.525) -- (-0.8,2.675);
  \draw[tline, midarrow] (0,-0.25) -- (0,2.55);
  \draw[tline, midarrow] (0.36,-0.242) -- (0.36,2.558);
  \node at (-0.77,1.848) {$SU(2)_1$};
  \node[anchor=west, inner sep=0.5pt] at (0.45,1.008) {$\mathcal{L}^{W}_{\alpha}$};
  \node[below=1pt] at (0,-0.25) {$\mathcal{L}_{\mathbf{g}(x_L,x_R,\vec{\theta})}$};
\end{tikzpicture}
    \caption{Twisted torus partition functions of the non-compact boson can be inferred from the twisted torus partition functions of $SU(2)_1$ using gauging interfaces.}
    \label{fig:XtwistedHS}
\end{figure}

We denote the twisted Hilbert space for the generic line defects $\mathcal{X}_{(x_L,x_R)}^{\vec\theta}$ by $\mathcal{H}_{\NCB}^{\mathcal{X}_{(x_L,x_R)}^{\vec\theta}}$, and the corresponding twisted torus partition function as 
\begin{align}
    \mathcal{Z}^{\mathcal{X}_{(x_L,x_R)}^{\vec\theta}}_{\NCB}(\tau,\bar\tau)=\mathrm{Tr}_{\mathcal{H}_{\NCB}^{\mathcal{X}_{(x_L,x_R)}^{\vec\theta}}}q^{L_0-c/24}\bar{q}^{\bar L_0-\bar c/24}, \ \ \ \ \ \ q=e^{2\pi i \tau}, \ \bar q=e^{-2\pi i \bar\tau}.
\end{align}
We can use the $U(1)$ gauging interface to relate the twisted torus partition functions of the non-compact boson to twisted torus partition functions of the $SU(2)_1$ WZW model. Indeed, using \eqref{eqn:IastI} and \eqref{eqn:Xsandwich} to perform the manipulations in Figure \ref{fig:XtwistedHS}, we obtain that\footnote{Similar to the cases in footnotes \ref{ncb_part} and \ref{ncb_part2}, the states in the Hilbert space \eqref{eqn:NCBtwistedHilbertspace1} are delta-function normalized, and an overall factor of $\mathcal{N}$ that appears in the fusion rule \eqref{eqn:IastI} is absorbed into the normalization of the states.} 
\begin{align}\label{eqn:NCBtwistedHilbertspace1}
\begin{split}
\mathcal{Z}_{\NCB}^{\mathcal{X}_{(x_L,x_R)}^{\vec\theta}} &= \mathcal{N}\int_{0}^{2\pi} \frac{d\alpha}{2\pi } ~ \mathcal{Z}_{SU(2)_1}^{\mathcal{L}_{\mathbf{g}(x_L,x_R,\vec\theta)\cdot (e^{i\alpha\sigma_3/2},e^{i\alpha\sigma_3/2})}} = \mathcal{N}\int_{0}^{2\pi}\frac{d\alpha}{2\pi }~\mathcal{Z}_{SU(2)_1}^{\mathcal{L}_{(g_L(\alpha),g_R(\alpha))}},
    \end{split}
\end{align}
where we have defined 
\begin{align}
   g_L(\alpha)=g(x_Le^{i(\theta_a+\alpha)/2},(\theta_b-\alpha)/2), \ \  \ \ \ g_R(\alpha)=g(x_Re^{-i(\theta_a-\alpha)/2},-(\theta_b+\alpha)/2),
\end{align}
with $g(a,\psi)$ defined as in \eqref{eqn:gapsi}.

This expression for the partition function can be made more explicit by plugging in the known twisted partition functions for the $SU(2)_1$ WZW model. (See Equation \eqref{eq:su2_twpart}.) Specifically, recall that the spectrum of the twisted Hilbert space of $(g_L,g_R)\in SO(4)$ depends only on the conjugacy class of $(g_L,g_R)$, and that any element is always conjugate to an element in the Cartan, 
\begin{align}
\begin{split}
    &\hspace{.45in}(g_L(\alpha),g_R(\alpha))\sim_{\mathrm{conj}} (e^{i\theta_L(\alpha)\sigma_3/2},e^{i\theta_R(\alpha)\sigma_3/2}), \\
    &\cos \frac{\theta_L(\alpha)}{2}\equiv x_L\cos\frac{\alpha+\theta_a}{2}, \ \ \ \cos\frac{\theta_R(\alpha)}{2}\equiv x_R\cos \frac{\alpha-\theta_a}{2}.
\end{split}
\end{align}
Then, substituting $\mathcal{Z}_{SU(2)_1}^{\mathcal{L}_{(g_L(\alpha),g_R(\alpha))}}$ with $\mathcal{Z}_{SU(2)_1}^{\mathcal{L}_{(e^{i\theta_L(\alpha)\sigma_3/2},e^{i\theta_R(\alpha)\sigma_3/2})}}$, we have that 
\begin{align}
    \mathcal{Z}_{\NCB}^{\mathcal{X}_{(x_L,x_R)}^{\vec\theta}}(\tau,\bar\tau) = \frac{\mathcal{N}}{2\pi|\eta(\tau)|^2} \int_0^{2\pi}d\alpha\sum_{\substack{n_L,n_R\in\mathbb{Z} \\ n_L\equiv n_R~\mathrm{mod}~2}}q^{\frac{p_L(\alpha)^2}{4}}\bar q^{\frac{p_R(\alpha)^2}{4}} ,
\end{align}
where $p_{L,R}(\alpha)\equiv n_{L,R}+\frac{\theta_{L,R}(\alpha)}{2\pi}$. 
As expected, the spectrum is continuous. 

\begin{figure}
    \centering
    \input{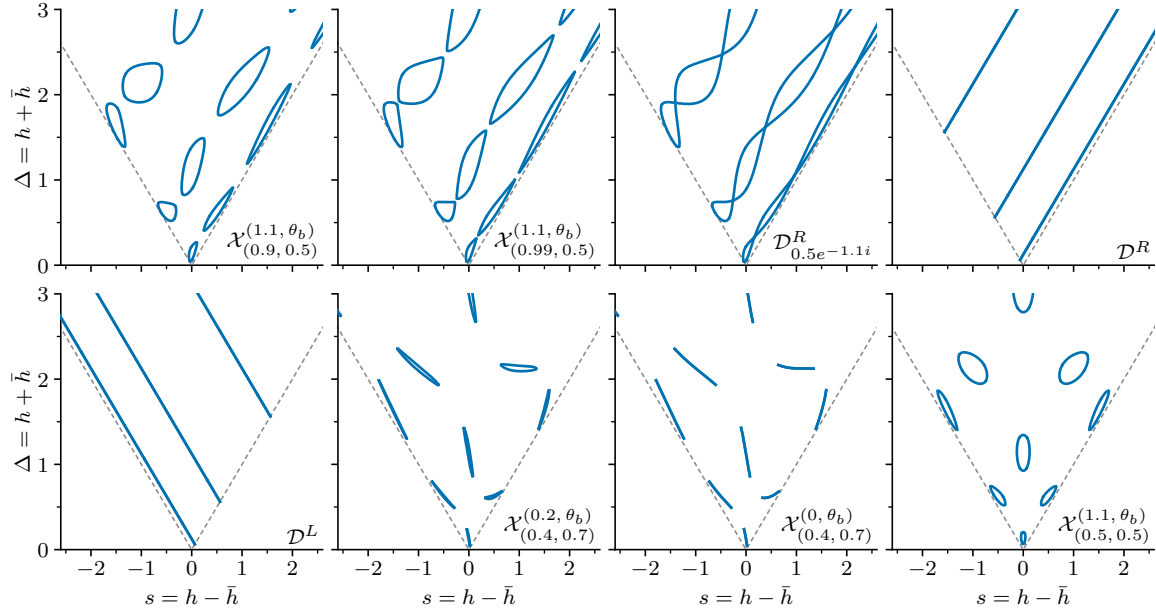}
    \caption{The twisted spectrum for a generic topological line defect $\mathcal{X}^{\vec\theta}_{(x_L,x_R)}$ consists of isolated bubbles in the $(s,\Delta)$ plane. As $(x_L,x_R)$ approaches the boundary of the square $[0,1]\times [0,1]$ these bubbles connect. In the limit that $(x_L,x_R)$ approaches one of the corners $(0,1)$ or $(1,0)$, the spectrum collapses to a tower of straight lines.}
    \label{fig:twistedspectrum}
\end{figure}

There is a useful rewriting of this twisted torus partition function, which is similar to our rewriting of the annulus partition functions of the non-compact boson in Section \ref{subsubsec:intervalannulus}. Namely, if we rewrite the sum over $n_L$ and $n_R$ as a sum over $n=(n_L+n_R)/2$ and $w=(n_L-n_R)/2$ and combine the sum over $n$ with the integral over $\alpha$ to produce an integral over $\mathbb{R}$, we obtain that 
\begin{align}
  \mathcal{Z}_{\NCB}^{\mathcal{X}_{(x_L,x_R)}^{\vec\theta}}(\tau,\bar\tau)=\frac{\mathcal{N}}{|\eta(\tau)|^2}\int_{-\infty}^\infty dp \sum_{w\in\mathbb{Z}}q^{\frac14 F_L(p,w)^2}\bar{q}^{\frac14F_R(p,w)^2},
\end{align}
where we have defined 
\begin{align}
\begin{gathered}
    F_L(p,w)=f_{x_L}(p+\theta_a/2\pi)+w, \ \ \ \ F_R(p,w)=f_{x_R}(p-\theta_a/2\pi)-w, \\
    f_x(u) = \lfloor u \rfloor +\frac1\pi \mathrm{cos}^{-1} \left[x\cos (\pi (u-\lfloor u \rfloor ))\right].
\end{gathered}
\end{align}
From this expression, we can readily extract the decomposition of the twisted Hilbert space into Virasoro representations: 
\begin{align}\label{eqn:XtwistedHS}
    \mathcal{H}_{\NCB}^{\mathcal{X}_{(x_L,x_R)}^{\vec\theta}}\cong \int_{(-\infty,\infty)}^{\oplus}dp \,\bigoplus_{w\in\mathbb{Z}}\widetilde{\mathcal{V}}_{\frac12 |F_L(p,w)|}\otimes \overline{\widetilde{\mathcal V}_{\frac12 |F_R(p,w)|}},
\end{align}
where $\widetilde{\mathcal{V}}_\mu$ are defined as in \eqref{eqn:HNaNpma}. The twisted spectra for various topological line defects are plotted in Figure \ref{fig:twistedspectrum}.

Notice that the spectrum of the $\mathcal{X}^{(\theta_a,\theta_b)}_{(x_L,x_R)}$-twisted Hilbert space does not depend on $\theta_b$. This is because conjugating $\mathcal{X}^{(\theta_a,\theta_b)}_{(x_L,x_R)}$ by an invertible momentum line $\mathcal{M}_\xi$ has the effect of shifting $\theta_b$ (cf.\ \eqref{eqn:MXfusion}), but keeps the spectrum invariant.

As before, we can obtain the twisted Hilbert spaces associated to the rest of the non-invertible topological lines of the non-compact boson by taking various limits of \eqref{eqn:XtwistedHS} in accordance with \eqref{eqn:limitingidNCBlines}. For example, the twisted sector Hilbert space of the $\mathbb{R}$ Tambara-Yamagami duality lines $\mathcal{D}^L$ and $\mathcal{D}^R$ are obtained by taking $(x_L,x_R)=(0,1)$ and $(x_L,x_R)=(1,0)$, respectively, whence we obtain
\begin{align}
\begin{split}
\mathcal{H}_{\NCB}^{\mathcal{D}^L}&\cong \int_{(-\infty,\infty)}^{\oplus}dp ~\mathcal{F}_{\mathrm{tw}}\otimes \overline{\mathcal{V}^{\widehat{\mathfrak{u}}(1)}_p} ,\\
    \mathcal{H}_{\NCB}^{\mathcal{D}^R}&\cong \int_{(-\infty,\infty)}^{\oplus}dp~\mathcal{V}^{\widehat{\mathfrak{u}}(1)}_p\otimes \overline{\mathcal{F}_{\mathrm{tw}}}.
\end{split}
\end{align}
Here, we have reorganized the answer using the fact that e.g.\ $\mathcal{D}^L$ preserves a $\widehat{\mathfrak{u}}(1)^+\otimes \overline{\widehat{\mathfrak{u}}(1)}$ chiral algebra, where $\widehat{\mathfrak{u}}(1)^+$ consists of the charge conjugation fixed points of the $\widehat{\mathfrak{u}}(1)$ Kac-Moody algebra. In particular, $\mathcal{F}_{\mathrm{tw}}$ is the unique $\mathbb{Z}_2^{\mathrm{C}}$-twisted module of the $\widehat{\mathfrak{u}}(1)$ Kac-Moody algebra. It decomposes into Virasoro representations as 
\begin{align}
    \mathcal{F}_{\mathrm{tw}}\cong \bigoplus_{n\in\mathbb{Z}_{\geq 0}}\mathcal{V}_{(2n+1)/4}.
\end{align}

The twisted sectors for the remaining non-invertible lines can be obtained similarly. The twisted-sector Hilbert spaces for the invertible lines $\mathcal{M}_\xi$ and $\mathcal{M}_\xi^C$ are known by more elementary means to be 
\begin{align}
\begin{split}
    \mathcal{H}_{\NCB}^{\mathcal{M}_\xi} &\cong \int_{(-\infty,\infty)}^{\oplus}dp~ \mathcal{V}^{\widehat{\mathfrak{u}}(1)}_{p+\frac{\xi}{2\pi}} \otimes \overline{\mathcal{V}}^{\widehat{\mathfrak{u}}(1)}_{p-\frac{\xi}{2\pi}}, \\
    \mathcal{H}_{\NCB}^{\mathcal{M}_\xi^C} &\cong \mathcal{F}_{\mathrm{tw}}\otimes \overline{\mathcal{F}_{\mathrm{tw}}}.
\end{split}
\end{align}

In principle, one could check that these twisted Hilbert spaces are consistent with modular covariance, analogously to our calculations in Section \ref{subsubsec:intervalannulus} for boundary conditions. Since the calculation is similar, we omit the details here.

\section*{Acknowledgements}

We sincerely thank Giovanni Galati, Igor Klebanov, Conghuan Luo, Juan Maldacena, Dmitri Nikshych, Sylvain Ribault, Hubert Saleur, Nathan Seiberg, Sahand Seifnashri, Yifan Wang,  Nicholas Warner, and Edward Witten for useful conversations.
YC was supported by NSF Grant PHY-2514611 and the Fund for Memberships in Natural Sciences at the Institute for Advanced Study.
HTL was supported in part by the Simons
Investigator Award No.~926198, the DOE grant DE-SC0011687 and the University of Southern California. BR gratefully acknowledges the Leinweber Physics Membership, the Fund for Memberships in Natural Sciences at the Institute for Advanced Study, and the DOE grant DE-SC0009988.

\section*{Statement of AI Use}

This work benefited from conversations with ChatGPT and Claude. Many of the figures were created by AI. Any calculations assisted by AI were rederived and written up by the authors, who take complete responsibility for the contents of this manuscript. In particular, every sentence of this manuscript was written by a human.

\appendix

\section{More on the \texorpdfstring{$SU(2)_1$}{SU(2)1} WZW Model} \label{app:su2}

In this appendix, we recall various details about the $SU(2)_1$ Wess-Zumino-Witten model. 

\subsection{Spectrum} \label{sec:su2_spec}

We begin with the spectrum of local operators.
Equivalently, through the state-operator correspondence, we discuss the Hilbert space of the theory quantized on a circle.
We will express the Hilbert space in three different ways, in terms of the representations of various (extended) chiral algebras of the theory.

\paragraph{In terms of $\widehat{\mathfrak{su}}(2)_1$ representations}

Let us first describe the Hilbert space of the theory in terms of the representations of the $\widehat{\mathfrak{su}}(2)_1$ current algebra (which is the maximally extended chiral algebra of this theory).
The $\widehat{\mathfrak{su}}(2)_1$ current algebra admits two integrable highest-weight representations labeled by the spin $j = 0, \frac{1}{2}$ \cite{DiFrancesco:1997nk}.
We denote these representations by $\mathcal{V}^{\widehat{\mathfrak{su}}(2)_1}_j$ for the left-moving $\widehat{\mathfrak{su}}(2)_1$ algebra, and by $\overline{\mathcal{V}}^{\widehat{\mathfrak{su}}(2)_1}_j$ for the right-moving $\widehat{\mathfrak{su}}(2)_1$ algebra.

The Hilbert space of the $SU(2)_1$ WZW model quantized on a circle, which we denote by $\mathcal{H}_{SU(2)_1}$, is given by
\begin{equation} \label{eq:Hilb_SU2}
    \mathcal{H}_{SU(2)_1} \cong \bigoplus_{j=0,\frac{1}{2}} \mathcal{V}^{\widehat{\mathfrak{su}}(2)_1}_j \otimes \overline{\mathcal{V}}^{\widehat{\mathfrak{su}}(2)_1}_j \,.
\end{equation}
In particular, the spectrum is diagonal, and the theory admits two primary operators with respect to its $\widehat{\mathfrak{su}}(2)_1$ current algebra.

\paragraph{In terms of $c=1$ Virasoro representations}

Let us denote an irreducible representation of the left-moving $c=1$ Virasoro algebra with conformal weight $h= P^2$ by $\mathcal{V}_{P}$, where $P \in \mathbb{R}_{\geq 0}$.
Similarly, an irreducible representation of the right-moving $c=1$ Virasoro algebra with conformal weight $\bar{h} = P^2$ will be denoted by $\overline{\mathcal{V}}_P$.
We denote the character of the representation $\mathcal{V}_{P}$ by $\chi_P (q)$, where
\begin{equation} \label{eq:vir_char}
    \chi_P (q) = \begin{cases}
        \frac{1}{\eta (q)} q^{P^2} \,, ~~~~ &P \notin \frac{1}{2} \mathbb{Z}_{\geq 0} \,,\\
        \frac{1}{\eta (q)} \left( q^{P^2} - q^{(P+1)^2} \right) \,, ~~~~ &P \in \frac{1}{2} \mathbb{Z}_{\geq 0} \,,
    \end{cases}
\end{equation}
and $\eta (q)=q^{1/24}\prod_{n=1}^\infty (1-q^n)$ is the Dedekind eta function.

The representation $\mathcal{V}^{\widehat{\mathfrak{su}}(2)_1}_j$ of the $\widehat{\mathfrak{su}}(2)_1$ current algebra decomposes as
\begin{equation} \label{eq:su2_to_vir}
    \mathcal{V}^{\widehat{\mathfrak{su}}(2)_1}_j \cong \bigoplus_{J \in j + \mathbb{Z}_{\geq 0}} R_J \otimes \mathcal{V}_J \,.
\end{equation}
On the right-hand side, $R_J$ denotes the $(2J+1)$-dimensional spin-$J$ irreducible representation of the Lie group $SU(2)$.
Similarly, for the right-moving representations, we have
\begin{equation} \label{eq:su2_to_vir_2}
    \overline{\mathcal{V}}^{\widehat{\mathfrak{su}}(2)_1}_j \cong \bigoplus_{\bar{J} \in j + \mathbb{Z}_{\geq 0}} R_{\bar{J}} \otimes \overline{\mathcal{V}}_{\bar{J}} \,.
\end{equation}
Substituting \eqref{eq:su2_to_vir} and \eqref{eq:su2_to_vir_2} into \eqref{eq:Hilb_SU2}, we obtain
\begin{equation} \label{eq:SU2_hilb_vir}
    \mathcal{H}_{SU(2)_1} \cong \bigoplus_{\substack{J,\bar{J}\in \frac{1}{2}\mathbb{Z}_{\geq 0}\\J-\bar{J}\in\mathbb{Z}}} R_J \otimes R_{\bar{J}} \otimes \mathcal{V}_{J}\otimes \overline{\mathcal{V}}_{\bar{J}} \,,
\end{equation}
as in \eqref{eqn:SU(2)1HSdecomp}.
This makes the action of the $c=1$ Virasoro algebra as well as that of the $SO(4) \cong [SU(2)_L \times SU(2)_R]/\mathbb{Z}_2$ global symmetry of the theory manifest. 

We denote the $c=1$ Virasoro primary states of the $SU(2)_1$ WZW model by
\begin{equation}\label{eq:cb_vir_primary}
\begin{aligned}
&\ket{J,\bar{J},m,\bar{m}}_{SU(2)_1}\,,
\qquad
J,\bar{J}\in \frac{1}{2}\mathbb{Z}_{\geq 0}\,,
\qquad
J-\bar{J}\in\mathbb{Z}\,,
\\
&m=-J,-J+1,\cdots,J\,,
\qquad
\bar m=-\bar J,-\bar J+1,\cdots,\bar J \,.
\end{aligned}
\end{equation}
Here, $m$ and $\bar{m}$, which are $SU(2)_L$ and $SU(2)_R$ quantum numbers, respectively, label basis states of $R_J$ and $R_{\bar{J}}$.
The state $\ket{J,\bar{J},m,\bar{m}}_{SU(2)_1}$ has conformal weights $h = J^2$ and $\bar{h} = \bar{J}^2$.

\paragraph{In terms of $\widehat{u}(1)$ representations}

As we discussed, the $SU(2)_1$ WZW model is equivalent to the free compact boson theory at the self-dual radius.
In that presentation (see \eqref{eq:cb_action}), the $\widehat{\mathfrak{u}}(1)$ current algebras generated by
the holomorphic and anti-holomorphic current operators are manifest,\footnote{By taking linear combinations of the (anti-)holomorphic currents \eqref{eq:u1_currents}, we obtain the momentum and winding symmetry currents.}
\begin{equation} \label{eq:u1_currents}
    j(z) = \sqrt{2}i\partial \phi (z) = \sum_{r\in\mathbb{Z}} a_r z^{-r-1} \,, \quad \bar{j}(\bar{z}) = \sqrt{2}i\bar{\partial} \phi (\bar{z}) = \sum_{r\in\mathbb{Z}} \bar{a}_r \bar{z}^{-r-1} \,,
\end{equation}
where the modes satisfy
\begin{equation} \label{eq:u1_algebra}
    [a_r, a_s] = [ \bar{a}_r, \bar{a}_s ] =  r \delta_{r+s , 0} \,.
\end{equation}
In the free boson presentation, the $\widehat{\mathfrak{su}}(2)_1$ current algebras of the theory are generated by the currents \eqref{eq:u1_currents} together with the (anti-)holomorphic vertex operators $e^{\pm i (\phi \pm \tilde{\phi} )}$, where $\tilde\phi$ is the T-dual boson.

An irreducible highest-weight representation of the left-moving $\widehat{\mathfrak{u}}(1)$ current algebra is labeled by the eigenvalue of $a_0$ acting on the primary state,
\begin{equation} \label{eq:a0}
    a_0 \ket{p} = \frac{p}{\sqrt{2}}  \ket{p} \,.
\end{equation}
(Note our convention for the eigenvalue $p$.) The $\widehat{\mathfrak{u}}(1)$ primary state $\ket{p}$ is annihilated by the action of $a_r$ with $r > 0$. 
The irreducible representation of the current algebra with the highest-weight state $\ket{p}$, which we denote by $\mathcal{V}_{p}^{\widehat{\mathfrak{u}}(1)}$, is the Fock space built on it by the action of oscillators $a_{-r}$ with $r >0$.
We will denote the irreducible representations of the right-moving $\widehat{u}(1)$ current algebra by $\overline{\mathcal{V}}_{\bar{p}}^{\widehat{\mathfrak{u}}(1)}$, which are obtained similarly.

The character of the representation $\mathcal{V}_{p}^{\widehat{\mathfrak{u}}(1)}$ is 
\begin{equation} \label{eq:u1_char}
    \chi^{\widehat{\mathfrak{u}}(1)}_p ( q) = \frac{q^{\frac{p^2}{4}}}{\eta (q)} \,.
\end{equation}
The $\widehat{\mathfrak{u}}(1)$ representations decompose into $c=1$ Virasoro representations as
\begin{equation} \label{eq:u1_to_vir}
    \mathcal{V}_{p}^{\widehat{\mathfrak{u}}(1)} = \begin{cases}
        \mathcal{V}_{\frac{|p|}{2}} &~~~ \text{if $p \notin \mathbb{Z}$} \,,\\
        \bigoplus_{\ell =0}^\infty \mathcal{V}_{\frac{|p|}{2} + \ell } &~~~ \text{if $p \in \mathbb{Z}$} \,,
    \end{cases} 
\end{equation}
and similarly for the right-moving representations.

In terms of the $\widehat{\mathfrak{u}}(1)$ representations, the Hilbert space of the $SU(2)_1$ WZW model is
\begin{equation} \label{eq:su2_hilb_u1}
    \mathcal{H}_{SU(2)_1} \cong \bigoplus_{n,w \in \mathbb{Z}} \mathcal{V}_{p_{n,w}}^{\widehat{\mathfrak{u}}(1)} \otimes \overline{\mathcal{V}}_{\bar{p}_{n,w}}^{\widehat{\mathfrak{u}}(1)} \,,
\end{equation}
where
\begin{equation}
    p_{n,w} = n+w \,, \quad \bar{p}_{n,w} = n-w\,.
\end{equation}
The integers $n,w \in \mathbb{Z}$ correspond to the charges under the 
$U(1) \times U(1)$ momentum and winding symmetries of the compact boson theory.
We denote the $\widehat{\mathfrak{u}}(1)$ primary states of the theory by $|n,w\rangle_{SU(2)_1}$.
Under the state-operator correspondence, they correspond to the vertex operators
\begin{equation} \label{eq:cb_u1_primaries}
    |n,w\rangle_{SU(2)_1} \longleftrightarrow e^{in\phi + iw \tilde{\phi}} \,,
\end{equation}
where $\tilde{\phi}$ is the dual compact boson.
The conformal weights of the state $\ket{n,w}_{SU(2)_1}$ are $h = \frac{1}{4} \left( n+w \right)^2$ and $\bar{h} = \frac{1}{4} \left( n-w \right)^2$.

The $\widehat{\mathfrak{u}}(1)$ primary states $\ket{n,w}_{SU(2)_1}$ are in particular also Virasoro primary states.
In \eqref{eq:cb_vir_primary}, we have denoted the Virasoro primary states of the theory in terms of their $SU(2)_L$ and $SU(2)_R$ quantum numbers.
We take our conventions such that the $SU(2)_L$ and $SU(2)_R$ quantum numbers of the $\widehat{\mathfrak{u}}(1)$ primary state $\ket{n,w}$ is
\begin{equation}
    J = \frac{|n+w|}{2} \,, \quad \bar{J} = \frac{|n-w|}{2} \,, \quad m =  \frac{n+w}{2} \,, \quad \bar{m} = \frac{n-w}{2} \,.
\end{equation}
In other words,
\begin{equation}
    \ket{n,w}_{SU(2)_1} \propto \Big| \frac{|n+w|}{2}, \frac{|n-w|}{2}, \frac{n+w}{2} , \frac{n-w}{2} \Big\rangle_{SU(2)_1} \,.
\end{equation}

\subsection{Boundary Conditions}\label{app:boundaries}

Next, we review details behind the boundary conditions of the theory. 

\subsubsection{Boundary States}

Let us recall the explicit conformal boundary states \cite{Cardy:2004hm,Cardy:1989ir} corresponding to the $SU(2)$ family of conformal boundary conditions $\mathbf{B}_{a,b} \equiv \mathbf{B}_g$ (with $g \in SU(2)$ as in \eqref{eq:gab}), which we discussed in Section \ref{subsec:CBbdy}.
Throughout, we will follow \cite{Gaberdiel:2001xm,Gaberdiel:2001zq}.

\paragraph{Virasoro Ishibashi States}

We first discuss the Ishibashi states of the theory \cite{Ishibashi:1988kg}.
Recall the spectrum of the Virasoro primary states of the $SU(2)_1$ WZW model given in \eqref{eq:cb_vir_primary}.
Each Virasoro primary state with a vanishing Lorentz spin, that is, those with $h=\bar{h} \Leftrightarrow J=\bar{J}$, gives rise to an Ishibashi state.
We denote them by
\begin{equation} \label{eq:cb_vir_ishibashi}
    |J , m, \bar{m} \rrangle_{SU(2)_1} \,, \quad J \in \frac{1}{2} \mathbb{Z}_{\geq 0} \,, \quad m,\bar{m} = -J ,-J+1, \cdots , J \,.
\end{equation}
They satisfy
\ie \label{eq:vir_is_overlap}
    {}_{SU(2)_1}\llangle J' , m', \bar{m}' |q^{\frac{1}{2} \left( L_0 + \bar{L}_0  - \frac{1}{12} \right)} |J , m, \bar{m}\rrangle_{SU(2)_1} = \delta_{JJ'}\delta_{mm'}\delta_{\bar{m}\bar{m}'}\chi_J (q) \,,
\fe
where $L_0$ and $\bar{L}_0$ are Virasoro generators, and $\chi_J (q)$ is the character of the Virasoro representation with $h= J^2$, as in \eqref{eq:vir_char}.

\paragraph{$SU(2)$ Family of Boundary States}

In terms of the Virasoro Ishibashi states \eqref{eq:cb_vir_ishibashi}, the conformal boundary states for the boundary conditions $\mathbf{B}_g$, which we denote by $| \mathbf{B}_g \rangle_{SU(2)_1}$, are
\begin{equation} \label{eq:su2_boundaries}
    | \mathbf{B}_g \rangle_{SU(2)_1} = 2^{-1/4} \sum_{J \in \frac{1}{2}\mathbb{Z}_{\geq 0}} \sum_{m,\bar{m} = -J}^J D^J_{m \bar{m}} (g) |J,m, \bar{m} \rrangle_{SU(2)_1} \,, \quad g \in SU(2) \,.
\end{equation}
Here, $D^J_{m \bar{m}} (g)$ are matrix elements of the Wigner D-matrices in the spin-$J$ representation.
For $g$ as in \eqref{eq:gab}, they are given by
\begin{align} \label{eq:su2_matrix_rep}
\begin{split}
    D^J_{m m'} (a,b) \equiv D^J_{m m'} (g) = \sum_{l=\mathrm{max}(0,m'-m)}^{\mathrm{min}(J-m,J+m')}&\frac{\sqrt{(J+m)!(J-m)!(J+m')!(J-m')!}}{l!(J-m-l)!\,(J+m'-l)!(m-m'+l)!}\\&~~~~~~~~\times a^{J+m'-l}(\bar{a})^{J-m-l}b^{m-m'+l}(-\bar{b})^{l}\,.
\end{split}
\end{align}
Some special cases that are used in the main text are
\begin{equation}
    D^J_{mm'} ( e^{i\theta} , 0 ) = e^{2i\theta m} \delta_{m, m'} \,, \quad D^J_{mm'} ( 0, e^{i\theta} ) = (-1)^{J-m} e^{2i\theta m} \delta_{m, -m'} \,.
\end{equation}

\paragraph{$\widehat{\mathfrak{u}}(1)$ Ishibashi States}

In the compact boson presentation of the theory \eqref{eq:cb_action}, it is common to express the conformal boundary states for the Dirichlet and Neumann boundary conditions using the Ishibashi states with respect to the $\widehat{\mathfrak{u}}(1)$ current algebra \cite{Ishibashi:1988kg}.
Let us describe these $\widehat{\mathfrak{u}}(1)$ Ishibashi states explicitly.

Recall the decomposition \eqref{eq:su2_hilb_u1} of the Hilbert space of the $SU(2)_1$ WZW model in terms of the representations of the $\widehat{\mathfrak{u}}(1)$ current algebra. 
For each $\widehat{\mathfrak{u}}(1)$ representation $\mathcal{V}_{p_{n,w}}^{\widehat{\mathfrak{u}}(1)} \otimes \overline{\mathcal{V}}_{\bar{p}_{n,w}}^{\widehat{\mathfrak{u}}(1)}$ satisfying $p_{n,w} = -\Omega \bar{p}_{n,w}$ with $\Omega = \pm$, there is a corresponding $\widehat{\mathfrak{u}}(1)$ Ishibashi state \cite{Ishibashi:1988kg,Cardy:2004hm}, which we denote by $|n,w\rrangle^{\widehat{\mathfrak{u}}(1),\Omega}_{SU(2)_1}$.
The superscript $\widehat{\mathfrak{u}}(1)$ is to emphasize that these are Ishibashi states with respect to the $\widehat{\mathfrak{u}}(1)$ current algebra. 

The condition $p_{n,w} = -\Omega \bar{p}_{n,w}$ is satisfied in the following cases: 
\begin{enumerate}[label=\arabic*)]
    \item $n \neq 0$, $w=0$, and $\Omega = -$,
    \item  $w\neq 0$, $n=0$, and $\Omega = +$,
    \item  $n=w=0$ with both signs $\Omega = \pm$ allowed.
\end{enumerate} 
Explicitly, the corresponding Ishibashi states are
\begin{align} \label{eq:u1_ishibashi}
\begin{split}
    &|n,0\rrangle^{\widehat{\mathfrak{u}}(1),-}_{SU(2)_1} = \exp \left( \sum_{r=1}^\infty \frac{1}{r} a_{-r}\bar{a}_{-r} \right) |n,0\rangle_{SU(2)_1} \,, \quad n \in \mathbb{Z}\setminus\{0\} \,, \\
    &|0,w\rrangle^{\widehat{\mathfrak{u}}(1),+}_{SU(2)_1} =  \exp \left( - \sum_{r=1}^\infty \frac{1}{r} a_{-r}\bar{a}_{-r} \right)|0,w\rangle_{SU(2)_1} \,,\quad w \in \mathbb{Z}\setminus\{0\}\,,\\
    &|0,0\rrangle^{\widehat{\mathfrak{u}}(1),\Omega}_{SU(2)_1} =  \exp \left( - \Omega \sum_{r=1}^\infty \frac{1}{r} a_{-r}\bar{a}_{-r} \right)|0,0\rangle_{SU(2)_1} \,,\quad \Omega = \pm \,.
\end{split}
\end{align}
Under the action of the $\widehat{\mathfrak{u}}(1)$ current algebra \eqref{eq:u1_algebra}, we have
\begin{equation}
    \left( a_r - \bar{a}_{-r} \right) |n,0;\alpha\rrangle^{\widehat{\mathfrak{u}}(1),-}_{SU(2)_1} = 0 \,, \quad \left( a_r + \bar{a}_{-r} \right) |0,w\rrangle^{\widehat{\mathfrak{u}}(1),+}_{SU(2)_1} = 0 \,, \quad \left( a_r + \Omega\bar{a}_{-r} \right) |0,0\rrangle^{\widehat{\mathfrak{u}}(1),\Omega}_{SU(2)_1} = 0 \,,
\end{equation}
for all $r \in \mathbb{Z}$.

The $\widehat{\mathfrak{u}}(1)$ Ishibashi states satisfy
\begin{align} \label{eq:u1_ish_overlap}
\begin{split}
    {}^{\widehat{\mathfrak{u}}(1),-}_{SU(2)_1}\llangle n,0 | q^{\frac{1}{2} \left( L_0 + \bar{L}_0  - \frac{1}{12} \right)} |n,0\rrangle^{\widehat{\mathfrak{u}}(1),-}_{SU(2)_1}& = \frac{q^{\frac{n^2}{4} }}{\eta (q)} = \chi^{\widehat{\mathfrak{u}}(1)}_n ( q) \,,\\
    {}^{\widehat{\mathfrak{u}}(1),+}_{SU(2)_1}\llangle 0,w | q^{\frac{1}{2} \left( L_0 + \bar{L}_0  - \frac{1}{12} \right)} |0,w\rrangle^{\widehat{\mathfrak{u}}(1),+}_{SU(2)_1}&= \frac{q^{\frac{w^2}{4}}}{\eta (q)}  = \chi^{\widehat{\mathfrak{u}}(1)}_w ( q) \,,\\
    {}^{\widehat{\mathfrak{u}}(1),\Omega}_{SU(2)_1}\llangle 0,0 | q^{\frac{1}{2} \left( L_0 + \bar{L}_0  - \frac{1}{12} \right)} |0,0\rrangle^{\widehat{\mathfrak{u}}(1),\Omega'}_{SU(2)_1}&= \begin{dcases}
        &\frac{1}{\eta (q)} ~~ \text{if $\Omega = \Omega'$} \,,\\
        &\frac{\eta(q)}{\eta(q^2)} ~~ \text{if $\Omega \neq \Omega' $} \,,
    \end{dcases}
\end{split}
\end{align}
where $\chi^{\widehat{\mathfrak{u}}(1)}_p ( q)$ is the character of the $\widehat{\mathfrak{u}}(1)$ current algebra representation as in \eqref{eq:u1_char}.
Ishibashi states built on distinct primary operators are orthogonal to each other.

The $\widehat{\mathfrak{u}}(1)$ Ishibashi states \eqref{eq:u1_ishibashi} decompose into Virasoro Ishibashi states \eqref{eq:cb_vir_ishibashi} as
\begin{align} \label{eq:u1_to_vir_ishibashi}
\begin{split}
    |n,0\rrangle^{\widehat{\mathfrak{u}}(1),-}_{SU(2)_1} &= \sum_{\ell = 0}^\infty |\tfrac{1}{2}|n| + \ell , \tfrac{1}{2} n, \tfrac{1}{2} n\rrangle_{SU(2)_1} \,,\\
    |0,w\rrangle^{\widehat{\mathfrak{u}}(1),+}_{SU(2)_1} &= \sum_{\ell = 0}^\infty (-1)^\ell |\tfrac{1}{2}|w| + \ell , \tfrac{1}{2} w, -\tfrac{1}{2} w\rrangle_{SU(2)_1} \,,
\end{split}
\end{align}
for all $n,w\in\mathbb{Z}$.

\paragraph{Dirichlet and Neumann Boundary States}

The compact boson theory, at any radius, admits an $S^1$-family of Dirichlet boundary conditions, parametrized by the boundary value of the compact boson,
as well as an $S^1$-family of Neumann boundary conditions, parametrized by the 
boundary theta angle 
as in \eqref{eq:cb_lag_bdy} (with $\mu = 0$).

In our conventions, the Dirichlet boundary conditions correspond to $\mathbf{B}_{e^{i\theta},0}$, and the Neumann boundary conditions correspond to $\mathbf{B}_{0,e^{i\eta}}$.
In terms of the $\widehat{\mathfrak{u}}(1)$ Ishibashi states \eqref{eq:u1_ishibashi}, they are given as
\begin{align} \label{eq:cb_DN}
\begin{split}
    \ket{\mathbf{B}_{e^{i\theta},0}}_{SU(2)_1} &= 2^{-1/4} \sum_{n\in\mathbb{Z}} e^{in\theta} |n,0 \rrangle^{\widehat{\mathfrak{u}}(1),-}_{SU(2)_1} \,,\\
    \ket{\mathbf{B}_{0,e^{i\eta}}}_{SU(2)_1} &= 2^{-1/4} \sum_{w\in\mathbb{Z}} e^{iw\eta} |0,w\rrangle^{\widehat{\mathfrak{u}}(1),+}_{SU(2)_1} \,.
\end{split}
\end{align}

\subsubsection{Interval Hilbert Spaces}
\label{subsubsec:SU(2)1interval}

Here, we recall the Hilbert space of the $SU(2)_1$ WZW model quantized on a finite interval, where at the two ends of the interval, we impose conformal boundary conditions $\mathbf{B}_{g_1}$ and $\mathbf{B}_{g_2}$, respectively, for two potentially different $g_1 , g_2 \in SU(2)$.

Let us denote the interval Hilbert space by $\mathcal{H}_{SU(2)_1}^{\mathbf{B}_{g_1} \mathbf{B}_{g_2}}$.
As is standard, we can infer the structure of $\mathcal{H}_{SU(2)_1}^{\mathbf{B}_{g_1} \mathbf{B}_{g_2}}$ by computing the partition function of the theory on $S^1 \times I$, where $I$ is an interval, in two different ways \cite{Cardy:2004hm}.

\paragraph{Closed String Channel}

Let us set the length of the interval $I$ to be 1, and the circumference of the circle $S^1$ to be $\delta >0$.
We write $q = e^{-\pi \delta}$, and $\tilde{q} = e^{-\frac{4\pi}{\delta}}$.
Denote the partition function on $S^1 \times I$ by ${\cal Z}_{SU(2)_1}^{\mathbf{B}_{g_1} \mathbf{B}_{g_2}} (q)$.
We first compute it from the overlap between two conformal boundary states,
\begin{equation}
    {\cal Z}_{SU(2)_1}^{\mathbf{B}_{g_1} \mathbf{B}_{g_2}} (q) = {}_{SU(2)_1}\langle \mathbf{B}_{g_1} | \tilde{q}^{\frac{1}{2} \left( L_0 + \bar{L}_0 - \frac{1}{12} \right)} | \mathbf{B}_{g_2} \rangle_{SU(2)_1} \,.
\end{equation}
Substituting the expressions for the conformal boundary states in \eqref{eq:su2_boundaries}, and the overlaps between the Ishibashi states in \eqref{eq:vir_is_overlap}, we obtain (see \cite{Gaberdiel:2001xm,Gaberdiel:2001zq} for more details)
\begin{align}
\begin{split}
    {\cal Z}_{SU(2)_1}^{\mathbf{B}_{g_1} \mathbf{B}_{g_2}} (q) &= \frac{1}{\sqrt{2}} \sum_{J \in \frac{1}{2} \mathbb{Z}} \mathrm{cos} (2J\alpha) \frac{\tilde{q}^{J^2}}{\eta (\tilde{q})} \,.
\end{split}
\end{align}
Here, $\alpha \sim \alpha+2\pi$ is defined by
\begin{equation}
    \mathrm{cos} (\alpha) = \frac{1}{2} \mathrm{Tr} \left[ g_1^{-1} g_2 \right] \,,
\end{equation}
with the trace on the right-hand side taken in the spin-$\frac{1}{2}$ defining representation of $SU(2)$.

Using the identity,
\begin{equation} \label{eq:modular_S}
    \frac{\tilde{q}^{\frac{1}{2}s^2}}{\eta (\tilde{q} )} = \int_{-\infty}^\infty dt \, e^{2\pi its} \, \frac{q^{\frac{1}{2}t^2}}{\eta (q )} \,,
\end{equation}
we can further rewrite the partition function as
\begin{equation} \label{eq:annulus_CB}
{\cal Z}_{SU(2)_1}^{\mathbf{B}_{g_1} \mathbf{B}_{g_2}} (q) = \sum_{n\in\mathbb{Z}} \frac{q^{\left( n- \frac{\alpha}{2\pi} \right)^2}}{\eta (q)} = \sum_{n\in\mathbb{Z}} \chi^{\widehat{\mathfrak{u}}(1)}_{2\left( n- \frac{\alpha}{2\pi} \right)} ( q) \,,
\end{equation}
where $\chi^{\widehat{\mathfrak{u}}(1)}_{2\left( n- \frac{\alpha}{2\pi} \right)} ( q)$ are the $\widehat{\mathfrak{u}}(1)$ characters \eqref{eq:u1_char}.

\paragraph{Open String Channel}

The partition function ${\cal Z}_{SU(2)_1}^{\mathbf{B}_{g_1} \mathbf{B}_{g_2}} (q)$ can alternatively be computed by taking a trace over the Hilbert space ${\cal H}_{SU(2)_1}^{\mathbf{B}_{g_1} \mathbf{B}_{g_2}}$,
\begin{equation}
    {\cal Z}_{SU(2)_1}^{\mathbf{B}_{g_1} \mathbf{B}_{g_2}} (q) = \mathrm{Tr}_{{\cal H}_{SU(2)_1}^{\mathbf{B}_{g_1} \mathbf{B}_{g_2}}} \left( q^{L_0 - \frac{1}{24}} \right) \,.
\end{equation}
Comparing this with \eqref{eq:annulus_CB}, we conclude that the interval Hilbert space can be decomposed into representations of the $\widehat{\mathfrak{u}}(1)$ current algebra as\footnote{For any $g_1 , g_2 \in SU(2)$, there exists a $\widehat{\mathfrak{u}}(1)$ subalgebra of the bulk left- and right-moving $\widehat{\mathfrak{su}}(2)_1$ algebras which acts on the interval Hilbert space ${\cal H}_{SU(2)_1}^{\mathbf{B}_{g_1} \mathbf{B}_{g_2}}$. For different choices of $g_1$ and $g_2$, we get different $\widehat{\mathfrak{u}}(1)$ subalgebras.}
\begin{equation} \label{eq:su2_int_hilb_u1}
    {\cal H}_{SU(2)_1}^{\mathbf{B}_{g_1} \mathbf{B}_{g_2}} \cong \bigoplus_{n \in \mathbb{Z}} \mathcal{V}^{\widehat{\mathfrak{u}}(1)}_{2\left( n- \frac{\alpha}{2\pi} \right)} \,.
\end{equation}
Using \eqref{eq:u1_to_vir}, we can further decompose the interval Hilbert space into $c=1$ Virasoro representations,
\begin{equation} \label{eq:su2_int_hilb}
    {\cal H}_{SU(2)_1}^{\mathbf{B}_{g_1} \mathbf{B}_{g_2}} \cong \begin{dcases}
        &\bigoplus_{n \in \mathbb{Z}} \mathcal{V}_{\left| n- \frac{\alpha}{2\pi} \right|} \,, ~~~~ \alpha \notin \pi\mathbb{Z} \,,\\
        &\bigoplus_{n \in \mathbb{Z}}\bigoplus_{\ell =0}^\infty \mathcal{V}_{\left| n- \frac{\alpha}{2\pi} \right| + \ell } \,, ~~~~ \alpha \in \pi\mathbb{Z} \,.
    \end{dcases} 
\end{equation}

\subsection{Topological Lines}

\subsubsection{Action on States}

Let us discuss how the $SO(4)$ global symmetry of the theory, generated by the topological lines $\mathcal{L}_{(g_L , g_R)}$ as in Section \ref{subsec:SU(2)1lines}, acts on the Hilbert space (on a circle).
When the topological line is supported on a spatial slice and acts on the Hilbert space as an operator, we denote it by $\widehat{\mathcal{L}}_{(g_L,g_R)}$.

On the Virasoro primary states given in \eqref{eq:cb_vir_primary}, the topological line operator $\widehat{\mathcal{L}}_{(g_L,g_R)}$ acts by
\begin{equation} \label{eq:so4_action}
    \widehat{\mathcal{L}}_{(g_L,g_R)} |J, \bar{J},m,\bar{m} \rangle_{SU(2)_1} = \sum_{m' = -J}^J \sum_{\bar{m}' = -\bar{J}}^{\bar{J}} D^J_{m' m} (g_L ) D^{\bar{J}}_{\bar{m}\bar{m}'} (g_R^{-1}) |J, \bar{J},m',\bar{m}' \rangle_{SU(2)_1} \,,
\end{equation}
where $D^J_{mm'} (g)$ were defined in \eqref{eq:su2_matrix_rep}.
We have $\widehat{\mathcal{L}}_{(g_L,g_R)} \widehat{\mathcal{L}}_{(g'_L,g'_R)} = \widehat{\mathcal{L}}_{(g_L g'_L,g_R g'_R)}$.

Our conventions are such that, in the compact boson description of the theory, the $U(1)$ momentum symmetry is generated by the lines $\mathcal{L}_\alpha^M \equiv \mathcal{L}_{(e^{\frac{i\alpha \sigma_3}{2}},e^{-\frac{i\alpha \sigma_3}{2}})}$, the $U(1)$ winding symmetry is generated by the lines $\mathcal{L}_\alpha^W \equiv \mathcal{L}_{(e^{\frac{i\alpha \sigma_3}{2}},e^{\frac{i\alpha \sigma_3}{2}})}$ (as in \eqref{eq:LW}), and the $\mathbb{Z}_2$ charge conjugation symmetry is generated by the line $\mathcal{L}^C\equiv \mathcal{L}_{(i\sigma_y , i \sigma_y)}$.

On the Virasoro primary states \eqref{eq:cb_vir_primary} and the $\widehat{\mathfrak{u}}(1)$ primary states \eqref{eq:cb_u1_primaries}, they act by
\begin{align}
\begin{split}
    &\widehat{\mathcal{L}}_\alpha^M|J, \bar{J},m,\bar{m} \rangle = e^{i\alpha (m + \bar{m} )} |J, \bar{J},m,\bar{m} \rangle \,, \quad \widehat{\mathcal{L}}_\alpha^M |n,w \rangle = e^{i\alpha n } |n,w \rangle\,,\\
    &\widehat{\mathcal{L}}_\alpha^W |J, \bar{J},m,\bar{m} \rangle = e^{i\alpha (m - \bar{m} )} |J, \bar{J},m,\bar{m} \rangle \,, \quad \widehat{\mathcal{L}}_\alpha^W |n,w \rangle = e^{i\alpha w } |n,w \rangle\,,\\
    &\widehat{\mathcal{L}}^C |J, \bar{J},m,\bar{m} \rangle = (-1)^{J+m} (-1)^{\bar{J} + \bar{m}} |J, \bar{J},-m,-\bar{m} \rangle \,,
\end{split}
\end{align}
which are special cases of \eqref{eq:so4_action}.

\subsubsection{Twisted Hilbert Spaces} \label{app:su2_twisted_hilb}

Let us denote the twisted Hilbert space of the theory on a circle, in the presence of a topological line defect $\mathcal{L}_{(g_L , g_R)}$ extending along the time direction,
by $\mathcal{H}_{SU(2)_1}^{\mathcal{L}_{(g_L , g_R)}}$.
Through the state-operator correspondence, states in $\mathcal{H}_{SU(2)_1}^{\mathcal{L}_{(g_L , g_R)}}$ are in correspondence with point operators attached to the topological line $\mathcal{L}_{(g_L , g_R)}$.

A standard way to understand such twisted Hilbert spaces is to consider the partition function of the theory on a torus $T^2 = S^1 \times S^1$, where the topological line defect $\mathcal{L}_{(g_L , g_R)}$ wraps around one of the $S^1$ factors \cite{Petkova:2000ip,Chang:2018iay}.

Let $\tau$ be the complex modulus of the torus. 
We write $q = e^{2\pi i \tau}$ and $\tilde{q} = e^{-\frac{2\pi i}{\tau}}$.
Denote the partition function of the theory on $T^2$ with the line defect $\mathcal{L}_{(g_L , g_R)}$ inserted by $\mathcal{Z}_{SU(2)_1}^{\mathcal{L}_{(g_L , g_R)}} (q, \bar{q})$.

In one way of computing the partition function, we take a trace over the twisted Hilbert space,
\begin{equation} \label{eq:su21_twisted_part_1}
    \mathcal{Z}_{SU(2)_1}^{\mathcal{L}_{(g_L , g_R)}} (q, \bar{q}) = \mathrm{Tr}_{\mathcal{H}_{SU(2)_1}^{\mathcal{L}_{(g_L , g_R)}} } \left( 
        q^{L_0 - \frac{1}{24}} \bar{q}^{\bar{L}_0 - \frac{1}{24}}
    \right) \,.
\end{equation}
On the other hand, swapping the role of the space and time directions, the same partition function can also be computed by
\begin{equation} \label{eq:su21_twisted_part_2}
    \mathcal{Z}_{SU(2)_1}^{\mathcal{L}_{(g_L , g_R)}} (q, \bar{q}) = \mathrm{Tr}_{\mathcal{H}_{SU(2)_1} } \left( 
        \widehat{\mathcal{L}}_{(g_L , g_R )} \tilde{q}^{L_0 - \frac{1}{24}} \bar{\tilde{q}}^{\bar{L}_0 - \frac{1}{24}}
    \right) \,,
\end{equation}
where now, the trace is over the untwisted Hilbert space $\mathcal{H}_{SU(2)_1}$, and the topological line is inserted as an operator $\widehat{\mathcal{L}}_{(g_L , g_R )}$ acting at a fixed time slice.

From the explicit action of the operator $\widehat{\mathcal{L}}_{(g_L , g_R )}$ on the Virasoro primary states given in \eqref{eq:so4_action}, we can compute the the partition function using the second expression \eqref{eq:su21_twisted_part_2}.
We obtain
\begin{equation} \label{eq:su21_twisted_trace}
     \mathcal{Z}_{SU(2)_1}^{\mathcal{L}_{(g_L , g_R)}} (q, \bar{q}) = 
        \sum_{\substack{J,\bar{J}\in \frac{1}{2}\mathbb{Z}_{\geq 0}\\J-\bar{J}\in\mathbb{Z}}} \mathrm{Tr}_{R_J} (g_L ) \mathrm{Tr}_{R_{\bar{J}}} (g_R^{-1} )  \chi_J (\tilde{q})
       \chi_{\bar{J}} (\bar{\tilde{q}})\,,
\end{equation}
where $R_J$ denotes the spin-$J$ representation of $SU(2)$, as before, and $\chi_J (q)$ is the character of the $c=1$ Virasoro representation in \eqref{eq:vir_char}.
Equation \eqref{eq:su21_twisted_trace} also follows directly from the decomposition of the untwisted Hilbert space in \eqref{eq:SU2_hilb_vir}.

Let us define $\alpha_L \sim \alpha_L + 2\pi$ and $\alpha_R \sim \alpha_R + 2\pi$ by
\begin{equation}
    \mathrm{cos}(\alpha_L ) = \frac{1}{2} \mathrm{Tr} (g_L ) \,, \quad \mathrm{cos}(\alpha_R ) = \frac{1}{2} \mathrm{Tr} (g_R^{-1} ) = \frac{1}{2} \mathrm{Tr} (g_R ) \,,
\end{equation}
where the traces are taken in the spin-$\frac{1}{2}$ representation of $SU(2)$.
Then, we have
\begin{equation}
    \mathrm{Tr}_{R_J} (g_L ) = \frac{\mathrm{sin}((2J+1)\alpha_L)}{\mathrm{sin}(\alpha_L )} \,, \quad \mathrm{Tr}_{R_{\bar{J}}} (g_R^{-1} ) = \frac{\mathrm{sin}((2\bar{J}+1)\alpha_R)}{\mathrm{sin}(\alpha_R )} \,.
\end{equation}
Substituting this and the expression for the Virasoro characters \eqref{eq:vir_char} into \eqref{eq:su21_twisted_trace}, we obtain
\begin{equation}
    \mathcal{Z}_{SU(2)_1}^{\mathcal{L}_{(g_L , g_R)}} (q, \bar{q}) = \frac{1}{|\eta(\tilde{q})|^2}
        \sum_{\substack{J,\bar{J}\in \frac{1}{2}\mathbb{Z}\\J-\bar{J}\in\mathbb{Z}}} \mathrm{cos}(2J\alpha_L )\mathrm{cos}(2\bar{J}\alpha_R ) {\tilde{q}}^{J^2} \bar{\tilde{q}}^{\bar{J}^2} \,.
\end{equation}

Finally, using \eqref{eq:modular_S}, we rewrite the partition function as
\begin{align}
\begin{split}
    \mathcal{Z}_{SU(2)_1}^{\mathcal{L}_{(g_L , g_R)}} (q, \bar{q}) &= \int_{-\infty}^\infty dt_L \int_{-\infty}^\infty dt_R 
    \sum_{\substack{J,\bar{J}\in \frac{1}{2}\mathbb{Z}\\J-\bar{J}\in\mathbb{Z}}}
    \frac{q^{\frac{t_L^2}{2}} \bar{q}^{\frac{t_R^2}{2}}}{|\eta(q)|^2} \mathrm{cos}(2J\alpha_L )\mathrm{cos}(2\bar{J}\alpha_R ) e^{2\pi i t_L (\sqrt{2}J )} e^{2\pi i t_R (\sqrt{2}\bar{J} )} \\
    &= \int_{-\infty}^\infty dt_L \int_{-\infty}^\infty dt_R 
    \sum_{\substack{J,\bar{J}\in \frac{1}{2}\mathbb{Z}\\J-\bar{J}\in\mathbb{Z}}}
    \frac{q^{\frac{t_L^2}{2}} \bar{q}^{\frac{t_R^2}{2}}}{|\eta(q)|^2}e^{2\pi i t_L (\sqrt{2}J ) + 2i J \alpha_L} e^{2\pi i t_R (\sqrt{2}\bar{J} ) + 2i\bar{J}\alpha_R} \,.
\end{split}
\end{align}
Then, using $\sum_{\ell \in \mathbb{Z}} e^{2\pi i x \ell} = \sum_{n\in\mathbb{Z}} \delta (x-n)$, we obtain, after a few manipulations,
\begin{equation} \label{eq:su2_twpart}
    \mathcal{Z}_{SU(2)_1}^{\mathcal{L}_{(g_L , g_R)}} (q, \bar{q}) = \sum_{n,w \in \mathbb{Z}} \frac{q^{\frac{1}{4}\left( n+w + \frac{\alpha_L}{\pi} \right)^2} \bar{q}^{\frac{1}{4}\left( n - w + \frac{\alpha_R}{\pi} \right)^2 }}{|\eta(q)|^2} = \sum_{n,w \in \mathbb{Z}} \chi^{\widehat{\mathfrak{u}}(1)}_{n+w + \frac{\alpha_L}{\pi}} (q) \chi^{\widehat{\mathfrak{u}}(1)}_{n - w + \frac{\alpha_R}{\pi}} (\bar{q}) \,.
\end{equation}

Comparing with \eqref{eq:su21_twisted_part_1}, we learn that the twisted Hilbert space decomposes into representations of the $\widehat{\mathfrak{u}}(1)$ current algebra as\footnote{For any choice of $g_L , g_R$, the twisted Hilbert space $\mathcal{H}_{SU(2)_1}^{\mathcal{L}_{(g_L , g_R)}}$ admits the action by one left-moving $\widehat{\mathfrak{u}}(1)$ algebar and one right-moving $\widehat{\mathfrak{u}}(1)$ algebra, which are subalgebras of the $\widehat{\mathfrak{su}}(2)_1$ current algebras of the theory. Depending on $g_L , g_R$, we get different $\widehat{\mathfrak{u}}(1)$ subalgebras.}
\begin{equation}
    \mathcal{H}_{SU(2)_1}^{\mathcal{L}_{(g_L , g_R)}} \cong \bigoplus_{n,w \in \mathbb{Z}} {\cal V}^{\widehat{\mathfrak{u}}(1)}_{n+w + \frac{\alpha_L}{\pi}} \otimes \overline{{\cal V}}^{\widehat{\mathfrak{u}}(1)}_{n-w + \frac{\alpha_R}{\pi}} \,.
\end{equation}
Using \eqref{eq:u1_to_vir}, it is straightforward to decompose the twisted Hilbert space into representations of the $c=1$ Virasoro algebra,
\begin{equation}
    \mathcal{H}_{SU(2)_1}^{\mathcal{L}_{(g_L , g_R)}} \cong \begin{dcases}
        &\bigoplus_{n,w \in \mathbb{Z}} {\cal V}_{\frac{1}{2} \left|n+w + \frac{\alpha_L}{\pi}\right|} \otimes \overline{{\cal V}}_{\frac{1}{2}\left|n-w + \frac{\alpha_R}{\pi}\right|} \,, ~~~~ \alpha_L \notin \pi \mathbb{Z} \,, \alpha_R \notin \pi \mathbb{Z} \,,\\
        &\bigoplus_{n,w \in \mathbb{Z}}\bigoplus_{\ell =0}^\infty {\cal V}_{\frac{1}{2} \left|n+w + \frac{\alpha_L}{\pi}\right| + \ell} \otimes \overline{{\cal V}}_{\frac{1}{2}\left|n-w + \frac{\alpha_R}{\pi}\right|} \,, ~~~~ \alpha_L \in \pi \mathbb{Z} \,, \alpha_R \notin \pi \mathbb{Z} \,,\\
        &\bigoplus_{n,w \in \mathbb{Z}}\bigoplus_{\bar{\ell} =0}^\infty {\cal V}_{\frac{1}{2} \left|n+w + \frac{\alpha_L}{\pi}\right|} \otimes \overline{{\cal V}}_{\frac{1}{2}\left|n-w + \frac{\alpha_R}{\pi}\right|+\bar{\ell}} \,, ~~~~ \alpha_L \notin \pi \mathbb{Z} \,, \alpha_R \in \pi \mathbb{Z} \,,\\
        &\bigoplus_{n,w \in \mathbb{Z}}\bigoplus_{\ell , \bar{\ell} =0}^\infty {\cal V}_{\frac{1}{2} \left|n+w + \frac{\alpha_L}{\pi}\right|+\ell} \otimes \overline{{\cal V}}_{\frac{1}{2}\left|n-w + \frac{\alpha_R}{\pi}\right|+\bar{\ell}} \,, ~~~~ \alpha_L \in \pi \mathbb{Z} \,, \alpha_R \in \pi \mathbb{Z} \,.
    \end{dcases}
\end{equation}

Particularly important for our purposes will be the twisted Hilbert spaces of the $U(1)$ winding symmetry topological lines, $\mathcal{L}_{\alpha}^W \equiv \mathcal{L}_{(e^{i\frac{\alpha}{2}\sigma_3 }, e^{i\frac{\alpha}{2}\sigma_3 } )}$.
This corresponds to $\alpha_L = \alpha_R = \frac{\alpha}{2}$, and we have
\begin{align} \label{eq:winding_tw}
\begin{split}
    \mathcal{H}_{SU(2)_1}^{\alpha} \equiv \mathcal{H}_{SU(2)_1}^{\mathcal{L}_{\alpha}^W} &\cong \bigoplus_{n,w \in \mathbb{Z}} {\cal V}^{\widehat{\mathfrak{u}}(1)}_{n+w + \frac{\alpha}{2\pi}} \otimes \overline{{\cal V}}^{\widehat{\mathfrak{u}}(1)}_{n-w + \frac{\alpha}{2\pi}}\\
    &\cong
        \bigoplus_{n,w \in \mathbb{Z}} {\cal V}_{\frac{1}{2} \left|n+w + \frac{\alpha}{2\pi}\right|} \otimes \overline{{\cal V}}_{\frac{1}{2}\left|n-w + \frac{\alpha}{2\pi}\right|} \,, ~~~~ \alpha \notin 2\pi \mathbb{Z} \,.
\end{split}
\end{align}
For $\alpha \in 2\pi\mathbb{Z}$, the line $\mathcal{L}_{\alpha}^W$ is trivial, and $\mathcal{H}_{SU(2)_1}^{\alpha} \cong \mathcal{H}_{SU(2)_1}$ is simply the untwisted Hilbert space discussed in Appendix \ref{sec:su2_spec}.
For $\alpha \notin 2\pi \mathbb{Z}$, we will denote the primary states in $\mathcal{H}_{SU(2)_1}^{\alpha}$ by
\begin{equation} \label{eq:tw_primaries}
    \ket{n,w;\alpha}_{SU(2)_1} \,, \quad n ,w \in \mathbb{Z} \,.
\end{equation}
They are primary states with respect to both the $\widehat{\mathfrak{u}}(1)$ current algebra as well as the Virasoro algebra, as in \eqref{eq:winding_tw}.
The conformal weights of the primary state $\ket{n,w;\alpha}_{SU(2)_1}$ are $h = \frac{1}{4}\left( n+w +\frac{\alpha}{2\pi}\right)^2$ and $\bar{h} = \frac{1}{4}\left( n - w +\frac{\alpha}{2\pi}\right)^2$.

\paragraph{Action of $[U(1) \times U(1) ]\rtimes \mathbb{Z}_2$ Symmetry on Twisted Hilbert Spaces}

Let us denote the action of a topological line $\mathcal{L}_{(e^{i\theta_L \sigma_3},e^{i\theta_R \sigma_3})}$, for some $\theta_{L,R} \sim \theta_{L,R} + 2\pi$, on the twisted Hilbert space $\mathcal{H}_{SU(2)_1}^{\alpha}$ by
\begin{equation}
    \widehat{\mathcal{L}}_{(e^{i\theta_L \sigma_3},e^{i\theta_R \sigma_3})}^\alpha : \mathcal{H}_{SU(2)_1}^\alpha \rightarrow \mathcal{H}_{SU(2)_1}^\alpha \,,
\end{equation}
with $\alpha \in (0,2\pi)$.
Explicitly, this is given by
\begin{equation} \label{eq:twisted_symm}
    \widehat{\mathcal{L}}_{(e^{i\theta_L \sigma_3},e^{i\theta_R \sigma_3})}^\alpha \ket{n,w;\alpha}_{SU(2)_1} = \exp\left[
        i\theta_L \left(
            n + w + \frac{\alpha}{2\pi}
        \right) - i \theta_R \left(
            n - w + \frac{\alpha}{2\pi}
        \right)
    \right] \ket{n,w;\alpha}_{SU(2)_1}\,.
\end{equation}
In particular, the operators $\widehat{\mathcal{L}}_{(e^{i\theta_L \sigma_3},e^{i\theta_R \sigma_3})}^\alpha$ with $\theta_L = - \theta_R$ define the action of $U(1)$ momentum symmetry on the twisted sectors of the $U(1)$ winding symmetry.

The $\mathbb{Z}_2$ charge conjugation symmetry maps a state in the twisted Hilbert space $\mathcal{H}_{SU(2)_1}^\alpha$ to a state in another twisted Hilbert space $\mathcal{H}_{SU(2)_1}^{-\alpha} \cong \mathcal{H}_{SU(2)_1}^{2\pi -\alpha}$, for $\alpha \notin 2\pi \mathbb{Z}$.
Let us denote this map by 
\begin{align}
\begin{split}
    \widehat{\mathcal{L}}^{C;\alpha} : \mathcal{H}_{SU(2)_1}^\alpha \rightarrow \mathcal{H}_{SU(2)_1}^{2\pi - \alpha} \,.
\end{split}
\end{align}
Explicitly, we have (for $0<\alpha < 2\pi$)
\begin{equation} \label{eq:C_twisted}
    \widehat{\mathcal{L}}^{C;\alpha}\ket{n,w;\alpha}_{SU(2)_1}= \ket{-n-1,-w;2\pi - \alpha}_{SU(2)_1} \,.
\end{equation}
If we compose \eqref{eq:twisted_symm} and \eqref{eq:C_twisted}, we obtain
\begin{equation}
    \widehat{\mathcal{L}}_{(e^{i\theta_L \sigma_3}g_C ,e^{i\theta_R \sigma_3}g_C )}^\alpha \equiv \widehat{\mathcal{L}}_{(e^{i\theta_L \sigma_3} ,e^{i\theta_R \sigma_3} )}^{2\pi - \alpha} \circ \widehat{\mathcal{L}}^{C;\alpha} : \mathcal{H}_{SU(2)_1}^\alpha \rightarrow \mathcal{H}_{SU(2)_1}^{2\pi - \alpha} \,,
\end{equation}
which acts on states by
\begin{align} \label{eq:twisted_symm_C}
\begin{split}
    &\widehat{\mathcal{L}}_{(e^{i\theta_L \sigma_3}g_C ,e^{i\theta_R \sigma_3}g_C )}^\alpha \ket{n,w;\alpha}_{SU(2)_1} \\
    &~~~= \exp\left[
        -i\theta_L \left(
            n + w + \frac{\alpha}{2\pi}
        \right) + i \theta_R \left(
            n - w + \frac{\alpha}{2\pi}
        \right)
    \right] \ket{-n-1,-w;2\pi - \alpha}_{SU(2)_1}\,.
\end{split}
\end{align}

\subsubsection{Action on Boundaries}

The topological lines act on the conformal boundary conditions by parallel fusion as in \eqref{eqn:SO(4)bdyact}.
More explicitly, from the action of the lines on the Hilbert space in \eqref{eq:so4_action} and the expression of the boundary states in \eqref{eq:su2_boundaries}, we have
\begin{equation} \label{eq:action_on_bdy_states}
    \widehat{\cal L}_{(g_L , g_R )} \ket{\mathbf{B}_g}_{SU(2)_1} =  \ket{\mathbf{B}_{g_L g g_R^{-1}}}_{SU(2)_1} \,.
\end{equation}

\subsection{Twisted-Sector Boundary States}

In the compact boson description of the $SU(2)_1$ WZW model, the Dirichlet boundary conditions $\mathbf{B}_{e^{i\theta},0}$, $\theta \sim \theta + 2\pi$, are invariant under the action of the $U(1)$ winding symmetry.
As was discussed in Section \ref{subsec:NCBconformalbdy}, this implies that the winding symmetry lines $\mathcal{L}_{\alpha}^W$ can topologically end on the Dirichlet boundary conditions \cite{Choi:2023xjw}.

In particular, we may consider the conformal boundary state of the Dirichlet boundary condition $\mathbf{B}_{e^{i\theta},0}$ in the twisted Hilbert space $\mathcal{H}_{SU(2)_1}^\alpha$ \eqref{eq:winding_tw} of the winding symmetry line $\mathcal{L}_{\alpha}^W$ \cite{Choi:2024tri,Fukusumi:2021zme,Ebisu:2021acm,Gaberdiel:2002qa,Fuchs:1997kt,Fuchs:2000cm,Fuchs:2004dz,Collier:2021ngi}.
We will denote such a twisted-sector boundary state by $\ket{\mathbf{B}_{e^{i\theta},0};\alpha}_{SU(2)_1}$.

Similar to ordinary boundary states, twisted-sector boundary states can be expressed in terms of linear combinations of twisted-sector Ishibashi states.

Recall the spectrum of the primary states \eqref{eq:tw_primaries} in the twisted Hilbert space $\mathcal{H}_{SU(2)_1}^\alpha$.
Just like ordinary Ishibashi states, for each twisted-sector primary state with a vanishing Lorentz spin (i.e., $h= \bar{h}$), there is a corresponding twisted-sector Ishibashi state.
In particular, among the primary states \eqref{eq:tw_primaries}, $\ket{n,0;\alpha}_{SU(2)_1}$ with arbitrary $n \in \mathbb{Z}$ satisfies the condition.

Let us denote the corresponding twisted-sector Ishibashi states by $|n,0;\alpha \rrangle_{SU(2)_1}$.
They are given by\footnote{The states $\ket{n,w;\alpha}$ for $\alpha \notin 2\pi \mathbb{Z}$, as we mentioned around \eqref{eq:tw_primaries}, are primaries with respect to both the Virasoro algebra and the $\widehat{\mathfrak{u}}(1)$ current algebra. Relatedly, the twisted-sector Ishibashi states \eqref{eq:tw_ish} are Ishibashi states with respect to both the $\widehat{\mathfrak{u}}(1)$ current algebra as well as the Virasoro algebra.}
\begin{equation} \label{eq:tw_ish}
    |n,0;\alpha \rrangle_{SU(2)_1} = \exp \left( \sum_{r=1}^\infty \frac{1}{r} a_{-r}\bar{a}_{-r} \right) |n,0;\alpha\rangle_{SU(2)_1} \,.
\end{equation}
They satisfy
\begin{equation}
    \left( a_r - \bar{a}_{-r} \right) |n,0;\alpha\rrangle_{SU(2)_1} = 0
\end{equation}
for all $r \in \mathbb{Z}$, and
\begin{equation}
    {}_{SU(2)_1}\llangle n,0;\alpha | q^{\frac{1}{2} \left( L_0 + \bar{L}_0  - \frac{1}{12} \right)} |n,0;\alpha\rrangle_{SU(2)_1}= \frac{q^{\frac{1}{4}\left( n + \frac{\alpha}{2\pi}\right)^2 }}{\eta (q)} = \chi_{\frac{1}{2}\left( n + \frac{\alpha}{2\pi}\right)} (q) \,.
\end{equation}

We claim that the twisted-sector boundary states $\ket{\mathbf{B}_{e^{i\theta},0};\alpha}_{SU(2)_1}$ are given by\footnote{The overall normalization of a twisted sector boundary state can be tuned by rescaling the topological junction between the topological line and the boundary \cite{Choi:2024tri}. This is unlike ordinary boundary states, where the overall normalization is important \cite{Cardy:2004hm}.}
\begin{equation} \label{eq:twisted_Dir}
    \ket{\mathbf{B}_{e^{i\theta},0};\alpha}_{SU(2)_1} = 2^{-1/4} \sum_{n\in\mathbb{Z}} e^{i\left(n + \frac{\alpha}{2\pi}\right)\theta} |n,0;\alpha \rrangle_{SU(2)_1} \,.
\end{equation}
In particular, for $\alpha = 0$, they reduce to the ordinary Dirichlet boundary states in the first line of \eqref{eq:cb_DN}.

Let us show that \eqref{eq:twisted_Dir} satisfies a twisted version of the usual Cardy condition \cite{Choi:2024tri}.
That is, we have
\begin{align} \label{eq:twisted_cardy}
\begin{split}
    {}_{SU(2)_1} \langle \mathbf{B}_{e^{i\theta_1},0};\alpha | \tilde{q}^{\frac{1}{2} \left( L_0 + \bar{L}_0 - \frac{1}{12} \right)} | \mathbf{B}_{e^{i\theta_2},0};\alpha \rangle_{SU(2)_1} = \mathrm{Tr}_{\mathcal{H}_{SU(2)_1}^{\mathbf{B}_{e^{i\theta_1},0}\mathbf{B}_{e^{i\theta_2},0}}} \left( \widehat{\mathcal{L}}_{\alpha}^{W;\theta_1 , \theta_2} \, q^{L_0 - \frac{1}{24}} \right) \,,
\end{split}
\end{align}
where $q = e^{-\pi \delta}$ and $\tilde{q} = e^{-4\pi /\delta}$.
On the right-hand side, $\mathcal{H}_{SU(2)_1}^{\mathbf{B}_{e^{i\theta_1},0}\mathbf{B}_{e^{i\theta_2},0}}$ is the interval Hilbert space of the theory quantized on an interval, with the Dirichlet boundary conditions $\mathbf{B}_{e^{i\theta_1},0}$ and $\mathbf{B}_{e^{i\theta_2},0}$ imposed on the two ends of the interval, and $\widehat{\mathcal{L}}_{\alpha}^{W;\theta_1 , \theta_2}$ denotes the action of the winding symmetry line $\mathcal{L}_{\alpha}^W$ on that Hilbert space.

From \eqref{eq:su2_int_hilb_u1}, we have
\begin{equation} \label{eq:DD_hilb}
    \mathcal{H}_{SU(2)_1}^{\mathbf{B}_{e^{i\theta_1},0}\mathbf{B}_{e^{i\theta_2},0}} \cong \bigoplus_{w\in\mathbb{Z}} \mathcal{V}^{\widehat{\mathfrak{u}}(1)}_{2\left( w+\frac{\theta_2 -\theta_1}{2\pi} \right)} \,,
\end{equation}
and $w$ labels the charge under the winding symmetry.

The left-hand side of \eqref{eq:twisted_cardy} is given by
\begin{align} \label{eq:twisted_Cardy2}
\begin{split}
    {}_{SU(2)_1} \langle \mathbf{B}_{e^{i\theta_1},0};\alpha | \tilde{q}^{\frac{1}{2} \left( L_0 + \bar{L}_0 - \frac{1}{12} \right)} | \mathbf{B}_{e^{i\theta_2},0};\alpha \rangle_{SU(2)_1} &= 2^{-1/2} \sum_{n \in \mathbb{Z}} e^{i\left( n + \frac{\alpha}{2\pi}\right) (\theta_1 - \theta_2 )} \, \frac{\tilde{q}^{\frac{1}{4}\left( n + \frac{\alpha}{2\pi}\right)^2 }}{\eta (\tilde{q})} \\
    &= \frac{1}{\eta (q)}\sum_{w\in\mathbb{Z}} e^{i\alpha w} q^{\left( w + \frac{\theta_2 -\theta_1}{2\pi} \right)^2} \,,
\end{split}
\end{align}
where we used \eqref{eq:twisted_Dir}, and performed a Poisson resummation from the first line to the second line.
On the other hand, from \eqref{eq:DD_hilb}, we see that the second line agrees with the right-hand side of \eqref{eq:twisted_cardy}.
Therefore, the twisted-sector boundary states \eqref{eq:twisted_Dir} satisfy the twisted Cardy condition \eqref{eq:twisted_cardy}.

\section{More on the Non-Compact Boson} \label{app:ncb}

In this appendix, we provide further details on the non-compact boson conformal field theory.

\subsection{Spectrum} \label{app:ncb_spec}

The non-compact boson theory has a continuous spectrum, and its Hilbert space, when quantized on a circle, is given by 
\begin{equation} \label{eq:ncb_u1_hilbert}
    \mathcal{H}_{\NCB} = \int^{\oplus}_{(-\infty, \infty)} dp \,\,\mathcal{V}_{p}^{\widehat{\mathfrak{u}}(1)} \otimes \overline{\mathcal{V}}_{p}^{\widehat{\mathfrak{u}}(1)} \,.
\end{equation}
The primary states with respect to the $\widehat{\mathfrak{u}}(1)$ current algebra will be denoted by $\ket{p}_{\NCB}$, with $p \in \mathbb{R}$.
The conformal weights of the state $\ket{p}_{\NCB}$ are $h= \bar{h} = \frac{1}{4}p^2  $.
Under the state-operator correspondence, we have
\begin{equation} \label{eq:ncb_p}
    \ket{p}_{\NCB} \longleftrightarrow e^{ip\Phi} \,.
\end{equation}
The states are normalized such that
\begin{equation} \label{eq:p_normalization}
    {}_{\NCB}\langle p' | p \rangle_{\NCB} = \delta (p - p' ) \,.
\end{equation}

We find it convenient to define
\begin{equation} \label{eq:R_volume}
    \mathcal{N} = {}_{\NCB}\langle p | p \rangle_{\NCB} =  \delta (0) \,,
\end{equation}
where $2\pi \mathcal{N}$ corresponds to the volume of the target space of the non-compact boson.
Although $\mathcal{N}$ is divergent, formally keeping track of it turns out to be useful.

From the decomposition of the $\widehat{\mathfrak{u}}(1)$ representations into Virasoro representations given in \eqref{eq:u1_to_vir}, we find that for $p\in\mathbb{Z}$, there are additional Virasoro primary states which are $\widehat{\mathfrak{u}}(1)$ descendants.
We denote them by
\begin{align} \label{eq:ncb_vir_primaries}
\begin{split}
    \ket{p,\ell,\bar{\ell}}_{\NCB} \,, \quad p \in \mathbb{Z} \,, \quad \ell,\bar{\ell}\in\mathbb{Z}_{\geq 0} \,, \quad h= \left( \frac{|p|}{2} + \ell \right)^2 \,, \quad \bar{h}= \left( \frac{|p|}{2} + \bar{\ell} \right)^2 \,.
\end{split}
\end{align}

\subsection{Ishibahi States}

Among the Virasoro primary states of the non-compact boson theory, $\ket{p}_{\NCB}$ with $p \notin \mathbb{Z}$ and $\ket{p,\ell , \bar{\ell}}_{\NCB}$ with $p \in \mathbb{Z}$ and $\ell=\bar \ell \in \mathbb{Z}_{\geq 0}$ have vanishing Lorentz spins, and they give rise to Virasoro Ishibashi states.

We denote these Ishibashi states by
\begin{align} \label{eq:vir_ishibashi_ncb}
\begin{split}
    &|p\rrangle_{\NCB} \,, \quad p \notin \mathbb{Z}\,, \\ 
    &|p,\ell\rrangle_{\NCB} \,, \quad p\in\mathbb{Z} \,, \quad \ell \in \mathbb{Z}_{\geq 0 }\,.
\end{split}
\end{align}
Their overlaps are given by
\begin{align}
\begin{split}
    {}_{\NCB}\llangle p | p' \rrangle_{\NCB} &= \delta(p - p' ) \frac{q^{\frac{p^2}{4}}}{\eta (q)} \,, \quad p, p' \notin \mathbb{Z}\,, \\
    {_{\NCB}} \llangle p ,\ell | p' ,\ell' \rrangle_{\NCB} &= 
    \mathcal{N} \delta_{p  p'} \delta_{\ell \ell'} \frac{1}{\eta (q)}\left(q^{\left( \frac{|p|}{2} + \ell\right)^2} - q^{\left( \frac{|p|}{2} + \ell + 1 \right)^2} \right)\,, \quad p,p'\in\mathbb{Z} \,, \quad \ell,\ell' \in \mathbb{Z}_{\geq 0 } \,.
\end{split}
\end{align}
Moreover, similar to \eqref{eq:u1_ishibashi}, we can also define Ishibashi states with respect to the $\widehat{\mathfrak{u}}(1)$ current algebra.
We write them as
\begin{align} \label{eq:u1_ishibashi_ncb}
\begin{split}
    &|p\rrangle^{\widehat{\mathfrak{u}}(1),-}_{\NCB} = \exp \left( \sum_{r=1}^\infty \frac{1}{r} a_{-r}\bar{a}_{-r} \right) |p\rangle_{\NCB} \,, \quad p \in \mathbb{R}\setminus\{0\} \,,\\
    &|0\rrangle^{\widehat{\mathfrak{u}}(1),\Omega}_{\NCB} =  \exp \left( - \Omega \sum_{r=1}^\infty \frac{1}{r} a_{-r}\bar{a}_{-r} \right)|0\rangle_{\NCB} \,,\quad \Omega = \pm \,.
\end{split}
\end{align}
For $p \notin \mathbb{Z}$, we have $|p\rrangle^{\widehat{\mathfrak{u}}(1),-}_{\NCB} = |p\rrangle_{\NCB}$, and for $p \in \mathbb{Z}$, $|p\rrangle^{\widehat{\mathfrak{u}}(1),-}_{\NCB}$ decompose into an infinite linear combination of $|p,\ell\rrangle_{\NCB}$ with $\ell \in \mathbb{Z}_{\geq 0}$.

\bibliographystyle{ytphys}
\bibliography{ref}

\end{document}